\documentclass[a4paper]{PoS}
\usepackage[italian,english]{babel}
\usepackage{slashed}
\usepackage{amssymb}
\usepackage{amsfonts}
\usepackage{amsmath}
\usepackage{fancyhdr}
\usepackage{lineno}
\usepackage{subfigure}

\newcommand{\be}{\begin{equation}}
\newcommand{\ee}{\end{equation}}
\newcommand{\beq}{\begin{equation}}
\newcommand{\eeq}{\end{equation}}
\newcommand{\beqa}{\begin{eqnarray}}
\newcommand{\eeqa}{\end{eqnarray}}

\newcommand{\bea}{\begin{eqnarray}}
\newcommand{\eea}{\end{eqnarray}}
\newcommand{\nn}{\nonumber}
\newcommand{\eps}{\epsilon}
\newcommand{\cS}{\mathcal{S}}
\newcommand{\veps}{\varepsilon}
\newcommand{\lsim}{\mathrel{\mathop{\kern 0pt \rlap
  {\raise.2ex\hbox{$<$}}}
  \lower.9ex\hbox{\kern-.190em $\sim$}}}
\newcommand{\gsim}{\mathrel{\mathop{\kern 0pt \rlap
  {\raise.2ex\hbox{$>$}}}
  \lower.9ex\hbox{\kern-.190em $\sim$}}}

\newcommand{\la}{\lambda}

\newcommand{\rra}{\rangle }
\newcommand{\lla}{\langle}

\newcommand{\pd}{\partial}

\usepackage{pstricks}
\usepackage{color}
\usepackage{multirow}
\usepackage{cite}
\usepackage{braket}

\let\a=\alpha   \let\b=\beta      \let\d=\delta
         
    \let\k=\kappa  \let\l=\lambda  \let\m=\mu
\let\n=\nu           \let\p=\pi      
\let\s=\sigma  \let\t=\tau      
         
\let\D=\Delta
\let\d=\delta
\let\s=\sigma
\renewcommand{\a}{\alpha}

\renewcommand{\t}{\tau}
\title{ Exact Correlators from Conformal Ward Identities  in Momentum Space and Perturbative Realizations\\
{\small Open issues on conformal anomaly actions}}

\ShortTitle{Exact Correlators from Conformal Ward Identities }

\author{\speaker{Claudio Corian\`o}\\
INFN Lecce, 
        Dipartimento di Matematica e Fisica "Ennio De Giorgi", \\ Universit\`a del Salento and INFN-Lecce, \\ Via Arnesano, 73100 Lecce, Italy\\
E-mail: \email{claudio.coriano@le.infn.it}}

\author{ Matteo Maria Maglio, Alessandro Tatullo, Dimosthenis Theofilopoulos\\
   INFN Lecce, 
        Dipartimento di Matematica e Fisica "Ennio De Giorgi", \\ Universit\`a del Salento and INFN-Lecce, \\ Via Arnesano, 73100 Lecce, Italy\\
        E-mail: \email{matteomaria.maglio@le.infn.it, alessandro.tatullo@le.infn.it,dimosthenis.theofilopoulos@le.infn.it \\}}

\abstract{ The general solution of the conformal Ward identities (CWI's) in momentum space, and their matching to perturbation theory, allows to uncover some specific characteristics of the breaking of conformal symmetry, induced by the anomaly. It allows to compare perturbative features of the 1-particle irreducible (1PI, nonlocal) anomaly action with the prediction of a similar (but exact) nonlocal action identified by the CWI's. The two predictions can be exactly matched at the level of 3-point functions. The analysis of the $TJJ$ and $TTT$ shows that both approaches - based either on 1PI or on the exact solutions of the CWI's - predict massless (dynamical) scalar exchanges in 3-point functions as the signature of the conformal anomaly. In a local formulation such 1PI actions exhibit a ghost in the spectrum which may induce ghost condensation. 
We also discuss alternative approaches, which take to Wess-Zumino forms of the action with an asymptotic dilaton, which should be considered phenomenological alternatives to the exact nonlocal action. If derived by a Weyl gauging, they also include a ghost in the spectrum. The two formulations, nonlocal and of WZ type, can be unified under the assumption that they describe the same anomaly phenomenon at two separate (UV/IR) ends of the renormalization group flow, possibly separated by a vacuum rearrangement at an intermediate scale. A similar analysis is presented for an $\mathcal{N}=1$ supersymmetric Yang-Mills theory. We comment on the possibile cosmological implications of such quasi Nambu-Goldstone modes as ultralight dilatons and axions. }

\FullConference{Proceedings of the Corfu Summer Institute 2018 "School and Workshops on Elementary Particle Physics and Gravity"\\

                 1-27 September 2018\\

                 Corfu, Greece}

\begin{document}

\section{Introduction}

The breaking of conformal symmetry by the conformal anomaly is a fascinating topic which has played a key role in $d=2$ as well as in higher spacetime dimensions. It is an open issue whether this phenomenon acquires a significant phenomenological meaning, indicating possible directions in the search for physics beyond the Standard Model (SM). In this case the issue is far from being trivial since a dilaton should be part of the spectrum of the scalars, while all the current data seem to indicate that the Higgs field found at the LHC is correctly accounted for by a single fundamental scalar. Therefore, the generation of the masses of the SM, which obviously break the conformal symmetry of a purely quartic Higgs potential, seems to be correctly accounted for by the Higgs mechanism. \\
However, there are several issues which remain unexplained, even with the tangible success of the SM.
For example, spontaneous symmetry breaking  requires the introduction of a scale in the theory, the Higgs vev $v$, but there is no specific theoretical reason, except phenomenological, why the electroweak scale lays around 246 GeV.  This is not the only open issue in the SM, as there are others which remain unsolved. For instance, there is no simple explanation of the fact that the number of fermion families is three, unless one considers very special extensions. A rare example is the 331 model \cite{Frampton:1992wt}, which is quite simple and unique, where 
the embedding of the third generation and the constraints from anomaly cancelation select the number of families to be exactly three, at the cost of breaking universality.\\
\subsection{Conformal extension of the SM: the Higgs plus a fundamental dilaton}
 Given such shortcomings and the puzzle raised by the gauge hierarchy problem, the inclusion of conformal symmetry may provide an alternative approach for answering at least some of these puzzles.
In a conformal extension motivated at a lower (TeV) scale, one can still envision the SM with its current field content, preserving the fundamental nature of the Higgs field, but with an electroweak scale generated by the vev of a second field $(\Sigma)$, whose role is to enforce a larger (conformal) symmetry in the classical Lagrangian. \\
 In fact, it is possible - and quite simple -  at tree level at least, to reconcile conformal symmetry and the Higgs mechanism by the introduction of an extra scalar field $\Sigma(x)$ in such a way to restore this symmetry. In this case the role of the Higgs remains the usual one, but the new scalar can mix with the Higgs, giving an ordinary mass eigenstate which would correspond to the SM Higgs, and to a dilaton. The real problem, in this scenario,
 is how to break the new symmetry in a simple way. We recall at this point that the dilaton $(\tau(x))$ is related to $\Sigma$ via a nonlinear realization with 
\begin{equation}
\Sigma\sim \Lambda \exp(\tau(x)/\Lambda),
\end{equation}
 where $\Lambda$ denotes the conformal symmetry breaking scale.
While this is one possibility, in which the dilaton is generated by enlarging the degrees of freedom of the SM, it is not the only one. A dynamical solution is also possible, as we are going to elaborate, where the dilaton emerges from the conformal dynamics in a specific way, as an effective degree of freedom. 
The arguments that we bring forward towards a resolution of some conflicting issues related to this topic are not necessarily advocated around the TeV scale, but may also reach very large scales, being generic and probably more of cosmological relevance than anything else. \\
In absentia of new physics at the LHC, and with the success of the SM, a way to explain the naturality of the Higgs mass, according to 't Hooft' s principle, is to invoke a larger symmetry which protects the small masses 
present in the SM from the large quadratic divergences of the scalar sector \cite{Bardeen:1995kv,Kawamura:2013kua}. \\  
In order to touch ground with the ordinary S-matrix formalism, it is necessary to promote the analysis of such conformal extensions to momentum space, where several non-perturbative tools, such as the conformal Ward identities, the operator product expansion and the conformal algebra at operatorial level, may allow to progress towards the analysis of multi-point correlation functions in a systematic way.
One of the advantages of a momentum space analysis is the possibility of identifying new effective degrees of freedom in such theories, and this brings us directly to investigate the form of the anomaly action, which is usually addressed within perturbation theory.\\
Given the obvious limitations of the perturbative approach, it is necessary to compare perturbative and non-perturbative methods in order to shed light on the issue of the nonlocality of the conformal anomaly action, which plays a key role in this context. This provides the main motivation for turning to a non-perturbative discussion of 3-point functions in momentum space.
 In particular, the statement that anomaly poles, which are the perturbative signature of the anomaly, do not correspond to physical states, should be taken with extreme care and correctly interpreted in a wider context, being a perturbative analysis probably insufficient to come to conclusions, given the complex nature of the phenomenon. \\
 We will argue that in the presence of conformal anomaly poles (and the same occurs for chiral anomaly poles) a certain theory rearranges its vacuum structure in such a way that the dynamics of such nonlocal interactions will generate an ordinary asymptotic state. Such a state would be a quasi Nambu-Goldstone mode in a local effective action and would correspond to a dilaton. \\
 We believe that in this way we can reconcile the two main formulations of the conformal anomaly action - the nonlocal and the local one - the latter incorporating an asymptotic dilaton, which otherwise appear to be unrelated. However, it is naturally expected that an intermediate potential will provide a small mass for such Nambu-Goldstone mode.
 A similar hypothesis has been formulated in the case of a St\"uckelberg field, 
 which carries analogous properties, and that may be rendered massive by a mechanism of misalignment as for an ordinary Peccei-Quinn axion. Our analysis is driven by this analogy. 
 
The goal of this review is to clarify some of the issues raised when comparing perturbative and non-perturbative approaches in theories affected by the conformal anomaly. For 3-point functions, the nonlocal structure of such action and its expression in terms of anomaly poles has been worked out directly from the solution of the conformal Ward identities (CWI's) in momentum space. For this reason, the comparison between the two descriptions, whenever possible, plays an important role. However, only in momentum space such combined analysis connect CFT's with their particle interpretations and help in clarifying these aspects. We are going to briefly summarise our arguments.
\section{The conformal anomaly action}
 In a conformal theory in even spacetime dimensions, the conformal symmetry can be broken by the conformal anomaly. Anomalies are related to the field content of a given theory and though apparently very different in their conformal and chiral versions, they are unified by the same phenomenon: the emergence of an anomaly pole, i.e. of a massless exchange in those diagrams which are held responsible for the origin of the anomaly, which are part of the 1PI (1-particle irreducible) effective action. They can be identified in a given anomaly vertex if we keep all the tensor structures of the same vertex uncontracted. By taking a trace or a divergence of an anomaly vertex, the pole is washed out, while only the residue at the pole remains in the interaction. 
 Given the perturbative nature of the 1PI action, and its simplicity, this behaviour may well be considered an artifact of perturbation theory, deprived of any specific meaning. \\
 One of the goals of our analysis will be to show that such effective interactions are present also if we move away from perturbation theory and discuss the same anomalous vertices using completely different methods.\\
 In a CFT the CWI's  fix the 3-point functions almost completely, modulo few constants, and it is in this case that the matching between the perturbative and non-perturbative approaches allows us to come to conclusions in regards to the emergence of such massless interactions. \\

\subsection{Wess-Zumino versus nonlocal actions}
An anomaly action modifies the classical action by the anomaly contribution. For instance, in the presence of an axial vector current, generated by an external gauge boson $B$, the anomaly contribution triggers the transition of such off-shell current into two photons, or into two gluons by a fermion loop, in QED and QCD respectively. Since an intermediate one-loop AVV interaction is involved, it is part of a simple 1PI action, and a direct computation shows that the interaction that ensues is mediated by the exchange of a massless pole \cite{Armillis:2010qk,Armillis:2009im,Armillis:2009sm,Giannotti:2008cv} which takes to a nonlocal action, as we will discuss in section \ref{ccff}.  \\
However, on the other hand, it is sufficient to couple linearly an axion $b(x)$ to the anomaly in order to obtain an effective action which now includes a scale $(M)$ and a dimension-5 operator 
\be
(b(x)/M) F\tilde{F}
\label{stuck}
\ee
to account for the anomaly, which takes to a local action. Operatorial terms of this type can also be introduced as counterterms in order to restore a gauge symmetry if an anomalous gauge boson is also part of the spectrum and not an external source for an axial-vector current. \\
As in the conformal case, where a similar coupling is present for the dilaton, in the local formulation the axion field $b(x)$ shifts under the local gauge symmetry as a typical Nambu-Goldstone mode. 
In the case of an anomalous axial-vector coupling, this construction takes to St\"uckelberg Lagrangians 
\cite{Coriano:2005js,Coriano:2018uip}, if the axial-vector gauge field is part of the dynamics. The St\"uckelberg field, introduced to cancel a gauge anomaly, turns into a physical gauge invariant component after mixing with the CP-odd phases of the Higgs sector. This is obtained by a mechanism of vacuum misalignment. \\
There are some issues which need to be addressed in the analysis of a perturbative anomaly action which is related to the choice of the regularization scheme, but it is clear that dimensional regularization (DR) plays a special role in this context. Other (mass-dependent) regularization schemes do not preserve the conformal symmetry of the theory and as such are not optimal. \\
The appearance of an anomaly pole in perturbation theory in mass-dependent corrections for any value of a correcting mass parameter $m_i$ is a feature which is typical of DR and not of other schemes. Indeed, DR allows to separate the anomaly contribution from the explicit mass-dependent corrections of a diagram in a very natural way for any value of the external momentum $(p)$ running in the loop. This holds even for $p^2 < 4 m^2$, where $m$ is the mass of the fermion in the loop,  
giving a non-vanishing contribution to the $\beta$ function of a given theory, and hence to the anomaly for any value of the external momentum. 
\\
Wess-Zumino types of action, the other variant of the conformal anomaly actions, as clear from the discussion above and from Eq. \eqref{stuck}, enlarge the number of degrees of freedom by introducing an axion or a dilaton in order to generate the same anomalies of the original theory. Obviously, a dilaton is not part of the original Lagrangian, though it is part of the anomaly action, and one has to view such actions as effective actions which need to be related to UV descriptions of the same phenomenon using renormalization group/effective field theory arguments.  Connecting the two descriptions requires special care and while this is done by preserving the global symmetries of a theory in its UV and IR phases, some aspects of this transition may not be easy to disentangle, as in the case of QCD versus the chiral Lagrangian.

\subsection{ The compensator and a ghost}
There are various ways to generate such actions, typically by the Noether method, where they are obtained starting from a linear coupling of the dilaton to the anomaly (see for instance \cite{Coriano:2012dg,Coriano:2013nja}).\\
A second approach is to use field-enlarging transformations, where the dilaton is introduced as a compensator. A compensator is not a dynamical field, and for this reason usually, such actions are modified by introducing by hand an (extra) kinetic term for the dilaton field. It is important to observe that if we decided to generate such a term from the Weyl gauging \cite{Coriano:2013nja, Codello:2012sn} of the Einstein Lagrangian, the procedure would generate a kinetic term which is ghost-like. \\
Flipping the sign of this term is a standard procedure, which, however, needs to be motivated since the inclusion of a Goldstone mode  by hand, while possible, should be compared against the description of the same theory in the (UV) conformal phase. In the UV this degree of freedom is completely absent and the violation of the conformal symmetry (i.e. the anomaly) appears as a purely radiative effect. Clearly, the latter is a pure phenomenological approach which allows to derive a possible final expression of the Lagrangian in a broken conformal phase without addressing the nature of the breaking itself. The closest example is the QCD chiral Lagrangian, where the global symmetries of the two theories in the UV and IR are matched, but the intermediate dynamics is essentially non-perturbative.\\
In order to shed some light on this, one possibility is to turn to exact methods, if these are available.\\
Up to 3-point functions, CFT's fix the structure of their correlators in an essentially unique way, except for few constants. It is then clear that the effective action built by combining the correlators of 2- and 3-point functions - which are solutions of the conformal constraints - naturally determine the simplest expression of the anomaly action of a certain theory. This action is specifically nonlocal, but obviously, it is not unique.  As we are going to discuss next, also in the case of a nonlocal conformal anomaly action, as well as in the chiral case, once this is rewritten in a local form, it manifests a kinetic mixing between two scalar degrees of freedom. In the chiral case, the scalars are replaced by pseudoscalars. If we try to decouple the two states we need to define a scale $M$ at which the decoupling takes place. It is then easy to realize that the new decoupled Lagrangian is characterized by a ghost in its spectrum, at least at the level of trilinear interactions.\\
 As emphasized in previous work, a simple analysis of this Lagrangian - in the case of a chiral theory - \cite{Armillis:2011hj} shows that the resolution of the kinetic mixing requires the inclusion of two axions, with one of them being a ghost.  In the unmixed case, a standard Coleman-Weinberg analysis of the potential for the ghost term shows that the Lagrangian induces a ghost condensation, with the possibility of a redefinition of the vacuum. We are going to briefly review such features. The analysis has been done in the case of an external axial-vector coupling, although the result in the case of a dilaton pole we expect it to be quite similar.  
\section{Conformal symmetry and its classical breaking} 
Let's come to a brief description of the conformal invariant extension of the SM with a fundamental Higgs and a dilaton. \\
A scale invariant extension of a given Lagrangian can be obtained if we promote all the dimensionful constants to dynamical fields. 
It is natural to ask whether the new degree of freedom introduced to restore the conformal symmetry of the theory can be generated dynamically, emerging from the effective interactions which can be held responsible for the generation of an anomaly. We illustrate this point in the case of a simple interacting scalar field theory incorporating the Higgs mechanism. At a second stage, we will derive the structure of the dilaton interaction at order $1/\Lambda$, where 
$\Lambda$ is the scale characterizing the spontaneous breaking of the dilatation symmetry and discuss some possible phenomenological constraints on $\Lambda$. 

The two equivalent forms of the scalar Higgs potential
\beqa
V_1(H, H^\dagger)&=& - \mu^2 H^\dagger H +\lambda(H^\dagger H)^2 =
\lambda \left( H^\dagger H - \frac{\mu^2}{2\lambda}\right)^2 - \frac{\mu^4}{4 \lambda}\nonumber \\
V_2(H,H^\dagger)&=&\lambda \left( H^\dagger H - \frac{\mu^2}{2\lambda}\right)^2
\eeqa
generate two {\em different} scale-invariant extensions 
\beqa
\label{v2}
V_1(H,H^\dagger, \Sigma)&=&- \frac{\mu^2\Sigma^2}{\Lambda^2} H^\dagger H +\lambda(H^\dagger H)^2 \nonumber \\
V_2(H,H^\dagger, \Sigma)&=& \lambda \left( H^\dagger H - \frac{\mu^2\Sigma^2}{2\lambda \Lambda^2}\right)^2 \,,
\eeqa 
where $H$ is the Higgs doublet, $\lambda$ is its dimensionless coupling constant, while $\mu$ has the dimension of a mass. 
The constant $\mu^4$ term present in $V_1$ which in a non-scale invariant theory can be absorbed by a redefinition of the Lagrangian, is clearly insignificant in flat space and generates two different scale-invariant potential, where only the second one is stable.
\beq
\label{original}
\mathcal L = \frac{1}{2} (\partial \phi)^2 - V_2(\phi) =
\frac{1}{2} (\partial \phi)^2 + \frac{\mu^2}{2}\, \phi^2 - \lambda\, \frac{\phi^4}{4} - \frac{\mu^4}{4\,\lambda}\, ,
\eeq
obeying the classical equation of motion
\beq \label{scalarEOM}
\square \phi = \mu^2\,\phi - \lambda\, \phi^3\, .
\eeq
This theory is not scale invariant due to the appearance of the mass term $\mu$, as one can easily notice from the trace of the canonical ($c$)
energy-momentum (EMT) tensor 
\bea
T^{\mu\nu}_{c}(\phi) 
&=& 
\partial^\mu \phi\, \partial^\nu \phi 
- \frac{1}{2}\,\eta^{\mu\nu} \bigg[ (\partial \phi)^2 + \mu^2 \,\phi^2 
-  \lambda\,\frac{\phi^4}{2} -  \frac{\mu^4}{2\,\lambda} \bigg] \, ,
\nn \\
{T_{c}^\mu}_\mu(\phi) &=& 
- (\partial\phi)^2 - 2\, \mu^2 \,\phi^2 +\lambda\, \phi^4 + \frac{\mu^4}{\lambda} \, .
\eea
The EMT of a scalar field can be improved so that its trace is proportional only to the scale breaking parameter,
i.e. the mass $\mu$. This can be achieved by adding an extra contribution $T_I^{\mu\nu}(\phi, \chi)$ which is symmetric and conserved
\beq
T_I^{\mu\nu}(\phi,\chi)=\chi \left(\eta^{\mu\nu} \square \phi^2 - \partial^\mu \partial^\nu \phi^2 \right) \,,
\eeq
where the $\chi$ parameter is specifically choosen.
The combination of the canonical plus the improvement EMT, 
$T^{\mu\nu} \equiv T_c^{\mu\nu} + T_I^{\mu\nu}$ has the off-shell trace
\beq
{T^\mu}_\mu(\phi,\chi)= 
(\partial\phi)^2\, \left( 6 \chi - 1 \right) - 2\, \mu^2\, \phi^2 
+ \lambda\, \phi^4 + \frac{\mu^4}{\lambda} + 6 \chi \phi\, \square \phi\, .
\eeq
Using the equation of motion (\ref{scalarEOM}) and selecting $\chi=1/6$, the trace relation given in the expression above 
becomes proportional to the scale-breaking term $\mu$  
\beq \label{ImprovedTrace}
{T^\mu}_\mu(\phi,1/6) = - \mu^2 \phi^2 + \frac{\mu^4}{\lambda} \, .
\eeq
The scale-invariant extension of the Lagrangian given in Eq.(\ref{original}) is obtained by promoting the mass terms to dynamical fields using the replacement 
\beq
\label{rep}
\mu \to \frac{\mu}{\Lambda} \, \Sigma(x),
\eeq
obtaining
\beq
\label{sigmaphi}
\mathcal L = 
\frac{1}{2}\, (\partial \phi)^2 +\frac{1}{2} (\partial \Sigma)^2 
+  \frac{ \mu^2}{2\,\Lambda^2}\, \Sigma^2\, \phi^2 - \lambda \frac{\phi^4}{4}
-  \frac{\mu^4}{4\,\lambda \, \Lambda^4}\, \Sigma^4,
\eeq
where we have included a kinetic term for the dilaton $\Sigma$. The new Lagrangian is dilatation invariant, as shown from the trace of the EMT

\bea
&& {T^{\mu}}_{\mu}(\phi,\Sigma,\chi,\chi') = \left( 6\, \chi - 1 \right)\, (\partial\phi)^2
+ \left( 6 \chi^\prime -1\right)\, (\partial\Sigma)^2 
+ 6 \chi\, \phi\, \square \phi + 6 \chi^\prime\, \Sigma\, \square \Sigma 
- 2\, \frac{\mu^2}{\Lambda^2}\, \Sigma^2\, \phi^2 + \lambda\, \phi^4 \nn\\
& &\hspace{3cm}+ \frac{1}{\lambda}\,\frac{\mu^4}{\Lambda^4}\,\Sigma^4 \,,
\eea
an expression that vanishes upon using the equations of motion for the $\Sigma$ and $\phi$ fields,
\bea
\square \phi &=& \frac{\mu^2}{\Lambda^2}\, \Sigma^2\, \phi -  \lambda\, \phi^3\, ,
\nn \\
\square \Sigma &=& \frac{\mu^2}{\Lambda^2}\, \Sigma\, \phi^2 - \frac{1}{\lambda}\, \frac{\mu^4}{\Lambda^4}\,\Sigma^3  \, ,
\eea
and after setting the value of the $\chi, \chi'$ parameters at the special value $\chi=\chi^\prime=1/6$, corresponding to conformally coupled scalars. 
The scalar potential $V_2$ triggers
the spontaneous breaking of the scale symmetry around a stable point of minimum. 
Expanding around the vacuum, parameterized by the conformal scale $\Lambda$ and the Higgs vev $v$ respectively
\beq
\label{parm}
\Sigma  =  \Lambda + \rho \, , \quad \phi = v + h\, 
\eeq
one can describe the theory in the broken phase.\\
For our present purposes, it is enough to expand the Lagrangian (\ref{sigmaphi}) around the vev for the dilaton field, 
as we are interested in the structure of the couplings of its fluctuation $\rho$
\beq \label{Manifest}
\mathcal L = \frac{1}{2}\, (\partial\phi)^2 + \frac{1}{2}\, (\partial\rho)^2 + \frac{\mu^2}{2}\, \phi^2 
- \lambda \, \frac{\phi^4}{4} - \frac{\mu^4}{4\,\lambda}
- \frac{\rho}{\Lambda}\,\left(- \mu^2\, \phi^2  + \frac{\mu^4}{\lambda}\right) + \dots\, ,
\eeq
where we have neglected terms of higher order in $1/\Lambda$.
It is clear, from (\ref{ImprovedTrace}) and (\ref{Manifest}), that one can write a dilaton Lagrangian
at order $1/\Lambda$, as
\beq \label{RhoInteraction}
\mathcal L_{\rho} = (\partial\rho)^2 - \frac{\rho}{\Lambda}\, {T^{\mu}}_{\mu}(\phi,1/6) + \dots\, ,
\eeq
where we have used the equations of motion in order to re-express the trace of the energy momentum tensor.\\
It is clear, from this simple analysis, that a dilaton, in general, does not couple to the anomaly,
but only to the sources of the explicit breaking of scale invariance, which are proportional to the mass terms of the action.
In $V_2$ we parameterize the Higgs around the electroweak vev $v$ as in Eq. (\ref{parm}), 
and indicate with $\Lambda$ the vev of the dilaton field $\Sigma = \Lambda + \rho$, with $\phi^+ = \phi = 0$ in the unitary gauge. \\
Performing a diagonalization of the mass matrix we define the two mass eigenstates $\rho_0$ and $h_0$, which are given by 
with the potential $V_2$  exhibiting a massless mode due to the existence of a flat direction. The Higgs and the dilaton will mix according to the mass matrix

\beq
 \left( \begin{array}{c}
 {\rho_0}\\
  h_0 \\
  \end{array} \right)
 =\left( \begin{array}{cc}
\cos\alpha & \sin\alpha \\
-\sin\alpha & \cos\alpha  \\
 \end{array} \right)
 \left( \begin{array}{c}
  \rho\\
 {h} \\
  \end{array} \right)
\eeq
with 
\beq
\cos\alpha=\frac{1}{\sqrt{1 + v^2/\Lambda^2}}\qquad \qquad  \sin\alpha=\frac{1}{\sqrt{1 + \Lambda^2/v^2}}.
\eeq
We denote with ${\rho_0}$ the massless dilaton generated by this potential, while 
$h_0$ denotes the Higgs scalar whose mass is given by  
\beq 
m_{h_0}^2= 2\lambda v^2 \left( 1 +\frac{v^2}{\Lambda^2}\right) \qquad \textrm{with} \qquad v^2=\frac{\mu^2}{\lambda},
\eeq
and with $m_h^2=2 \lambda v^2$ being the mass of the Standard Model Higgs.
Notice that the Higgs mass, in  this case, is corrected by the new scale of the spontaneous breaking of the dilatation symmetry ($\Lambda$), 
which remains a free parameter.

Obviously, the presence of a massless dilaton in the spectrum is troublesome, a problem which send us back to the issue of how to select a single vacuum state from the underlying vacuum degeneracy.  This can be lifted by the introduction of extra (explicit breaking) terms which give a small mass to the dilaton field.
To remove such degeneracy, one can introduce, for instance, the term
\beq
\mathcal{L}_{break} 
= \frac{1}{2} m_{\rho}^2 {\rho}^2 + \frac{1}{3!}\, {m_{\rho}^2} \frac{{\rho}^3}{\Lambda} + \dots \, ,
\eeq
where $m_{\rho}$ represents the dilaton mass.
 The coupling of a dilaton to an anomaly is necessary, since the dilaton is the pseudo Nambu-Goldstone mode of the dilatation symmetry and the anomaly is a source of such breaking. Thus, this coupling has to be introduced by hand and the role of the conformal anomaly action is to account for it. \\

\begin{figure}[t]
\centering
\subfigure[]{\includegraphics[scale=0.7]{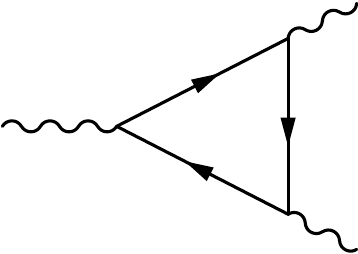}} \hspace{2cm}
\subfigure[]{\includegraphics[scale=0.7]{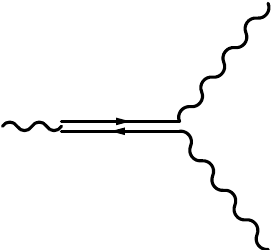}} \hspace{2cm}
\subfigure[]{\includegraphics[scale=0.7]{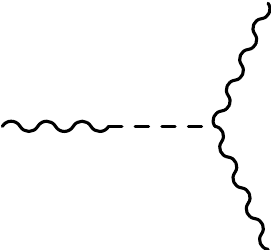}}
\caption{The triangle diagram in the fermion case (a), the collinear fermion configuration responsible for the anomaly (b) and a diagrammatic representation of the exchange via an intermediate state (dashed line) (c). \label{fig1}}
\end{figure}

\section{ Phenomenology of a classical scale invariant extension of the Standard Model}
\label{revv}
In the case of the SM, the dilaton interaction takes the form discussed above
\bea\label{tmunu}
\mathcal L_{int} = -\frac{1}{\Lambda}\rho T^\mu_{\mu\,SM}.
\eea
where $T$ is the EMT of the SM. As usual, it can be easily derived by embedding the 
SM Lagrangian in the background metric $g_{\mu\nu}$ 
\beq S = S_{SM} + S_{I}=  \int d^4 x
\sqrt{-g}\mathcal{L}_{SM} + \xi \int d^4 x \sqrt{-g}\, R \, \mathcal{H}^\dag \mathcal{H}      \, ,
\eeq
where $\mathcal{H}$ is the Higgs doublet and $R$ the scalar curvature of the same metric, and then defining 
\beq  T_{\mu\nu}(x)  = \frac{2}{\sqrt{-g(x)}}\frac{\d [S_{SM} + S_I ]}{\d g^{\mu\nu}(x)},
\eeq
or, in terms of the SM Lagrangian, as
\beq \label{TEI spaziocurvo}
\frac{1}{2} \sqrt{-g} T_{\mu\nu}{\equiv} \frac{\pd(\sqrt{-g}\mathcal{L})}
{\pd g^{\mu\nu}} - \frac{\pd}{\pd x^\s}\frac{\pd(\sqrt{-g}\mathcal{L})}{\pd(\pd_\s g^{\mu\nu})}\, .
\eeq
The complete expression of the energy-momentum tensor can be found in \cite{Coriano:2011zk}.
$S_{I}$ is responsible for generating a term of improvement $(I)$, which induces a mixing between the Higgs and the dilaton after spontaneous symmetry breaking. 
As usual, we parameterize
the vacuum $\mathcal H_0$ in the scalar sector in terms of the electroweak vev $v$ as
\beq \label{VEVHiggs}
\mathcal H_0 =
\left(\begin{array}{c} 0 \\ \frac{v}{\sqrt{2}} \end{array}\right)
\eeq
and we expand the Higgs doublet in terms of the physical Higgs boson $H$ and the two Goldstone bosons $\phi^{+}$, $\phi$ as
\bea
\mathcal H = \left(\begin{array}{c} -i \phi^{+} \\ \frac{1}{\sqrt{2}}(v + H + i \phi) \end{array}\right),
\eea
obtaining from the term of improvement of the stress-energy tensor the expression
\begin{figure}[t]
\centering
\subfigure[]{\includegraphics[scale=.6]{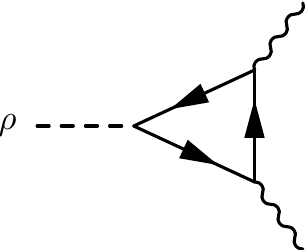}}
\hspace{.2cm}
\subfigure[]{\includegraphics[scale=.6]{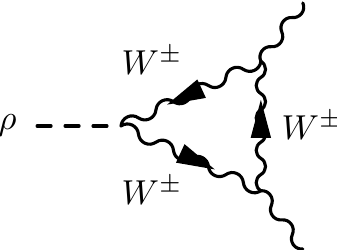}}
\hspace{.2cm}
\subfigure[]{\includegraphics[scale=.6]{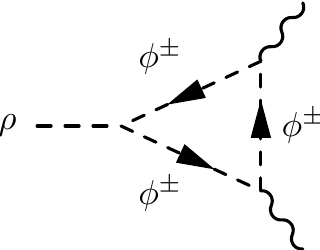}}
\hspace{.2cm}
\subfigure[]{\includegraphics[scale=.6]{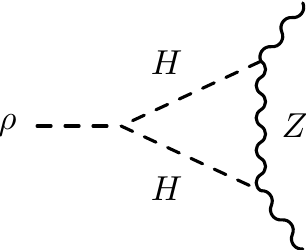}}
\centering
\subfigure[]{\includegraphics[scale=.6]{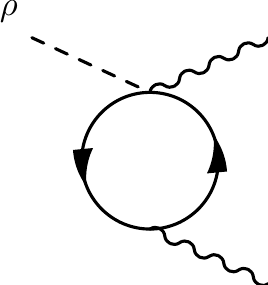}}
\hspace{.2cm}
\subfigure[]{\includegraphics[scale=.6]{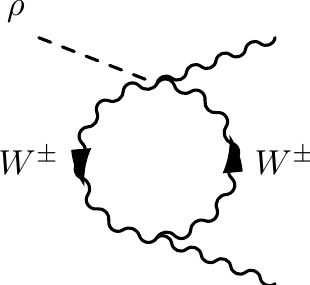}}
\hspace{.2cm}
\subfigure[]{\includegraphics[scale=.6]{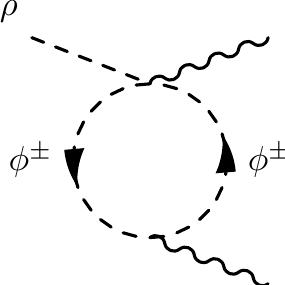}}
\hspace{.2cm}
\subfigure[]{\includegraphics[scale=.6]{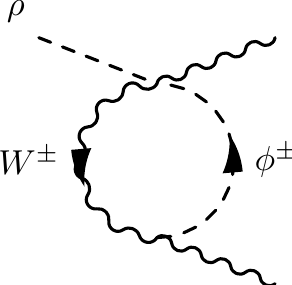}}
\subfigure[]{\includegraphics[scale=.6]{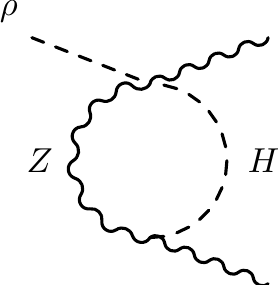}} 
\subfigure[]{\includegraphics[scale=.6]{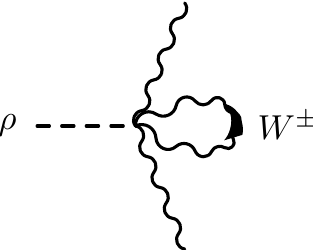}}
\caption{Typical amplitudes of triangle and bubble  topologies contributing to the $\rho \gamma\gamma$, $\rho \gamma Z$ and $\rho ZZ$ interactions. They 
include fermion $(F)$, gauge bosons $(B)$ and contributions from the term of improvement (I). Diagrams (a)-(g) contribute to all the 
three channels while (h)-(k) only in the $\rho ZZ$ case.}
\label{figuretriangle}
\end{figure}

\beq
T^I_{\mu\nu} = - 2 \xi \bigg[ \partial_{\mu} \partial_{\nu} - \eta_{\mu\nu} \, \Box \bigg] \mathcal H^\dag \mathcal H = - 2 \xi \bigg[ \partial_{\mu} \partial_{\nu} - \eta_{\mu\nu} \, \Box \bigg] \bigg( \frac{H^2}{2} + \frac{\phi^2}{2} + \phi^{+}\phi^{-} + v \, H \bigg),
\eeq
which is responsible for a bilinear vertex
\beq
V_{I,\, \rho H }(k)
= - \frac{i}{\Lambda} \frac{12\,\xi \, s_w M_W}{e} \, k^2 \nn
\eeq
where $s_W$ is the Weinberg angle.
 The trace takes contribution from the massive fields, the fermions and the electroweak gauge bosons, and from the conformal anomaly  in the massless gauge boson sector, through the $\beta$ functions of the corresponding coupling constants. \\
One can directly verify the the separation between the anomalous and the explicit mass-related terms in the expression of the correlators responsible of the conformal anomaly. It has been verified in QED and in the neutral sector of the SM \cite{Armillis:2009pq, Coriano:2011ti} by explicit computations. Such separation does not necessarily hold in other regularization schemes.  The reason for such difference is related to the special role of DR compared to other regularization schemes, which explicitly break the conformal symmetry by introducing a cutoff. This point has been discussed by Bardeen \cite{Bardeen:1995kv}  in the context of 't Hooft's naturalness principle applied to the Higgs mass. Also in this case there are strong arguments that convey a truly specials role to such regularization scheme compared to others. In the case of the $TVV$ vertex, for instance, by taking a trace one can derive an anomalous Ward identity of the form 
\beq
\Gamma^{\alpha\beta}(z,x,y) 
\equiv \eta_{\mu\nu} \left\langle T^{\mu\nu}(z) V^{\alpha}(x) V'^{\beta}(y) \right\rangle 
= \frac{\delta^2 \mathcal A(z)}{\delta A_{\alpha}(x) \delta A_{\beta}(y)} + \left\langle {T^\mu}_\mu(z) V^{\alpha}(x) V'^{\beta}(y) 
\right\rangle.
\label{traceid1}
\eeq
where $\mathcal A(z)$ is the anomaly functional, while $A_{\alpha}$ indicates the gauge fields coupled to the current $V^{\alpha}$.  $\Gamma^{\alpha\beta}$ is a generic 
dilaton/gauge/gauge vertex, whose Feynman expansion takes a form depicted in Fig. \ref{figuretriangle}. It is obtained from the $TVV'$ vertex by tracing the spacetime indices $\mu\nu$.  $\mathcal A(z)$ is derived from the renormalized expression 
of the vertex by tracing the gravitational counterterms in $4-\epsilon$ dimensions (see for instance 
\cite{Coriano:2012dg,Coriano:2013nja})
\beqa
\langle T_\mu^\mu \rangle=\mathcal A(z),
\eeqa
which in a curved background is given by the metric functional 
\beqa
\mathcal A(z)= -\frac{1}{8} \left[ 2 b \,C^2 + 2 b' \left( E - \frac{2}{3}\square R\right) + 2 c\, F^2\right],
\label{anomalyeq}
\eeqa
 where $b$, $b'$ and $c$ are parameters.  For the case of a single fermion in an abelian gauge theory they are given by $b = 1/320 \, \pi^2$,  $b' = - 11/5760 \, \pi^2$,
and $c= -e^2/24 \, \pi^2$. $C^2$ is the square of the Weyl tensor and $E$ is the Euler density given by
\beqa
C^2 &=& C_{\lambda\mu\nu\rho}C^{\lambda\mu\nu\rho} = R_{\lambda\mu\nu\rho}R^{\lambda\mu\nu\rho}
-2 R_{\mu\nu}R^{\mu\nu}  + \frac{R^2}{3}  \\
E &=& ^*\hskip-.1cm R_{\lambda\mu\nu\rho}\,^*\hskip-.1cm R^{\lambda\mu\nu\rho} =
R_{\lambda\mu\nu\rho}R^{\lambda\mu\nu\rho} - 4R_{\mu\nu}R^{\mu\nu}+ R^2.
\eeqa
In a flat metric background the expression of such functional reduces to the simple form
\beqa \label{TraceAnomaly}
\mathcal A(z)
&=& \sum_{i} \frac{\beta_i}{2 g_i} \, F^{\alpha\beta}_i(z) F^i_{\alpha\beta}(z), 
\eeqa
where $\beta_i$ are clearly the mass-independent $\beta$ functions of the gauge fields
and $g_i$ the corresponding coupling constants. Obviously, for a theory which which is quantum conformal invariant, the $\beta_i$ vanish. We refer to \cite{Coriano:2012nm,Bandyopadhyay:2016fad} for more details concerning the phenomenology of such models of direct LHC relevance. \\
It is possible, using such effective interaction which couples the dilaton to the anomaly, to address some phenomenological issues which are can be studied at the LHC. \\
We show in Figs \ref{brrhoh} and \ref{ggvbf} some results of a phenomenological analysis of the production and decay of a dilaton ($\rho$) as a function of its mass and branching ratios, together with some comparisons with the Higgs. In particular Fig. \ref{bzzww}  shows the results of an analysis of the bounds on $\Lambda$, the conformal scale by a comparison with experimental data. In general it is possible to select a value of $\Lambda$ in  the few TeV region in such a way that such additional interactions are in agreement with the SM results. We refer to \cite{Bandyopadhyay:2016fad} for further details.\\

\begin{figure}[t]
\begin{center}
\hspace*{-2cm}
\mbox{\subfigure[]{
\includegraphics[width=0.5\linewidth]{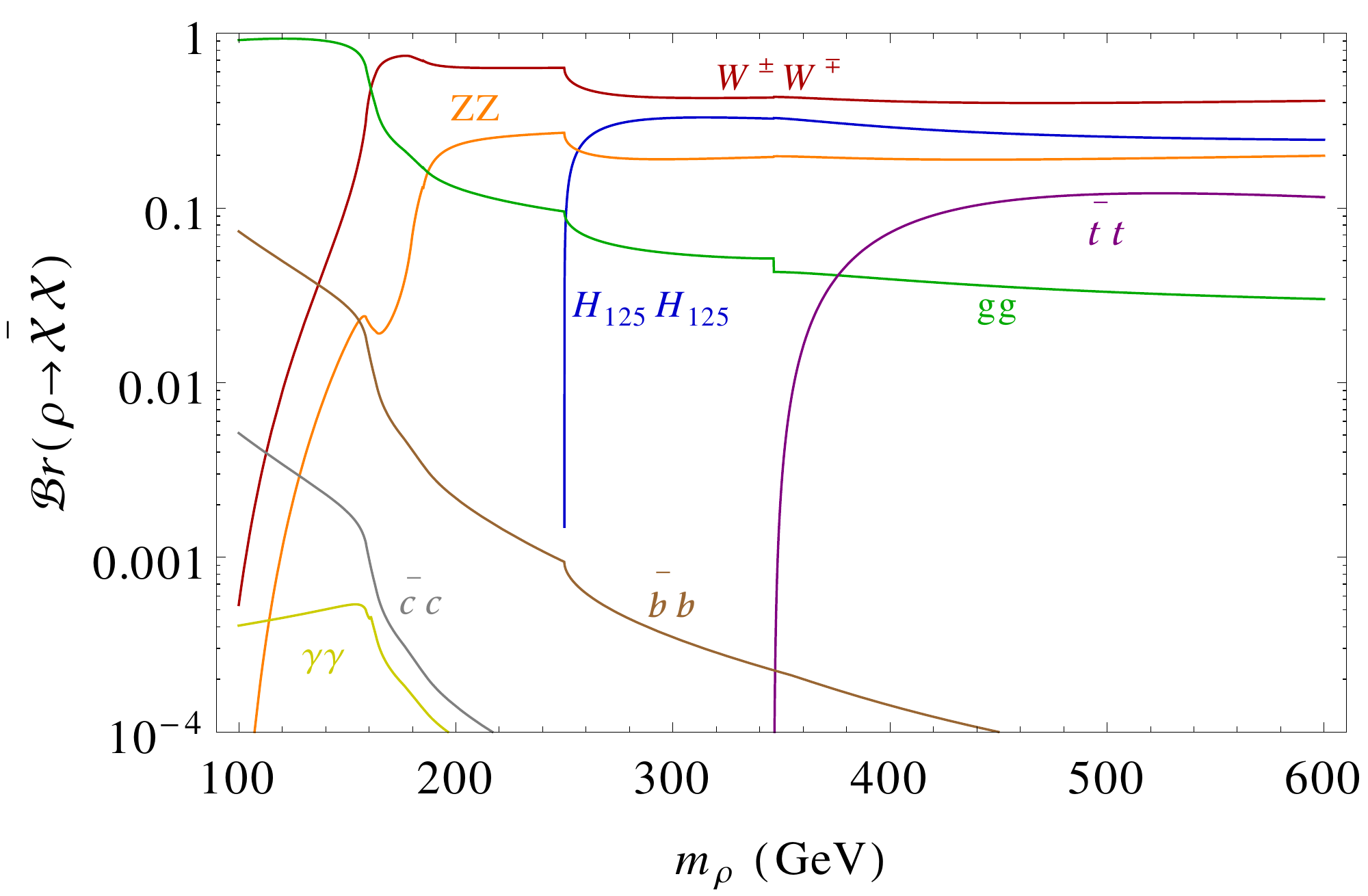}}\hskip 15pt
\subfigure[]{\includegraphics[width=0.5\linewidth]{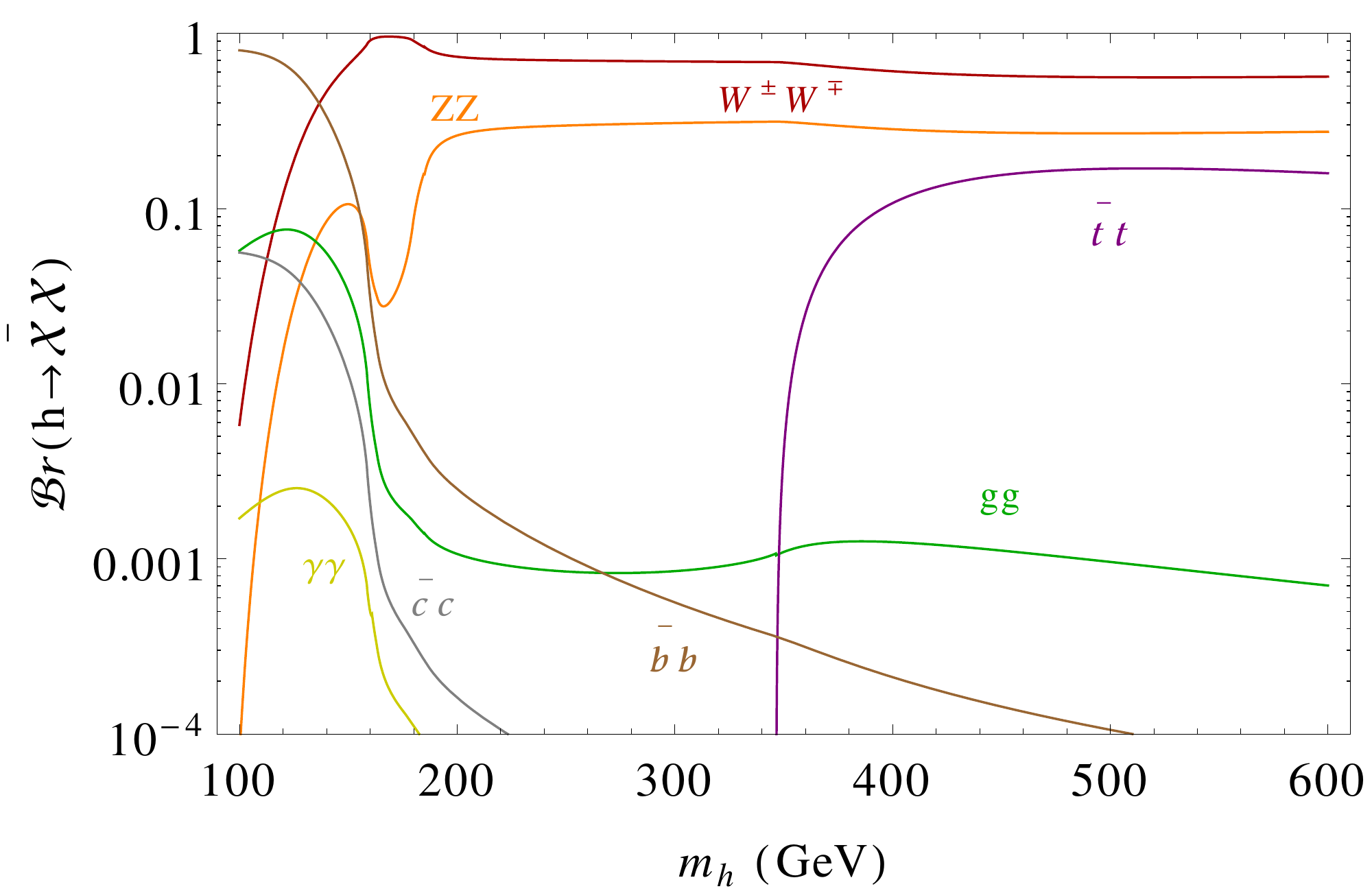}}}
\caption{The mass dependence of the branching ratios of the dilaton (a) and of the Higgs boson (b).}\label{brrhoh}
\end{center}
\end{figure}

\begin{figure}[thb]
\begin{center}
\hspace*{-2cm}
\mbox{\subfigure[]{
\includegraphics[width=0.4\linewidth]{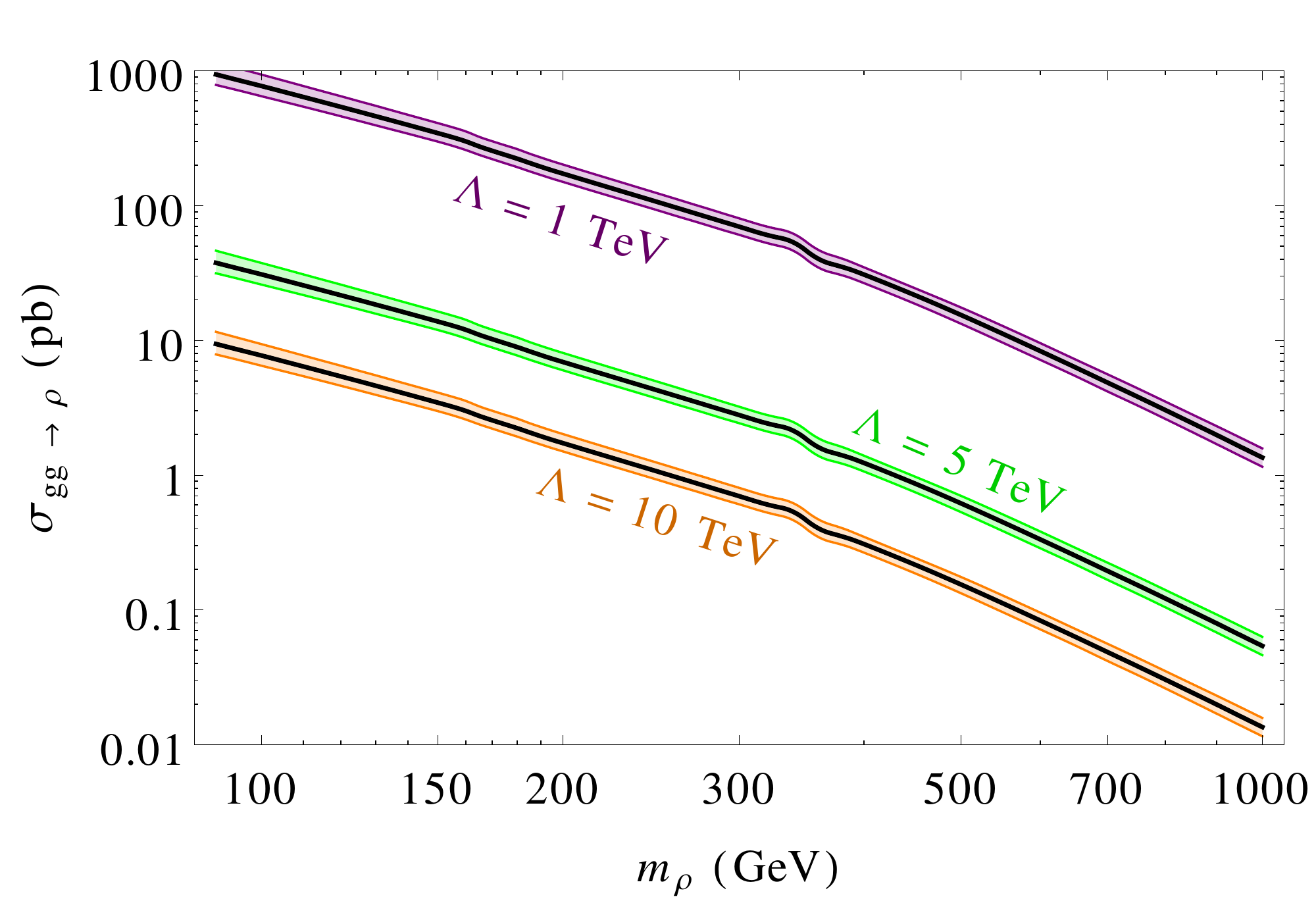}}\hskip 15pt
\subfigure[]{\includegraphics[width=0.4\linewidth]{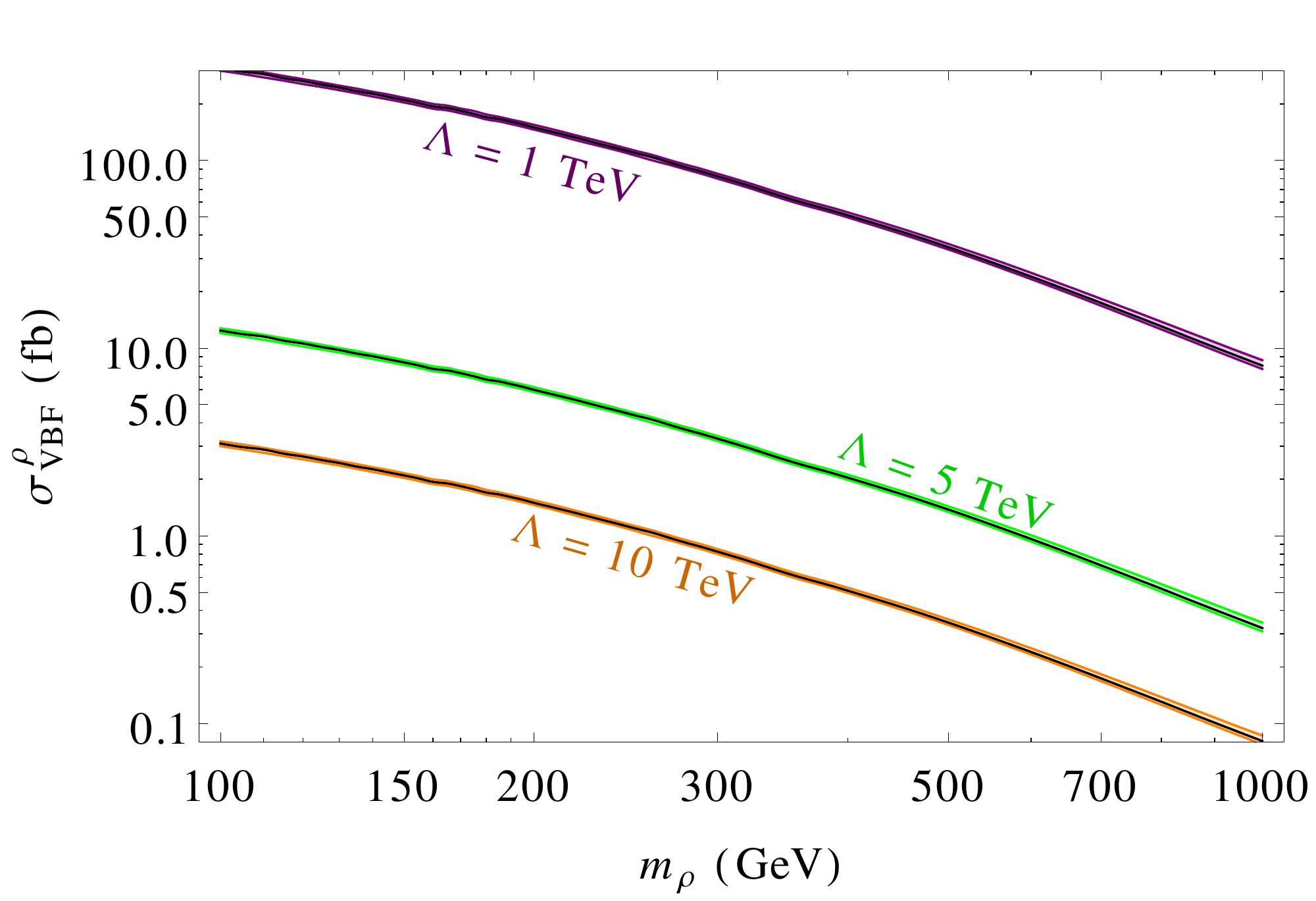}}}
\caption{The mass dependence of the dilaton cross-section via gluon fusion (a) and vector boson fusion (b) for three different choices of the conformal scale, $\Lambda=1, 5, 10$ TeV respectively.}\label{ggvbf}
\end{center}
\end{figure}

%%%%%%%%%%

%%%%%%%%%%%%%%%%%%%%%%%%%
\begin{figure}%[thb]
\begin{center}
\hspace*{-2cm}
\mbox{\subfigure[]{
\includegraphics[width=0.4\linewidth]{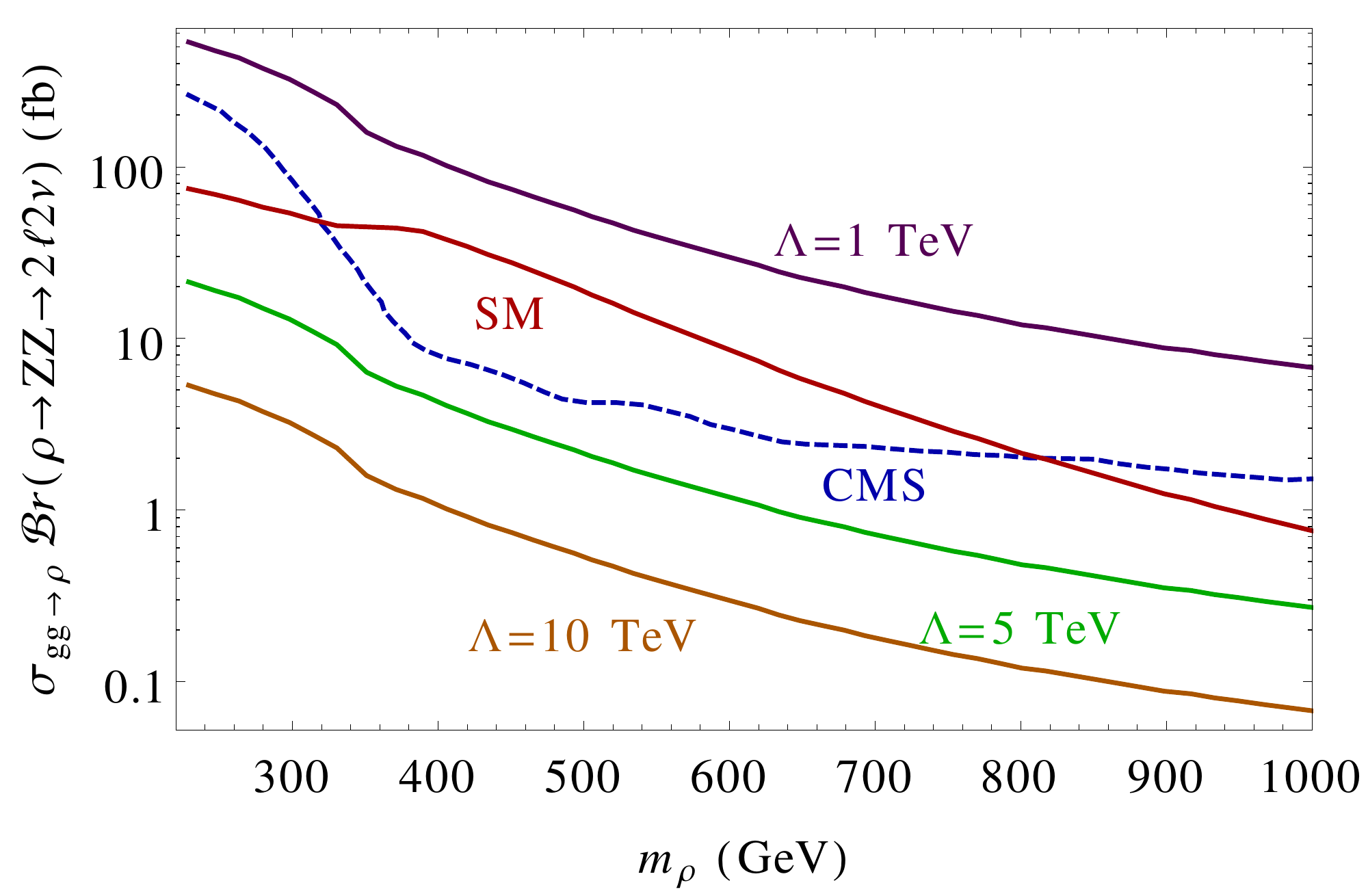}}\hskip 15pt
\subfigure[]{\includegraphics[width=0.4\linewidth]{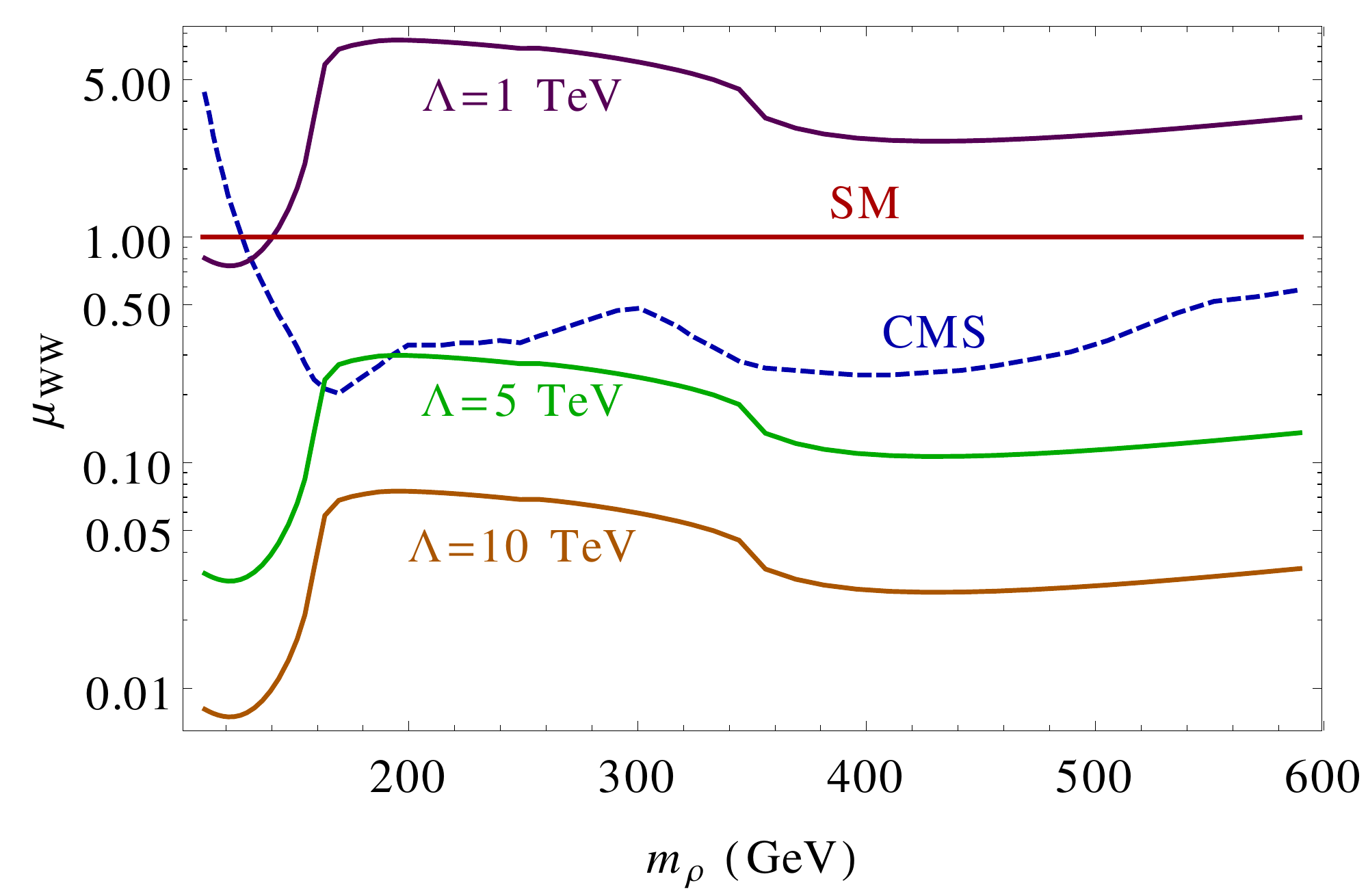}}}
\hspace*{-2cm}
\mbox{\subfigure[]{\includegraphics[width=0.4\linewidth]{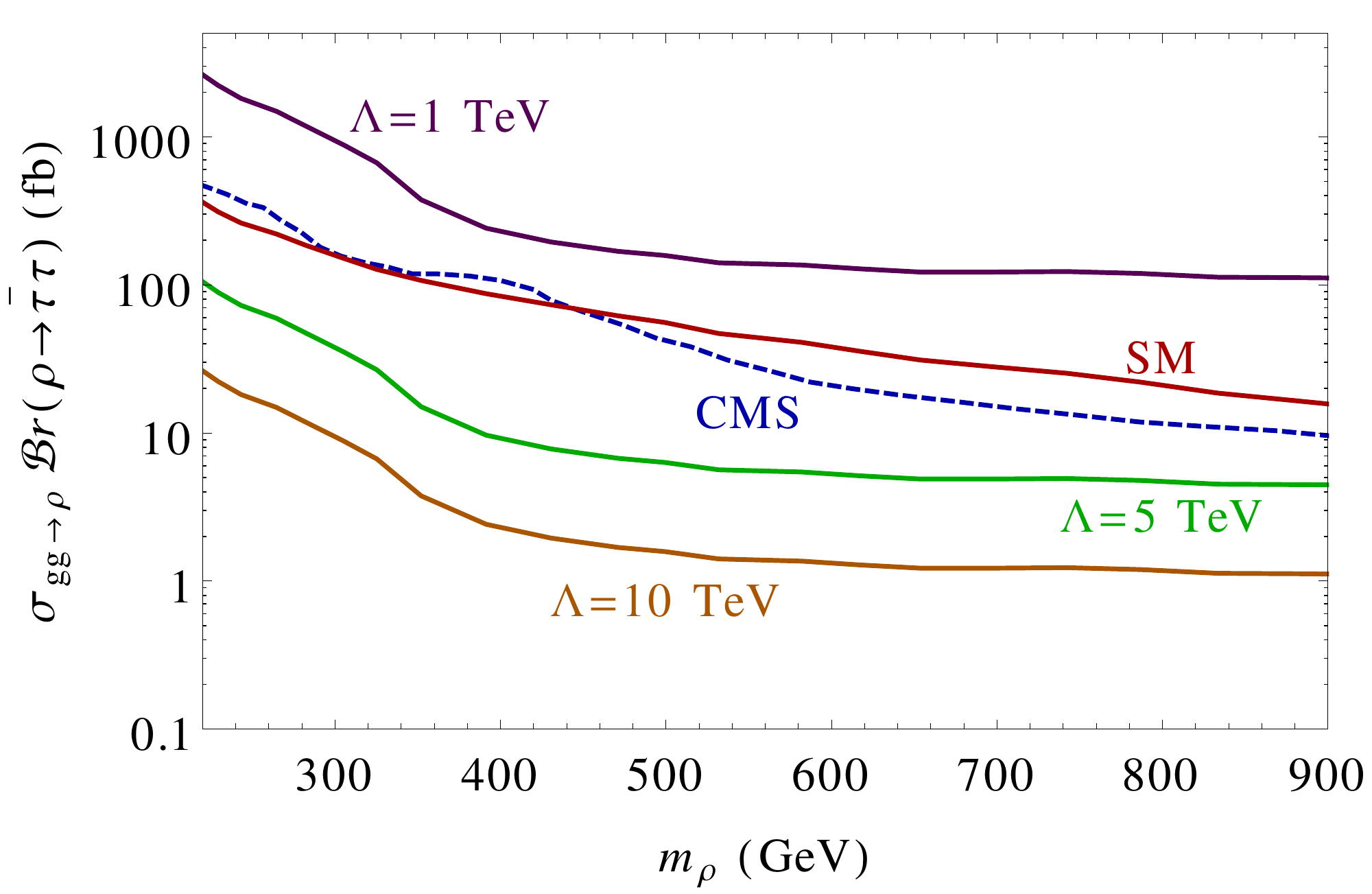}}\hskip 15pt
\subfigure[]{\includegraphics[width=0.4\linewidth]{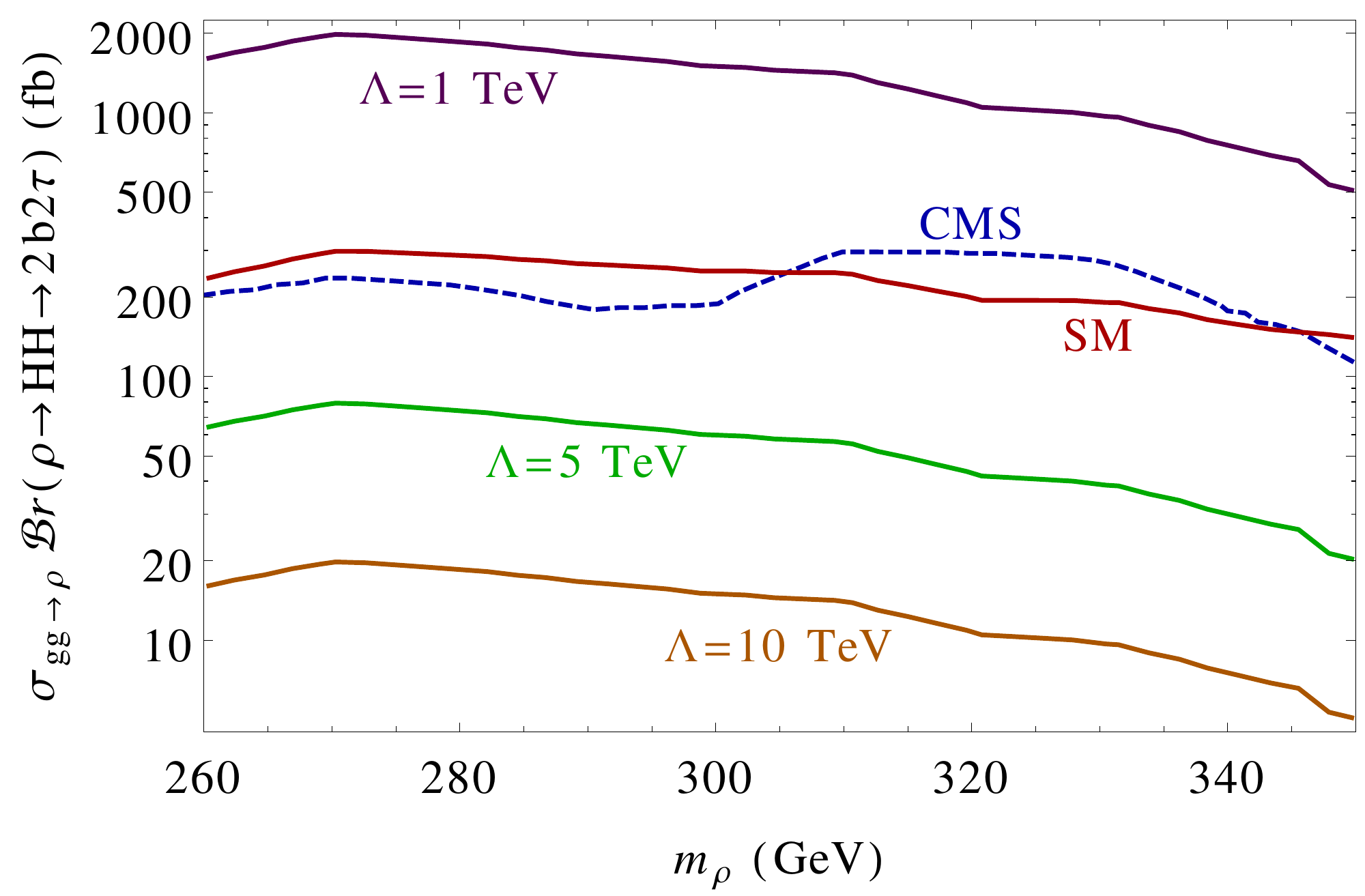}}}
\caption{The mass bounds on the dilaton from heavy scalar decays to (a) $ZZ$, (b) $W^\pm W^\mp$, (c) $\bar\tau\tau$ and (d) to $H\,H$  for three different choices of conformal scale, $\Lambda=1, 5, 10$ TeV respectively.}\label{bzzww}
\end{center}
\end{figure}
\section{Wess Zumino actions with an asymptotic dilaton from a kinetic flip}
The inclusion of an asymptotic dilaton, as done above, and the derivation of the corresponding anomaly action can be performed in various ways. One possibility is to use the procedure of Weyl gauging (see \cite{Codello:2012sn}).  It is possible to generate such an action by the application of this procedure to the anomaly counterterms. In this case one can also show that multiple traces of stress energy tensors are functionally dependent on the first 4 ones in $d=4$, the first 6 ones in $d=6$ and so on. More details can be found in \cite{Coriano:2012dg,Coriano:2013nja}. We briefly summarize the method in the case of $d=4$, before stressing one crucial aspect of this approach, i.e. the presence of a ghost in the procedure. 
The method amounts to a field-enlarging transformation, with the removal of the ghost by flipping the sign of the kinetic term. In the next sections, once we turn to the analysis of 1PI anomaly actions, which are not constructed by this formal procedure, but are directly computed either in perturbation theory or by solving the 
CWI's in momentum space, we will point out a similar feature. 

The metric tensor $g_{\mu\nu}(x)$, the vierbein $V_{a\,\rho}(x)$ 
and the fields $\Phi$ change under Weyl scalings according to
\beqa \label{WeylTransf}
g'_{\mu\nu}(x)     &=&  e^{2\, \sigma(x)}\, g_{\mu\nu}(x)\, ,  \nn \\
{V'}_{a\,\rho}(x)  &=&  e^{\sigma(x)}\, V_{a\,\rho}(x)\, , \nn \\
\Phi'(x)           &=&  e^{d_{\Phi}\, \sigma(x)}\, \Phi(x)\, ,
\eeqa
with $\sigma(x)$ being a dimensionless function parameterizing a local Weyl transformation.
Here $d_\Phi$ is the scaling dimension of the generic field $\Phi$, the latin suffix $a$ in $V_{a\rho}$ denotes the flat local index, 
while the Greek indices are the curved indices of the spacetime manifold.
A way to build a Weyl invariant theory containing the fields in (\ref{WeylTransf}) consists in making the metric tensor, the vierbein and the fields $\Phi$, Weyl invariant through the substitutions 
\bea \label{MetricGauge}
V_{a\,\rho}(x) &\rightarrow& \hat V_{a\,\rho}(x) \equiv  e^{- \frac{\tau(x)}{\Lambda}}\, V_{a\,\rho}(x) \, ,
\nn \\
\Phi(x)        &\rightarrow& \hat{\Phi}(x)       \equiv  e^{-d_{\Phi} \frac{\tau(x)}{\Lambda}}\, \Phi  \, 
\eea
and take the form of field-enlarging transformations.

Under a Weyl scaling (\ref{WeylTransf}), the  dilaton $\tau$ is required to shift as a Goldstone mode
\beq \label{DilTransf}
\t'(x) = \tau(x) + \Lambda\, \s(x) \, .
\eeq  

The Weyl invariant terms may take the form of any scalar contraction of $\hat R_{\mu\nu\rho\sigma}$, $\hat R_{\mu\nu}$ and $\hat R$
and can be classified by their mass dimension. Typical examples are
\beq
\mathcal{J}_n\sim \frac{1}{\Lambda^{2(n-2)}}\int d^4 x \sqrt{\hat{g}}\hat{R}^n,
\eeq
and so forth, with the case $n=1$ describing the relevant operator in the infrared which reproduces the kinetic term of the dilaton. 
These terms can be included into a local action of the metric, $\Gamma_0[\hat{g}]\equiv \Gamma_0[g,\tau]$, extended with the inclusion of $\tau(x)$.  
\beq
\Gamma_0[\hat{g}]\sim \sum_n \mathcal J_n[\hat{g}].
\eeq
The leading contribution to $\Gamma_0$ is the kinetic term for the dilaton, which can be obtained in two ways. The first method is to 
consider the Weyl-gauged Einstein-Hilbert term
\beqa
\int d^dx\, \sqrt{\hat g}\, \hat R 
&=& 
\int d^dx\, \sqrt{g}\, e^{\frac{(2-d)\,\tau}{\Lambda}}\, \bigg[
R - 2\, \left(d-1\right)\, \frac{\Box\tau}{\Lambda} +
\left(d-1\right)\,\left(d-2\right)\, \frac{\pd^\lambda\tau\,\pd_\lambda\tau}{\Lambda^2} \bigg] \nn \\
&=&
\int d^dx\, \sqrt{g}\, e^{\frac{(2-d)\,\tau}{\Lambda}}\, \bigg[R - 
\left(d-1\right)\,\left(d-2\right)\, \frac{\pd^\lambda\tau\,\pd_\lambda\tau}{\Lambda^2} \bigg]\, ,
\eeqa
with the inclusion of an appropriate normalization 
\beqa \label{EH}
\mathcal S^{(2)}_{\tau} = - \frac{\Lambda^{d-2}\, \left(d-2\right)}{8\,\left(d-1\right)}\, \int d^dx\, \sqrt{\hat g}\, \hat R  \, ,
\eeqa
{\em which reverses} the sign in front of the Einstein term. Indeed, the extraction of a conformal factor ($\tilde{\sigma}$) 
from the Einstein-Hilbert term from a fiducial metric $\bar{g}_{\mu\nu}$ ($g_{\mu\nu}=\bar{g}_{\mu\nu}e^{\tilde{\sigma}}$) generates 
a kinetic term for ($\tilde{\sigma}$) which is ghost-like.

An alternative method consists in writing down the usual conformal invariant action for a scalar field $\chi$ 
in a curved background
\beqa \label{ScalarImprovedChi}
\mathcal S^{(2)}_{\chi} = \frac{1}{2}\, \int d^dx\, \sqrt{g}\, \bigg( 
g^{\mu\nu}\,\pd_\mu \chi\, \pd_\nu\chi - \frac{1}{4}\,\frac{d-2}{d-1}\, R\,\chi^2 \bigg) \, .
\eeqa
By the field redefinition $\chi\equiv\Lambda^{\frac{d-2}{2}}\, e^{-\frac{(d-2)\,\tau}{2\Lambda}}$ Eq.  (\ref{ScalarImprovedChi}) 
becomes
\beq \label{DilatonKinetic}
\mathcal S^{(2)}_{\tau} = \frac{\Lambda^{d-2}}{2}\, \int d^dx\, \sqrt{g}\, e^{-\frac{(d-2)\,\tau}{\Lambda}}\, \bigg( 
\frac{(d-2)^2}{4\,\Lambda^2}\, g^{\mu\nu}\,\pd_\mu \tau\, \pd_\nu\tau - \frac{1}{4}\, \frac{d-2}{d-1}\, R \bigg) \, ,
\eeq
which, for $d=4$, reduces to the familiar form
\beq \label{KinTau}
\mathcal S^{(2)}_{\tau} = \frac{1}{2}\, \int d^4 x\, \sqrt{g}\, e^{-\frac{2\,\tau}{\Lambda}}\, \bigg(
g^{\mu\nu}\,\pd_\mu \tau\, \pd_\nu\tau - \frac{\Lambda^2}{6}\, R \bigg)\, 
\eeq
and coincides with the previous expression (\ref{EH}), obtained from the formal Weyl invariant construction.

In four dimensions we can build the following possible subleading contributions (in $1/\Lambda$) to the effective action which, when gauged, 
can contribute to the fourth order dilaton action 
\beq
\mathcal{S}^{(4)}_\tau = \int d^4x\, \sqrt{g}\, \bigg( \alpha\, R^{\mu\nu\rho\sigma}\, R_{\mu\nu\rho\sigma} + \beta\, R^{\mu\nu}\, R_{\mu\nu} 
                            + \gamma\, R^2 + \delta\, \Box R\bigg)\, .
\eeq
The fourth term ($\sim \Box R$) is just a total divergence, whereas two of the remaining three terms can be 
traded for the squared Weyl tensor $F$ and the Euler density $G$. 
As $\sqrt{g}\, F$ is Weyl invariant and $G$ is a topological term, neither of them contributes, 
when gauged according to (\ref{MetricGauge}), so that the only non vanishing four-derivative term
in the dilaton effective action in four dimensions is
\beq \label{UpToMarginal}
\mathcal{S}^{(4)}_{\tau} = \gamma\, \int d^4x\, \sqrt{\hat{g}}\, \hat{R}^2 = 
\gamma\, \int d^4x\, \sqrt{g}\, 
\bigg[ R - 6\, \bigg( \frac{\Box\tau}{\Lambda} - \frac{\pd^\lambda\tau\,\pd_\lambda\tau}{\Lambda^2} \bigg) \bigg]^2, \, 
\eeq
with $\gamma$ a dimensionless constant.
If we also include a possible cosmological constant term, $ S^{(0)}_{\tau}$, we get the final form 
of the dilaton effective action in $d=4$ up to order four in the derivatives of the metric tensor
\beqa
\label{tot}
\mathcal{S}_{\tau} &=& 
\mathcal{S}^{(0)}_{\tau} + \mathcal{S}^{(2)}_{\tau} + \mathcal{S}^{(4)}_{\tau} + \dots =
\int d^4x\, \sqrt{\hat{g}}\, \bigg\{ \alpha
- \frac{\Lambda^{d-2}\, \left(d-2\right)}{8\,\left(d-1\right)}\, \hat{R} + \gamma\, \hat{R}^2\, 
\bigg\} + \dots\, ,
\eeqa
where the ellipsis refer to additional operators which are suppressed in $1/\Lambda$. 
In flat space ($g_{\mu\nu}\rightarrow \delta_{\mu\nu}$), (\ref{UpToMarginal}) becomes
\beq
\mathcal{S}_{\tau} = \int d^4x\, \bigg[ e^{-\frac{4\,\tau}{\Lambda}}\, \alpha + 
\frac{1}{2}\, e^{-\frac{2\,\tau}{\Lambda}} \, \pd^\lambda\tau\,\pd_\lambda\tau +
36\,\gamma\, \bigg( \frac{\Box\tau}{\Lambda} - \frac{\pd^\lambda\tau\,\pd_\lambda\tau}{\Lambda^2} \bigg) \bigg] + \dots
\eeq
This approach can be extended to the anomaly counterterms as well, by considering the renormalized action $\Gamma_{\textrm{ren}}[g,\tau]$. This takes to a Wess-Zumino (WZ) action which consistently accounts for the anomaly, at the price of including an extra dynamical degree of freedom and a kinetic flip. The latter is obtained by the relation
\beqa \label{ExtractWZ}
\Gamma_{WZ}[g,\tau] =\Gamma_{\textrm{ren}}[g,\tau] - \hat\Gamma_{\textrm{ren}}[g,\tau]
\eeqa
and takes the WZ form 
\beqa
\Gamma_{WZ}[g,\tau] 
&=&
\int d^4x\, \sqrt{g}\, \bigg\{ \beta_a\, \bigg[ \frac{\tau}{\Lambda}\, \bigg( F - \frac{2}{3} \Box R \bigg) + 
\frac{2}{\Lambda^2}\, \bigg( \frac{R}{3}\, \pd^\lambda\tau\, \pd_\lambda\tau + \left( \Box \tau \right)^2 \bigg) - 
\frac{4}{\Lambda^3}\, \pd^\lambda\tau\,\pd_\lambda\tau\,\Box \tau + 
\frac{2}{\Lambda^4}\, \left( \pd^\lambda\tau\, \pd_\lambda\tau \right)^2 \bigg] \nn \\
&& 
\hspace{15mm}
+\, \beta_{b}\, \bigg[ \frac{\tau}{\Lambda}\,G  - 
\frac{4}{\Lambda^2}\, \bigg( R^{\alpha\beta} - \frac{R}{2}\,g^{\alpha\beta} \bigg)\, \pd_\alpha\tau\, \pd_\beta\tau -
\frac{4}{\Lambda^3} \, \pd^\lambda\tau\,\pd_\lambda\tau\,\Box \tau + 
\frac{2}{\Lambda^4}\, \left( \pd^\lambda\tau\, \pd_\lambda\tau \right)^2 \bigg]\bigg\} \, .\nn \\
\label{Effective4d}
\eeqa
This action is local and contains an asymptotic dilaton, but does not appear to be equivalent to the nonlocal action (the Riegert action)
\be
\cS_{\rm anom}^{^{NL}}[g] =\frac {1}{4}\!\int \!d^4 x\sqrt{-g_x}\, \Big(E - \frac{2}{3} \square R\Big)_{\!x} 
\int\! dx'\sqrt{-g_{x'}}\,D_4(x,x')\bigg[\frac{b'}{2}\, \big(E - \frac{2}{3} \square R\big) +  b\,C^2\bigg]_{x'}
\label{Snonl}
\ee
where $D_4(x,x') = (\D_4^{-1})_{xx'}$  and

\be
\D_4 \equiv \nabla_\mu \left(\nabla^\mu\nabla^\nu +2R^{\mu\nu} - \frac{2}{3} R g^{\mu\nu} \right)\nabla_\nu
=\square^2 + 2 R^{\mu\nu}\nabla_\mu\nabla_\nu - \frac{2}{3} R \square + \frac{1}{3} (\nabla^\mu R)\nabla_\mu
\label{elfdef}
\ee
is the unique fourth order scalar kinetic operator that is conformally covariant
\be
\sqrt{-g}\, \D_4 = \sqrt{-\bar g}\, \bar \D_4
\label{invfour}
\ee
under the local conformal reparameterization $g_{\m\n} = e^{2\s} \bar g_{\m\n}$, for an arbitrary rescaling $\s(x)$ \cite{Riegert:1987kt,Antoniadis:1992xu}. \\
Such second form of the anomaly action, without an asymptotic dilaton field, is what one rediscovers both from a perturbative analysis and from the solution of the CWI's for 3-point functions, as we will be discussing below. We are going to emphasize one key feature of such two actions.
\section{Turning to general issues: The perturbative structure of $1PI$ anomaly actions for chiral and conformal anomalies}
\label{ccff}
The type of actions discussed above, which enlarge the spectrum of the SM by the inclusion of an asymptotic dilaton field, are the most popular ones, and are justified within an ordinary phenomenological approach. Obviously, they do not introduce a dilaton dynamically, but are viable anyhow, as far as the issue of the explicit breaking of the dilaton mass is accepted on a purely phenomenological basis.
An alternative approach consists in starting from a given classically conformal invariant theory and investigate the quantum corrections which are responsible for the generation of the conformal anomaly. In this case, the basic procedure for chiral and conformal anomalies are quite similar, although for a chiral anomaly the approach is far simpler, since only a linear coupling of a Nambu-Goldstone mode to the anomaly is sufficient in order to generate the anomaly contribution. This type of approach is at the core of the St\"uckelberg mechanism for the cancelation of the gauge anomalies generated by a certain anomalous gauge interaction $(B_\mu)$. The method allows to obtain a physical axion only in the presence of a non-perturbative periodic potential, which can be easily justified under the assumption that at a phase transition the instanton sector can generate it. We refer to a recent review for a discussion of such a mechanism \cite{Coriano:2018uip}. \\ 
\subsection{The chiral vertex}
For a chiral anomaly, the main features of the 1PI effective actions have been discussed in detail in \cite{Giannotti:2008cv,Armillis:2009sm}. They are based on a perturbative representation of the anomaly vertex which is equivalent to the original description in terms of 6 form factors found on textbooks, but formulated in terms of longitudinal and transverse components, as we are going to illustrate. \\
We recall that the $AVV$ amplitude with off-shell external lines is parameterized in the form 

\bea
\Delta_0^{\la\mu\nu} &=& V_1 (k_1, k_2) \veps [k_1,\mu,\nu,\la] + V_2 (k_1, k_2)\veps [k_2,\mu,\nu,\la] +
V_3 (k_1, k_2) \veps [k_1,k_2,\mu,\la]{k_1}^{\nu} \nonumber \\
&+&  V_4 (k_1, k_2) \veps [k_1,k_2,\mu,\la]k_2^{\nu}
+ V_5 (k_1, k_2)\veps [k_1,k_2,\nu,\la]k_1^\mu
+ V_6 (k_1, k_2) \veps [k_1,k_2,\nu,\la]k_2^\mu\nonumber \\
\label{Ros}
\eea
where $\veps [k_1,\mu,\nu,\la]\equiv \veps^{\alpha\mu\nu\la} k_{1 \alpha}$, and so on, with $k_1$ and $k_2$ the momenta of the two vector lines. 
 The four invariant amplitudes $V_i$ for $i\geq3$ are finite and given by explicit parametric integrals \cite{Rosenberg:1962pp}
\bea
V_3(k_1, k_2) &=& - V_6 (k_2, k_1) =  - 16 \pi^2 I_{11}(k_1, k_2), \\
V_4(k_1,k_2) &=& - V_5 (k_2, k_1) = 16 \pi^2 \left[ I_{20}(k_1,k_2) - I_{10}(k_1,k_2) \right],
\eea
where the general massive $I_{st}$ integral is defined by
\bea
I_{st}(k_1,k_2) = \int_0^1 dw \int_0^{1-w} dz w^s z^t \left[ z(1-z) k_1^2 + w(1-w) k_2^2 + 2 w z (k_1 k_2) - m^2 \right]^{-1},
\eea
Both $A_1$ and $A_2$ are instead represented by formally divergent integrals, which can be rendered finite only by imposing the Ward identities on the two vector lines, giving
\bea
V_1 (k_1,k_2) &=& k_1 \cdot k_2 \, V_3 (k_1,k_2) + k_2^2 \, V_4 (k_1,k_2),
\label{WI1} \\
V_2 (k_1,k_2) &=& k_1^2 \, V_5 (k_1,k_2) + k_1 \cdot k_2 \, V_6 (k_1,k_2),
\label{WI2}
\eea
which allow to re-express the formally divergent amplitudes in terms of the convergent ones.
The  Bose symmetry on the two vector vertices with indices $\mu$ and $\nu$ is fulfilled thanks to the relations
\bea
V_5(k_1,k_2) &=& - V_4(k_2, k_1)\\
V_6(k_1, k_2) &=& - V_3 (k_2, k_1).
\eea
Coming to the second parameterization of the three-point correlator function, this is the one presented in \cite{Knecht:2003xy}.  One of the features of this parameterization is the presence of a longitudinal contribution (i.e. of an anomaly pole), apparently for generic virtualities of the external momenta of the two vector lines. 
For on-shell photons there is a single form factor with a $1/k^2$ behaviour $(k=k_1+k_2)$ in the only longitudinal structure of the vertex. In the general off-shell case several structures are affected by other poles in their transverse components as well, raising some doubts about the significance of the longitudinal pole, for general kinematics. The various contributions and parameterizations can be related one to the other by the Schoutens relations, as discussed in \cite{Armillis:2009sm}. \\ 
This second parameterization plays an important role in the description of the anomalous magnetic moment of the muon \cite{Knecht:2003xy}.
 For this reason we start by recalling the structure of such L/T parameterization, which separates the longitudinal (L) from the transverse (T) components of the anomaly vertex, which is given by

\beq
 \mathcal \, W ^{\lambda\mu\nu}= \frac{1}{8\pi^2} \left [  \mathcal \, W^{L\, \lambda\mu\nu} -  \mathcal \, W^{T\, \lambda\mu\nu} \right],
\label{long}
\eeq
where the longitudinal component
\beq
 \mathcal \, W^{L\, \lambda\mu\nu}= w_L  \, k^\lambda \veps[\mu,\nu,k_1,k_2]
\eeq
(with $w_L=- 4 i /k^2 $) describes the anomaly pole, while the transverse contributions take the form
\beqa
\label{calw}
{  \mathcal \, W^{T}}_{\lambda\mu\nu}(k_1,k_2) &=&
w_T^{(+)}\left(k^2, k_1^2, k_2^2 \right)\,t^{(+)}_{\lambda\mu\nu}(k_1,k_2)
 +\,w_T^{(-)}\left(k^2, k_1^2,k_2^2\right)\,t^{(-)}_{\lambda\mu\nu}(k_1,k_2) \nonumber \\
 && +\,\, {\widetilde{w}}_T^{(-)}\left(k^2, k_1^2, k_2^2 \right)\,{\widetilde{t}}^{(-)}_{\lambda\mu\nu}(k_1,k_2),
 \eeqa
with the transverse tensors given by
\beqa
t^{(+)}_{\lambda\mu\nu}(k_1,k_2) &=&
k_{1\nu}\, \veps[ \mu,\la, k_1,k_2]  \,-\,
k_{2\mu}\,\veps [\nu,\la, k_1, k_2]  \,-\, (k_{1} \cdot k_2)\,\veps[\mu,\nu,\la,(k_1 - k_2)]
\nonumber\\
&& \quad\quad+ \, \frac{k_1^2 + k_2^2 - k^2}{k^2}\, \, k_\la \, \,
\veps[\mu, \nu, k_1, k_2]
\nonumber \ , \\
t^{(-)}_{\lambda\mu\nu}(k_1,k_2) &=& \left[ (k_1 - k_2)_\la \,-\, \frac{k_1^2 - k_2^2}{k^2}\,\, k_\la \right] \,\veps[\mu, \nu, k_1, k_2]
\nonumber\\
{\widetilde{t}}^{(-)}_{\lambda\mu\nu}(k_1,k_2) &=& k_{1\nu}\,\veps[ \mu,\la, k_1,k_2] \,+\,
k_{2\mu}\,\veps [\nu,\la, k_1, k_2] \,
-\, (k_{1}\cdot k_2)\,\veps[ \mu, \nu, \la, k].
\label{tensors}
\eeqa
taking to a parameterization of the vertex in the form
\beq
 \Gamma^{(3)}=  \Gamma^{(3)}_{pole} + \tilde{\Gamma}^{(3)}
 \eeq
 with the pole part, coupled to the external axial vector field $B_\mu$ given by
\beq
\Gamma^{(3)}_{pole}= -\frac{1}{8 \pi^2} \int d^4 x \, d^4 y  \,\partial \cdot B(x) \square^{-1}_{x,y} F(y) \wedge F(y)
\label{gammapole}
\eeq
and the rest ($\tilde{\Gamma}^{(3)}$) given by a complicated nonlocal expression which contributes homogeneously to the Ward identify of the anomaly graph.\\
All the terms in this parameterization are linked together, and the isolation of a single contribution from the rest is only possible for specific kinematics. Nevertheless, around the light-cone $(k^2\to 0)$ the anomaly can be attributed to the pole-like behaviour of the longitudinal part. Notice that the absence of the pole structure in Rosenberg's formulation of the AVV \cite{Rosenberg:1962pp} is due to the redundancy of the latter. Poles can be removed and reinserted in a given parameterization by using the Schoutens relations, but there is no way one can remove the anomaly pole of $W_L$ consistently. One can naturally try to to unconver the meaning of such interactions. As we are going to see, at least from the analysis of 3-point functions, the effective action shows the emergence of some instabilities, signalling the possibility that the vacuum will be restructured in the presence of such interactions. This feature is probably shared by the effective actions of both chiral and conformal anomalies, which get unified in the context of supersymmetric theories. We are going to illustrate this phenomenon in the chiral case, where it has been worked out in some detail.

\subsection{A ghost in the spectrum }
As an example we consider a gauge boson $B$ coupled to a single chiral fermion which generates a 1PI effective action of the form 
 
\begin{eqnarray}
&&\emph{L}=\bar{\psi}(i\not{\partial}+g\not{B}\gamma_5)\psi-\frac{1}{4}F^2_B + \langle \Delta_{BBB} BBB\rangle + c_2 \partial B \frac{1}{\square} 
F_B\tilde{F}_B +\ldots
\label{WZ2}
\eeqa
where the ellipsis refer to additional transverse terms identified in the trilinear $BBB$ vertex, which is anomalous. We have isolated the longitudinal contribution related to the anomaly pole and we will focus in this contribution. It is easy to show that the $1/\square$  term can be generated by the introduction of two pseudoscalar fields $a$ and $b$ which allow to remove the nonlocal contribution of the action

\begin{eqnarray}
\mathcal{L} &=& \overline{\psi} \left( i \not{\partial}  + g\not{B} \gamma_5\right)\psi - \frac{1}{4} F_B^2  +\langle \Delta_{BBB}BBB\rangle + c_3 F_B\wedge F_B ( a + b) \nonumber \\
&& + \frac{1}{2}  \left( \partial_\mu b - M_1 B_\mu\right)^2 -
\frac{1}{2} \left( \partial_\mu a - M_1 B_\mu\right)^2, 
\label{fedeq}
\eeqa
where both $a$ and $b$ shift under the gauge symmetry 
\beq
\delta b = M_1 \theta_B(x) \qquad \qquad \delta B_\mu= \partial_\mu \theta_B(x)
\label{shifts}
  \eeq
and where $\theta_B(x)$ parameterizes a gauge transformation. The inclusion of two pseudoscalars which acquire St\"uckelberg mass terms, given by the second line of the equation above, is a specific feature of this reformulation, with a St\"uckelberg mass $M_1$.
The equivalence between (\ref{WZ2}) and (\ref{fedeq}) can be proven directly from the functional integral, integrating out both $a$ and $b$, which gives two gaussian integrations \cite{Coriano:2008pg}. Notice that $b$ has a positive kinetic term and $a$ is ghost-like. A similar behaviour obviously holds also in the case of an external axial-vector current coupled to 
a corresponding classical gauge field if the fermion spectrum is anomalous. In this case the appearance of a ghost in the effective action indicates the onset of an instability. Notice also that the inclusion of a St\"uckelberg mass term indicates that we need to introduce a suitable scale in order to be able to define a local action. \\
There is a third equivalent formulation of the same action (\ref{fedeq}) which can be defined with the inclusion of a kinetic mixing between the two pseudoscalars. This has been given for QED (with a single fermion) coupled to an external axial-vector field $\mathcal{B}_\mu$ \cite{Giannotti:2008cv} and takes the form 

\beq
\mathcal{L}=\partial_\mu\eta \partial^\mu\chi - \chi \partial \mathcal{B} + \frac{e^2}{8 \pi^2}\eta F \tilde{F},
\label{GM}
\eeq
where $F$ is the field strength of the photon $A_\mu$ while $\mathcal{B}_\mu$ takes the role of a source.  It is quite straightforward to relate (\ref{fedeq}) and (\ref{GM}). This can be obtained by the field redefinitions
\beqa
\eta &=& \frac{(a + b)}{M}, \nonumber \\
\chi &=& M(a - b),
\label{changes}
\eeqa
showing that indeed a mixing term is equivalent to the presence of either an anomaly pole or to two pseudoscalars in the spectrum of the theory, one of them being a ghost, and the inclusion of a scale at which to define their decoupling, which is the St\"uckelberg scale. Notice that in Eq. \eqref{changes} $\chi$ is gauge invariant while $\eta$ is not.

 \section{The Coleman-Weinberg potential and ghost condensation at trilinear level}
We have clarified that a Lagrangian containing a pole counterterm shows some nontrivial features.  In particular, the presence of the nonlocal $\partial B \square^{-1} F\tilde{F}$ interaction induced by an anomaly pole, rewritten  in  a local version, allows to proceed with some further perturbative analysis which sheds some light on the character of the effective potential of the ghost field $a$ defined in (\ref{fedeq}). \\
We can use the Coleman-Weinberg approach, which indicates the presence of an instability in the action, obtained after integration over all the remaining fields of the model. The instability is signalled by the presence of a ghost condensate at 1-loop level.  To illustrate this formal result  we follow closely the analysis of \cite{Armillis:2011hj}.

Consider the gauge-fixed version of the Lagrangian in (\ref{fedeq}) given by 

\begin{eqnarray}
&&\emph{L}=\bar{\psi}(i\not{\partial}+e\not{B}\gamma_5)\psi-\frac{1}{4}F^2_B -\frac{(\partial_\mu B^\mu)^2}{2\alpha} + \frac{e^3}{48\pi^2M_1}F_B\wedge F_B (a+b)\nonumber\\
&&+\frac{1}{2}(\partial_\mu b -M_1 B_\mu)^2 -\frac{1}{2}(\partial_\mu a -M_1 B_\mu)^2
\label{fed}
\end{eqnarray}
where $\alpha$ is the gauge parameter. 
We shift the ghost field, separating the classical 
ghost background (still denoted as $a(x))$, from its quantum fluctuating part on which we will integrate, $A(x)$
\begin{equation}
a(x) \longrightarrow a(x) + A(x).
\end{equation}
Dropping the linear terms in the quantum fluctuation field $A(x)$ and taking just the quadratic part of all the quantum fields we get the quadratic Lagrangian

\begin{eqnarray}
&&\mathcal{L}_{\textrm{quad}}= \bar{\psi}i\not{\partial}\psi -\frac{1}{4}F^2_B -\frac{(\partial_\mu B^\mu)^2}{2\alpha}+ \frac{e^3}{48\pi^2M_1}a F_B\wedge F_B \nonumber\\
&& \qquad \qquad \qquad +\frac{1}{2}(\partial b)^2 - M_1 \partial_\mu b B^\mu -\frac{1}{2}(\partial_\mu A)^2 +M_1 B_\mu \partial^\mu A,
\end{eqnarray} 
from which we can determine, after an integration by parts, the contribution which is quadratic in the anomalous gauge field $B$ 
\begin{equation}
\frac{1}{2}\int d^4 x B_\mu \Big[g^{\mu\nu} \Box -\left(1-\frac{1}{\alpha}\right)\partial_\mu\partial_\nu +\frac{e^3}{24\pi^2M_1}\,\partial_\alpha a\, \epsilon^{\mu\alpha\rho\nu}\,\partial_\rho\,\Big]B_\nu.
\end{equation}

The one loop effective action in the background of the ghost $a$ is obtained by integration over all the quantum fields in the form 
\begin{eqnarray}
&&e^{i\Gamma[a]}=\int[DA][D\psi][D\bar{\psi}][DB][Db]\times \nonumber\\
&& \qquad \qquad \exp  \{i\int d^4x \Big[\bar{\psi}i\not{\partial}\psi-\frac{1}{4}F_B^2-\frac{(\partial_\mu B^\mu)^2}{2\alpha}+\frac{1}{2}(\partial b)^2-M_1\partial_\mu b B^\mu-\frac{1}{2}(\partial_\mu A)^2+M_1B_\mu\partial^\mu A\nonumber\\
&&\qquad \qquad \qquad \qquad +\frac{e^3}{48\pi^2M_1}F_B\wedge F_B a\Big]\Big\}.
\end{eqnarray} 
The integration over the quantum fluctuations of the ghost field $A$ gives

\begin{eqnarray}
&&\int[DA]\exp\Big[ i\int d^4x\Big({\frac{1}{2}A\Box A-M_1\partial_\mu B^\mu A}\Big)\Big]\propto\nonumber\\
&&\exp\Big [-\frac{1}{2}\int d^4 x d^4 y\Big( M_1 \partial_\mu B^\mu (x) D_F(x-y) M_1 \partial_\nu B^\nu (y)\Big)\Big] 
\label{effect1}
\end{eqnarray}
where
\begin{equation}
D_F(x-y)=\int \frac{d^4 p}{(2\pi)^4}\frac{-i e^{ip(x-y)}}{p^2-i\epsilon}
\end{equation} 
is the propagator for the quantum fluctuations of the ghost field. The integration over the axion $b$ induces some cancelations of various terms giving

\begin{eqnarray}
\int[Db]\exp\Big[i\int d^4x\Big({-\frac{1}{2}b\Box b + M_1\partial_\mu B^\mu b}\Big)\Big]\propto\exp\Big[-\frac{1}{2}\int d^4 x d^4 y M_1 \partial_\mu B^\mu (x) D^{1}_F(x-y) M_1 \partial_\nu B^\nu (y)\Big], \nonumber \\ 
\label{effect2}
\end{eqnarray}
where we have introduced the propagator of the axion field
\begin{equation}
D^1_F(x-y)=\int \frac{d^4 p}{(2\pi)^4}\frac{i e^{ip(x-y)}}{p^2+i\epsilon}.
\end{equation}
Notice that $D_F(x-y)+D^1_F(x-y)$ vanishes in the limit $\epsilon\rightarrow 0$, thus the integration in $A$ and $b$ eliminates the terms (\ref{effect1}) and (\ref{effect2}) from the action.  Therefore we are just left with the expression
\begin{equation}
e^{i\Gamma[a]}\propto \int [DB]\exp\Big[-\frac{1}{4}i\int d^4 x(F_B)^2 -\int d^4 x \frac{(\partial_\mu B^\mu)^2}{2\alpha}+i\frac{e^3}{24\pi^2M_1}\int d^4 x a F_B\wedge F_B \Big].
\label{effective} 
\end{equation}

 Defining
\begin{equation}
l\equiv \frac{e^3}{24\pi^2M_1}\,\,\,\,\,\,\, {\rm and }\,\,\,\phi_\alpha\equiv\partial_\alpha a,
\end{equation}
the effective action of the classical background ghost field is then given by
\begin{equation}
i\Gamma[a]=-\frac{1}{2}{\rm Tr}\log\left(\Box {g^\mu}_\nu -\left(1-\frac{1}{\alpha}\right)\partial_\mu\partial_\nu + l {\epsilon^{\mu\alpha\rho}}_\nu \phi_\alpha\partial_\rho\right)
\label{finaleffect}
\end{equation}
where the trace ${\rm Tr}$, as usual, must be taken in the functional sense.

To perform the calculation of (\ref{finaleffect}) we use the heat kernel method and define the functional determinant in 
(\ref{finaleffect}) using a $\zeta$ function regularization. We take $\phi_\alpha$ to be constant. We have

\begin{equation}
\log \det Q = -\displaystyle\lim_{s\to 0}\frac{d}{ds}\frac{\mu^{2s}}{\Gamma(s)}\int^{+\infty}_{0} dt\, t^{s-1} {\rm Tr}(e^{-tQ})
\end{equation}
with the functional trace performed in the plane wave basis\begin{equation}
{\rm Tr e^{-tQ}}=\int d^4 x\, \textrm{tr} <x|e^{-tQ}|x> =\int d^4 x \, \textrm{tr}\int \frac{d^4 k}{(2\pi)^4}e^{-ikx}e^{-tQ}e^{ikx},
\label{zita}
\end{equation}
and with $\textrm{tr}$ denoting the trace on the Lorentz indices. Further manipulations give
\begin{eqnarray}
{(e^{-ikx}e^{-tQ}e^{ikx})^\mu}_\tau 
&& ={g^\mu}_\tau e^{tk^2}\exp\left({-itl{{\epsilon^\tau}_\nu}^{\alpha\rho}\phi_\alpha k_\rho}\right)+(1-\frac{1}{\alpha})\frac{k^\mu k^\nu}{k^2} (1-e^{t k^2}),
\label{matrix}
\end{eqnarray}
where we have used the relation
\beqa
e^{-t k^\mu k_\nu}&=& g^\mu_\nu+\frac{k^\mu k_\nu}{k^2}(e^{-t k^2}-1) \nonumber \\
k^\mu k_\tau e^{-itl{{\epsilon^\tau}_\nu}^{\alpha\rho}\phi_\alpha k_\rho}&=& k^\mu k_\nu. 
\eeqa
We need to consider in (\ref{matrix}) just the $\phi$-dependent part. As usual, the Coleman-Weinberg potential is gauge-dependent. 
In this case the dependence on the gauge-fixing parameter $\alpha$ can be assimilated to the constant terms.
The functional trace receives contributions only from the terms with $n$ even, and after some manipulations we obtain 
\begin{eqnarray}
tr (e^{-ikx}e^{-tQ}e^{ikx})
&=& -2e^{t k^2}\cosh{\,tl\sqrt{k^2\phi^2-(k\cdot\phi)^2}}+{\rm const}.
\label{traza}
\end{eqnarray}
Inserting (\ref{traza}) into (\ref{zita}) we get, apart from a constant factor of infinite volume, the expression of the trace 
\begin{equation}
\textrm{Tr} e^{-t Q} \sim -2\int\frac{d^4 k}{(2\pi)^4}e^{t k^2}\cosh{\,tl\sqrt{k^2\phi^2-(k\cdot\phi)^2}},
\label{integraldifficile}
\end{equation}
giving an effective potential for the background $\phi_\alpha$ of the form
\begin{equation}
V[\phi]=-\displaystyle\lim_{s\to 0}\frac{d}{ds}\frac{\mu^{2s}}{\Gamma(s)}\int^{+\infty}_{0} dt\, t^{s-1}\int\frac{d^4 k}{(2\pi)^4}e^{t k^2}\cosh{\,tl\sqrt{k^2\phi^2-(k\cdot\phi)^2}}. 
\label{effective}
\end{equation}
We can obtain the leading contribution of this effective potential by expanding the integrand in $l$, i.e. in $1/M_1$.
After performing the expansion and restoring the infinite space-time volume we obtain the effective action
\begin{equation}
S=\int{d^3 x d t }\left\{ \left(-\frac{1}{2}-\frac{3 l^2}{32 \pi^2}\right)(\partial a)^2 + \frac{5 l^4}{256 \pi^2}(\partial a)^4\right\}
\end{equation}
which obviously can be rewritten as
\begin{equation}
S=\int{d^3 x d t }\left\{-\frac{1}{2}(\partial a)^2 + \frac{5 l^4}{256 \pi^2}(\partial a)^4\right\}.
\label{effectivaction}
\end{equation}
Notice that the polynomial in the integrand 
\begin{equation}
P(\phi)=-\frac{1}{2}\phi^2+ \frac{5 l^4}{256 \pi^2}\phi^4
\label{polynomium}
\end{equation}
has a minimum at
\begin{equation}
\bar{\phi}^2\sim\frac{1}{l^2}\sqrt{\frac{128\pi^2}{5}}>0.
\end{equation}
To investigate the character of the minimum and of the fluctuations around this minimum, we select a time-like frame, where the background takes the form
\begin{equation}
\bar{a}=\bar{\phi}t.
\end{equation}
If we now consider small fluctuactions around this configuration of minimum, denoted as $\pi$
\begin{equation}
a=\bar{\phi}t+\pi
\end{equation}
and expanding (\ref{effectivaction}) we obtain the action

\begin{equation}
S=\int d^3 x d t \left\{
\dot{\pi}^2+\sqrt{\frac{5}{128\pi^2}}l^2\dot{\pi}^3+\frac{5 l^4}{512\pi^2}\dot{\pi}^4 + \frac{5 l^4}{512\pi^2}\dot{\pi}^4 |\nabla \pi|^4-\sqrt{\frac{5}{128\pi^2}} l^2 \dot{\pi}|\nabla\pi|^2+\cdots\right\}.
\label{arka}
\end{equation}
This action has the same form as in \cite{ArkaniHamed:2003uy} (see formula (4.2)). As in this previous analysis, we do not get the term $|\nabla\pi|^2$ since its coefficient is proportional to $P^\prime (\bar{\phi}) = 0$. Clearly, the Lorentz symmetry is broken, at least at 1-loop level, and is signalling an instability of the local model 
(\ref{WZ2}) generated in the infrared region.  Notice, in fact, that in the Coleman-Weinberg approach we are closing the gauge boson loop and we are taking the long wavelength limit of the external background ghost field. 
Finally, one should also notice that the dependence of the effective potential on $M_1$, in this approach, is recovered at higher orders. For instance, additional contributions, suppressed by $1/M_1^2$, are obtained by the insertion of the self-energy of the anomalous gauge boson on the lowest order contribution (the gauge boson loop). These features of actions containing Wess-Zumino terms have been studied in the past with similar results \cite{Andrianov:1998ay} \cite{Mavromatos:2010ar}.\\
There are some conclusions that one can draw from this simple analysis. There are some indications that the presence of an anomaly pole in a chiral theory, taken face value, generates a vacuum instability which leads to ghost condensation, and a similar feature is expected from the analysis of a conformal anomaly pole. It is natural to think that the vacuum gets redefined if the one loop analysis can be trusted.

\section{The 1PI for the $TJJ$ in QED and QCD}
The perturbative character of chiral and conformal anomalies is encoded in both cases in such singularities, that we will describe jointly in the case of a supersymmetric $\mathcal{N}=1$ theory where they appear in a single anomaly multiplet. Before coming to that point, it is convenient to summarize some basic findings concerning the structure of the 1PI anomaly action in the conformal case, which allow to extend the considerations presented before in the AVV diagram to the new case.\\
Once we move to the analysis of the conformal anomaly, the diagrammatic expansion induces at 1-loop level a trace contribution, identified at lowest order in the gravitational coupling in correlators with a single insertion of a stress-energy tensor. The $TJJ$ vertex, in QED, describes the coupling of a graviton to two photons and provides the simplest realization of this phenomenon. The correlator is shown in Fig. \ref{vertex} for QED.
\begin{figure}[t]
\begin{center}
\includegraphics[scale=0.9]{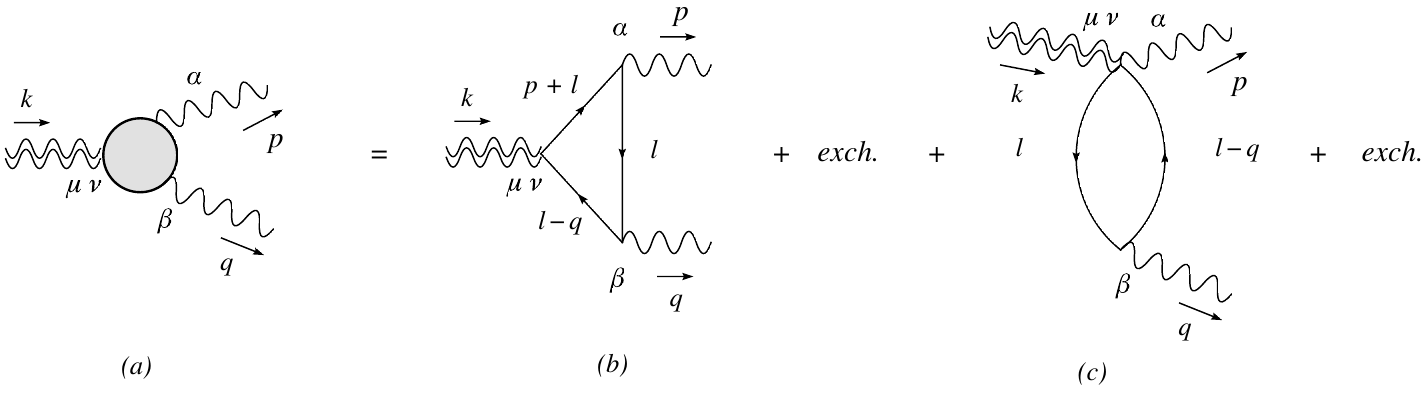}
\caption{\small The complete $TJJ$ one-loop vertex (a) given by the sum of the 1PI contributions with triangle (b) and pinched topologies (c). }
\label{vertex}
\end{center}
\end{figure}

\begin{table}
$$
\begin{array}{|c|c|}\hline
i & t_i^{\mu\nu\alpha\beta}(p,q)\\ \hline\hline
1 &
\left(k^2 g^{\mu\nu} - k^{\mu } k^{\nu}\right) u^{\alpha\beta}(p.q)\\ \hline
2 &
\left(k^2g^{\mu\nu} - k^{\mu} k^{\nu}\right) w^{\alpha\beta}(p.q)  \\ \hline
3 & \left(p^2 g^{\mu\nu} - 4 p^{\mu}  p^{\nu}\right)
u^{\alpha\beta}(p.q)\\ \hline
4 & \left(p^2 g^{\mu\nu} - 4 p^{\mu} p^{\nu}\right)
w^{\alpha\beta}(p.q)\\ \hline
5 & \left(q^2 g^{\mu\nu} - 4 q^{\mu} q^{\nu}\right)
u^{\alpha\beta}(p.q)\\ \hline
6 & \left(q^2 g^{\mu\nu} - 4 q^{\mu} q^{\nu}\right)
w^{\alpha\beta}(p.q) \\ \hline
7 & \left[p\cdot q\, g^{\mu\nu}
-2 (q^{\mu} p^{\nu} + p^{\mu} q^{\nu})\right] u^{\alpha\beta}(p.q) \\ \hline
8 & \left[p\cdot q\, g^{\mu\nu}
-2 (q^{\mu} p^{\nu} + p^{\mu} q^{\nu})\right] w^{\alpha\beta}(p.q)\\ \hline
9 & \left(p\cdot q \,p^{\alpha}  - p^2 q^{\alpha}\right)
\big[p^{\beta} \left(q^{\mu} p^{\nu} + p^{\mu} q^{\nu} \right) - p\cdot q\,
(g^{\beta\nu} p^{\mu} + g^{\beta\mu} p^{\nu})\big]  \\ \hline
10 & \big(p\cdot q \,q^{\beta} - q^2 p^{\beta}\big)\,
\big[q^{\alpha} \left(q^{\mu} p^{\nu} + p^{\mu} q^{\nu} \right) - p\cdot q\,
(g^{\alpha\nu} q^{\mu} + g^{\alpha\mu} q^{\nu})\big]  \\ \hline
11 & \left(p\cdot q \,p^{\alpha} - p^2 q^{\alpha}\right)
\big[2\, q^{\beta} q^{\mu} q^{\nu} - q^2 (g^{\beta\nu} q^ {\mu}
+ g^{\beta\mu} q^{\nu})\big]  \\ \hline
12 & \big(p\cdot q \,q^{\beta} - q^2 p^{\beta}\big)\,
\big[2 \, p^{\alpha} p^{\mu} p^{\nu} - p^2 (g^{\alpha\nu} p^ {\mu}
+ g^{\alpha\mu} p^{\nu})\big] \\ \hline
13 & \big(p^{\mu} q^{\nu} + p^{\nu} q^{\mu}\big)g^{\alpha\beta}
+ p\cdot q\, \big(g^{\alpha\nu} g^{\beta\mu}
+ g^{\alpha\mu} g^{\beta\nu}\big) - g^{\mu\nu} u^{\alpha\beta} \\
& -\big(g^{\beta\nu} p^{\mu}
+ g^{\beta\mu} p^{\nu}\big)q^{\alpha}
- \big (g^{\alpha\nu} q^{\mu}
+ g^{\alpha\mu} q^{\nu }\big)p^{\beta}  \\ \hline
\end{array}
$$
\caption{The basis of 13 fourth rank tensors satisfying the vector current conservation on the external lines with momenta $p$ and $q$. \label{genbasis}}
\end{table}
\noindent The anatomy of this vertex is due to Giannotti and Mottola \cite{Giannotti:2008cv}, who have classified its possible tensor structures in terms of 13 form factors. On this basis, which is built by imposing on the $TJJ$ vertex all the Ward identities derived from diffeomorphism invariance, gauge invariance and Bose symmetry, the original 43 tensor structures can be reduced to this smaller number (see also the discussion in  \cite{Armillis:2009pq}. 
We report them in Table \ref{genbasis}. The vertex can be written as
\begin{equation}
\Gamma^{\m_1\n_1\m_2\m_3}(p_2,p_3)=\sum_{i=1}^{13}\,F_i(s;s_1,s_2,0)\,t_i^{\m_1\n_1\m_2\m_3}(p_2,p_3),
\end{equation} 
where the invariant amplitudes $F_i$ are functions of the kinematic invariants $s=p_1^2=(p_2+p_3)^2$, $s_1=p_2^2$, $s_2=p_3^2$, and the $t_i^{\m_1\n_1\m_2\m_3}$ define the basis of the independent tensor structures. 
\begin{figure}[t]
\begin{center}
\includegraphics[scale=0.9]{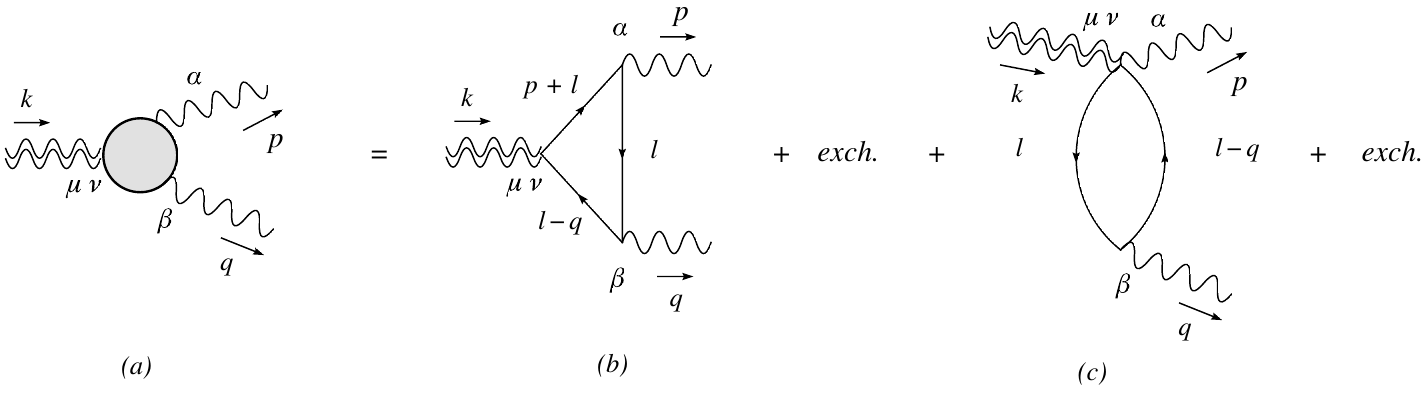}
\caption{\small The complete $TJJ$ one-loop vertex (a) given by the sum of the 1PI contributions with triangle (b) and pinched topologies (c). }
\label{vertex}
\end{center}
\end{figure}

The set of the $13$ tensors $t_i$ is linearly independent for generic $k^2, p^2, q^2$
different from zero. Five of the $13$ are Bose symmetric, 
\be
t_i^{\mu\nu\alpha\beta}(p,q) = t_i^{\mu\nu\beta\alpha}(q,p)\,,\qquad i=1,2,7,8,13\,,
\ee
while the remaining eight tensors are Bose symmetric pairwise
\bea
\label{pair}
&&t_3^{\mu\nu\alpha\beta}(p,q) = t_5^{\mu\nu\beta\alpha}(q,p)\,,\\
&&t_4^{\mu\nu\alpha\beta}(p,q) = t_6^{\mu\nu\beta\alpha}(q,p)\,,\\
&&t_9^{\mu\nu\alpha\beta}(p,q) = t_{10}^{\mu\nu\beta\alpha}(q,p)\,,\\
&&t_{11}^{\mu\nu\alpha\beta}(p,q) = t_{12}^{\mu\nu\beta\alpha}(q,p)\,.
\eea
In the set are present two tensor structures

\bea
&&u^{\alpha\beta}(p,q) \equiv (p\cdot q) g^{\alpha\beta} - q^{\alpha}p^{\beta}\,,\\
&&w^{\alpha\beta}(p,q) \equiv p^2 q^2 g^{\alpha\beta} + (p\cdot q) p^{\alpha}q^{\beta}
- q^2 p^{\alpha}p^{\beta} - p^2 q^{\alpha}q^{\beta}\,,
 \label{uwdef}
 \eea
which appear in $t_1$ and $t_2$ respectively.
Each of them satisfies the Bose symmetry  requirement,
\bea
&&u^{\alpha\beta}(p,q) = u^{\beta\alpha}(q,p)\,,\\
&&w^{\alpha\beta}(p,q) = w^{\beta\alpha}(q,p)\,,
\eea
and vector current conservation,
\bea
&&p_{\alpha} u^{\alpha\beta}(p,q) = 0 = q_{\beta}u^{\alpha\beta}(p,q)\,,\\
&&p_{\alpha} w^{\alpha\beta}(p,q) = 0 = q_{\beta}w^{\alpha\beta}(p,q)\,.
\eea
They are obtained from the variation of gauge invariant quantities
$F_{\mu\nu}F^{\mu\nu}$ and $(\partial_{\mu} F^{\mu}_{\ \,\lambda})(\partial_{\nu}F^{\nu\lambda})$

\bea
&&u^{\alpha\beta}(p,q) = -\frac{1}{4}\int\,d^4x\,\int\,d^4y\ e^{ip\cdot x + i q\cdot y}\ 
\frac{\delta^2 \{F_{\mu\nu}F^{\mu\nu}(0)\}} {\delta A_{\alpha}(x) A_{\beta}(y)} \,,
\label{one}\\
&&w^{\alpha\beta}(p,q) = \frac{1}{2} \int\,d^4x\,\int\,d^4y\ e^{ip\cdot x + i q\cdot y}\
\frac{\delta^2 \{\partial_{\mu} F^{\mu}_{\ \,\lambda}\partial_{\nu}F^{\nu\lambda}(0)\}} 
{\delta A_{\alpha}(x) A_{\beta}(y)}\,.\label{two}
\eea\label{three}
 All the $t_i$'s are transverse in their photon indices
\bea
q^\alpha t_i^{\mu\nu\alpha\beta}=0  \qquad p^\beta t_i^{\mu\nu\alpha\beta}=0.
\eea
$t_2\ldots t_{13}$ are traceless, $t_1$ and $t_2$ have trace parts in $d=4$. With this decomposition, the two vector Ward identities on the photon lines are automatically satisfied by all the amplitudes, as well as the Bose symmetry. \\ 
Diffeomorphism invariance, instead, is automatically satisfied (separately) by the two tensor structures $t_1$ and $t_2$, which are completely transverse, while it has to be imposed on the second set ($t_3\ldots t_{13}$). \\In this way it is possible to extract from the 9 traceless tensor structures a (completely) transverse and traceless set of 5 amplitudes, two of them related by the bosonic symmetry. \\
To summarize, from the original 13 tensor structures $t_i$, split into a set of two transverse and trace components and a remaining set of 11 partially transverse but traceless ones (in $d=4$), one is left with 7 form factors after imposing the pairing conditions \eqref{pair}. \\
Finally, by imposing the conservations WI's these are reduced to 4, which are related to the 4 form factors $A_i$'s  introduced in a completely independent reconstruction method \cite{Bzowski:2013sza}, based on the solutions of the CWI's, with no reference to the perturbative expansion. We will re-investigate this different decomposition in a following section once we turn to the analysis of the CWI's of 3-point functions for the same TJJ vertex and for the $TTT$. \\
The $F_i$'s are functions of the kinematical invariants $s=k^2=(p+q)^2$, $s_1=p^2$, $s_2=q^2$ and of the internal mass $m$. Explicit expressions of these form factors are given in \cite{Armillis:2009pq}. Also in the massive case, as already pointed out, in DR one finds a neat separation between the anomaly contribution and the mass dependent corrections. \\
In the massless case only few form factors survive and one gets
\bea
F_{1} (s, 0, 0, 0) &=& - \frac{e^2}{18 \pi^2  s}, \\
F_{3} (s, 0, 0, 0) &=&  F_{5} (s, 0, 0, 0) = - \frac{e^2}{144 \pi^2 \, s}, \\
F_{7} (s, 0, 0, 0) &=& -4 \, F_{3} (s, 0, 0, 0), \\
F_{13, R} (s, 0, 0, 0) &=& - \frac{e^2}{144 \pi^2} \, \left[ 12 \log \left(-\frac{s}{\mu^2}\right) - 35\right],
\eea
where $F_{13\, R}$ is the only renormalized form factor $(F_{13})$ of the entire amplitude.\\
 The anomaly is entirely given by $F_1$, {\em which indeed shows the presence of an anomaly pole}. Further details on the organization of the effective action mediated by the trace anomaly can be found 
in  \cite{Armillis:2009pq}.
Perturbative investigations of this correlator have shown that the pole contribution is described in the 1PI effective action by the term 
 \be
 \label{pole}
\mathcal{S}_{pole}= - \frac{e^2}{ 36 \pi^2}\int d^4 x d^4 y \left(\square h(x) - \partial_\mu\partial_\nu h^{\mu\nu}(x)\right)  \square^{-1}_{x\, y} F_{\alpha\beta}(x)F^{\alpha\beta}(y),
\ee
where 
\be
R^{(1)}=\square h(x) - \partial_\mu\partial_\nu h^{\mu\nu}(x)
\ee
is the linearized curvature and the background metric has been expanded as at first order in its fluctuations 
$h^{\mu\nu}$. 
\subsection{An example: The 1PI conformal anomaly action in perturbative QCD}
A similar phenomenon, in perturbation theory, occurs in each gauge invariant sector of the non-abelian $TVV$ correlator, as shown in the case of QCD.\\
In fact, coming to QCD, the on-shell vertex (the two gluons are taken on-shell), which is the sum of the quark and pure gauge contributions, can be decomposed by using three appropriate tensor structures $\phi_i^{\mu\nu\alpha\beta}$, given in \cite{Armillis:2010qk}. The anomaly pole appears in the expansion of quark ($\Gamma_q^{\mu\nu\a\b}(p,q)$) and gluon ($\Gamma_g^{\mu\nu\a\b}(p,q)$) subsets of diagrams
\bea
\Gamma^{\mu\nu\alpha\beta}(p,q) =  \Gamma^{\mu\nu\alpha\beta}_g(p,q) + \Gamma^{\mu\nu\alpha\beta}_q(p,q) =  \sum_{i=1}^{3} \Phi_{i} (s,0, 0)\, \delta^{ab}\, \phi_i^{\mu\nu\alpha\beta}(p,q)\,,
\eea
with form factors defined as
\bea
\Phi_i(s,0,0) = \Phi_{i,\,g}(s,0,0) + \sum_{j=1}^{n_f}\Phi_{i, \,q}(s,0,0,m_j^2),
\eea
where the sum runs over the $n_f$ quark flavors. They are given by
\bea
\Phi_{1}(s,0,0) &=& - \frac{g^2}{72 \pi^2 \,s}\left(2 n_f - 11 C_A\right) + \frac{g^2}{6 \pi^2}\sum_{i=1}^{n_f} m_i^2 \, \bigg\{ \frac{1}{s^2} \, - \, \frac{1} {2 s}\mathcal C_0 (s, 0, 0, m_i^2)
\bigg[1-\frac{4 m_i^2}{ s}\bigg] \bigg\}, \,
\label{Phi1}\\
\Phi_{2}(s,0,0) &=& - \frac{g^2}{288 \pi^2 \,s}\left(n_f - C_A\right) \nn \\
&&- \frac{g^2}{24 \pi^2} \sum_{i=1}^{n_f} m_i^2 \, \bigg\{ \frac{1}{s^2}
+ \frac{ 3}{ s^2} \mathcal D (s, 0, 0, m_i^2)
+ \frac{ 1}{s } \mathcal C_0(s, 0, 0, m_i^2 )\, \left[ 1 + \frac{2 m_i^2}{s}\right]\bigg\},
\label{Phi2} \\
\Phi_{3}(s,0,0) &=& \frac{g^2}{288 \pi^2}\left(11 n_f - 65 C_A\right) - \frac{g^2 \, C_A}{8 \pi^2} \bigg[ \frac{11}{6} \mathcal B_0^{\overline{MS}}(s,0) - \mathcal B_0^{\overline{MS}}(0,0) +  s  \,\mathcal C_0(s,0,0,0) \bigg] \nn \\
&& + \, \frac{g^2}{8 \pi^2} \sum_{i=1}^{n_f}\bigg\{  \frac{1}{3}\mathcal B_0^{\overline{MS}}(s, m_i^2) + m_i^2 \, \bigg[
\frac{1}{s}
 + \frac{5}{3 s}  \mathcal D (s, 0, 0, m_i^2) + \mathcal C_0 (s, 0,0,m_i^2) \,\left[1 + \frac{2 m_i^2}{s}\right]
\bigg]\bigg\} ,\nn \\
\label{Phi3}
\eea
with $C_A = N_C$. The scalar integrals $\mathcal B_0^{\overline{MS}}$, $\mathcal D$ and $\mathcal C_0$ are defined in \cite{Armillis:2010qk}.
Notice the appearance in the total amplitude of the $1/s$ pole in $\Phi_1$, which is present both in the quark and in the gluon sectors, and which saturates the contribution to the trace anomaly in the massless limit. In this case the entire trace anomaly is just proportional to this component, which becomes 
\beq
\Phi_{1}(s,0,0) = - \frac{g^2}{72 \pi^2 \,s}\left(2 n_f - 11 C_A\right).
\label{polepole}
\eeq
Further elaborations show that the effective action takes the form
\bea
S_{pole} &=&	- \frac{c}{6}\, \int d^4 x \, d^4 y \,R^{(1)}(x)\, \square^{-1}(x,y) \,  F^a_{\alpha \beta} \,  F^{a \, \alpha \beta} \nn\\
&=& \frac{1}{3} \, \frac{g^3}{16 \pi^2} \left (  - \frac{11}{3} \, C_A + \frac{2}{3} \, n_f \right)  \, \int d^4 x \, d^4 y \,R^{(1)}(x)\, \square^{-1}(x,y) \, F^a_{\alpha \beta}F^{a\,\alpha \beta}
\eea
and is in agreement with the same action derived from the nonlocal gravitational action proposed by Riegert long ago. Here $R^{(1)}$ denotes the linearized expression of the Ricci scalar 
\beq
 R^{(1)}_x\equiv \partial^x_\mu\, \partial^x_\nu \, h^{\mu\nu} - \square \,  h, \qquad h=\eta_{\mu\nu} \, h^{\mu\nu}
 \eeq
and the constant $c$ is related to the non-abelian $\beta$ function as
\beq
c= - 2 \, \, \frac{\beta (g)}{g}.
\eeq
The presence of such effective scalar interactions mediated by the anomaly diagrams show that only one component of an entire anomaly vertex is responsible for the generation of the anomaly, which, obviously, cannot be isolated from the entire vertex. It is not an artificial component introduced by a specific decomposition of the same vertex, but it is simply the signature of the same vertex. \\
We are going to see conclusively, at least from the point of perturbation theory, that such interactions are generated by a region in the loop integration where the two intermediate particle emerging from the stress-energy tensor $T$ (or from the axial- vector current  $J_A$, in the chiral case), move collinearly before reaching the final state, made of two gluons, two photons or, more generally, two stress-energy tensors, and describe a pairing. This pairing is what is identified Fig. \ref{fig1} and it is described by a spectral density whose support is on a single point in phase space, being proportional to $\delta(s)$, with $s$ being the invariant mass of the lines on which $T$ or $J_A$ are inserted. \\
A similar pairing has been noticed in the case of topological insulators \cite{Rinkel:2016dxo} and of Weyl semimetals \cite{Chernodub:2017jcp}, which are materials in which chiral and conformal anomalies play a significant role \cite{Landsteiner:2013sja}.

\section{Non-perturbative solutions of the TJJ and TTT from the CWI's} 
The same $TJJ$ vertex andt its anomaly can be completely determined, up to few constants, in a completely independent way just by solving the CWI's of a general CFT in $d=4$. The goal of this section will be to show how the results coming from the perturbative analysis and the perturbative approach can be merged completely, bringing to conclusive evidence that such massless interactions are genuinely present in any anomaly vertex. A similar result, in fact, will be shown to hold also in the $TTT$ case. The result is also in agreement with the prediction coming from a nonlocal version of the conformal anomaly action given in Eq. \eqref{Snonl}, recently discussed in \cite{Coriano:2017mux}, which accounts for the same anomaly structure which we are going to discuss here.  
\subsection{CWI's in momentum space}
An independent analysis of the features described above requires a momentum space approach in the solution of the CWI's. These have been discussed in several papers \cite{Coriano:2013jba}\cite{2014JHEP...03..111B} and several comparisons against the free field theory realizations of the same theories have been discussed in \cite{Coriano:2018zdo,Coriano:2018bbe,Coriano:2018bsy}. A systematic discussion of how to move to momentum space from coordinate space in the analysis of scalar and tensor correlators in CFT's can be found in \cite{Coriano:2018bbe}. Here we just recall that the dilatation and the special conformal transformations, in momentum space take the forms   

\be
\left(\sum_{j=1}^n\D_j-(n-1)d-\sum_{j=1}^{n-1}p_j^\a\frac{\partial}{\partial p_j^\a} \right)\Phi(p_1,\ldots p_{n-1},\bar{p}_n)=0.
\ee

\be
\sum_{j=1}^{n-1}\left(p_j^\kappa \frac{\partial^2}{\partial p_j^\alpha\partial p_j^\alpha} + 2(\Delta_j- d)\frac{\partial}{\partial p_j^\kappa}-2 p_j^\alpha\frac{\partial^2}{\partial p_j^\kappa \partial p_j^\alpha}\right)\Phi(p_1,\ldots p_{n-1},\bar{p}_n)=0.
\ee 
The latter is the momentum space version of the differential constraint
\be
\sum_{j=1}^{n} \left(- x_j^2\frac{\partial}{\partial x_j^\kappa}+ 2 x_j^\kappa x_j^\alpha \frac{\partial}
{\partial x_j^\alpha} +2 \Delta_j x_j^\kappa\right)\Phi(x_1,x_2,\ldots,x_n) =0.
\ee
Once we move to momentum space, one needs to select the independent momenta, which in 
our conventions will be the first $n-1$, with the $n-$th as a dependent one 
$\bar{p}_n=-(p_1+\ldots p_{n-1})$. 

\subsection{Example: the scalar solution of the CWI's and its hypergeometric origin}
The simplest case, investigated in the conformal approach is the $\Phi=\langle OOO\rangle$, corresponding to the scalar 3-point function. Here we will sketch the derivation of the equations in this simpler case. We refer to \cite{Coriano:2013jba} for more details. \\
In the case of a scalar correlator all the anomalous conformal WI's can 
be re-expressed in scalar form by taking as independent momenta the magnitude $ {p}_i=\sqrt{p_i^2}$ as the 
three independent variables, the dilatation equation becomes 
\be
\label{scale}
\left(\Delta -2 d - \sum_{i=1}^3    p_i \frac{ \partial}{\partial   p_i}\right)\Phi(p_1,p_2,\bar{p}_3)=0.\ee
The relation above is derived using the chain rule
\be
\label{chainr}
\frac{\partial \Phi}{\partial p_i^\mu}=\frac{p_i^\mu}{  p_i}\frac{\partial\Phi}{\partial  p_i} 
-\frac{\bar{p}_3^\mu}{  p_3}\frac{\partial\Phi}{\partial   p_3},
\ee
where $\Delta$ is the sum of of the scaling dimensions of each operator in the 3-point function. 
It is a straightforward but lengthy computation to show that the special (non anomalous) conformal transformation in 
$d$ dimensions takes the form, for  the scalar component  

\be
{{K}_{scalar}}^{\kappa}\Phi=0
\ee
with
\be
{{K}}^{\kappa}_{scalar}=\sum_{i=1}^3 p_i^\kappa \,{K}_i 
\label{kappa2}
\ee
\be
\label{ksk}
{ K}_i\equiv \frac{\partial^2}{\partial    p_i \partial    p_i} 
+\frac{d + 1 - 2 \Delta_i}{   p_i}\frac{\partial}{\partial   p_i}
\ee
with the expression (\ref{kappa2}) which can be split into the two independent equations
\be
\frac{\partial^2\Phi}{\partial   p_i\partial   p_i}+
\frac{1}{  p_i}\frac{\partial\Phi}{\partial  p_i}(d+1-2 \Delta_1)-
\frac{\partial^2\Phi}{\partial   p_3\partial   p_3} -
\frac{1}{  p_3}\frac{\partial\Phi}{\partial  p_3}(d +1 -2 \Delta_3)=0\qquad i=1,2.
\label{3k1}
\ee
The expressions of $K^\kappa$ as given in \eqref{ksk} has been first defined in \cite{2014JHEP...03..111B}. Similar results have been obtained in \cite{Coriano:2013jba} using a direct change of variables that reduces the special CWI to a hypergeometric system of equations. \\
In the scalar case, defining 
\be
\label{kij}
K_{ij}\equiv {K}_i-{K}_j
\ee
Eqs. (\ref{3k1}) take the homogeneous form 
\be
\label{3k2}
K^\kappa_{13}\Phi=0 \qquad \textrm{and} \qquad K^\kappa_{23}\Phi=0.
\ee
The solutions of such equations and their reduction to hypergeometric forms is obtained by the ansatz 
\begin{equation}
\label{ans}
\Phi(p_1,p_2,p_3)=p_1^{\Delta - 2 d} x^{a}y^{b} F(x,y)
\end{equation}

with $x=\frac{p_2^2}{p_1^2}$ and $y=\frac{p_3^2}{p_1^2}$. Here we are taking $p_1$ as "pivot" in the expansion, but we could equivalently choose any of the 3  momentum invariants. $\Phi$ is required to be homogenous of degree $\Delta-2 d$ under a scale transformation, according to (\ref{scale}), and in (\ref{ans}) this is taken into account by the factor $p_1^{\Delta - 2 d}$.
The use of the scale invariant variables $x$ and $y$ leads to the hypergeometric form of the solution. One obtains 
%\begin{equation}
%K_1\Phi=4 {p_3}^{\Delta -2 d - 2} x^{\alpha-1}y^\beta 
%\left( x^2\frac{\partial^2 }{\partial x\partial x} + 
%(2 a +\frac{d}{2} -\Delta_1 +1)\frac{\partial  }{\partial x} + 
%(a +\frac{d}{2}-\Delta_1)\right)F(x,y)
%\end{equation}
%and 
\begin{align}
K_{21}\phi &= 4 p_1^{\Delta -2d -2} x^a y^b
\left(  x(1-x)\frac{\partial }{\partial x \partial x}  + (A x + \gamma)\frac{\partial }{\partial x} -
2 x y \frac{\partial^2 }{\partial x \partial y}- y^2\frac{\partial^2 }{\partial y \partial y} + 
D y\frac{\partial }{\partial y} + (E +\frac{G}{x})\right) \notag\\
& \hspace{3cm}\times F(x,y)=0
\label{red}
\end{align}
with
\begin{align}
&A=D=\Delta_2 +\Delta_3 - 1 -2 a -2 b -\frac{3 d}{2} \qquad \gamma(a)=2 a +\frac{d}{2} -\Delta_2 + 1
\notag\\
& G=\frac{a}{2}(d +2 a - 2 \Delta_2)
\notag\\
&E=-\frac{1}{4}(2 a + 2 b +2 d -\Delta_1 -\Delta_2 -\Delta_3)(2 a +2 b + d -\Delta_3 -\Delta_2 +\Delta_1).
\end{align}
Similar constraints are obtained from the equation $K_{31}\Phi=0$, with the obvious exchanges $(a,b,x,y)\to (b,a,y,x)$
\begin{align}
K_{31}\phi &= 4 p_1^{\Delta -2 d -2} x^a y^b
\left(  y(1-y)\frac{\partial }{\partial y \partial y}  + (A' y + \gamma')\frac{\partial }{\partial y} -
2 x y \frac{\partial^2 }{\partial x \partial y}- x^2\frac{\partial^2 }{\partial x \partial x} + 
D' x\frac{\partial }{\partial x} + (E' +\frac{G'}{y})\right) \notag\\
& \hspace{3cm}\times F(x,y)=0
\label{red}
\end{align}
with
\begin{align}
&A'=D'= A   \qquad \qquad \gamma'(b)=2 b +\frac{d}{2} -\Delta_3 + 1
\notag\\
& G'=\frac{b}{2}(d +2 b - 2 \Delta_3)
\notag\\
&E'= E.
\end{align}
The exponents $a$ and $b$, as shown in \cite{Coriano:2018bbe}, are exactly those that 
allow to remove the $1/x$ and $1/y$ terms in \eqref{red}, which implies that
\begin{equation}
\label{cond1}
a=0\equiv a_0 \qquad \textrm{or} \qquad a=\Delta_2 -\frac{d}{2}\equiv a_1.
\end{equation}
From the equation $K_{31}\Phi=0$ we obtain a similar condition for $b$ by setting $G'/y=0$, thereby fixing the two remaining indices
\begin{equation}
\label{cond2}
b=0\equiv b_0 \qquad \textrm{or} \qquad b=\Delta_3 -\frac{d}{2}\equiv b_1.
\end{equation}
The four independent solutions of the CWI's will all be characterised by the same 4 pairs of indices $(a_i,b_j)$ $(i,j=1,2)$.
Setting 
\begin{equation}
\alpha(a,b)= a + b + \frac{d}{2} -\frac{1}{2}(\Delta_2 +\Delta_3 -\Delta_1) \qquad \beta (a,b)=a +  b + d -\frac{1}{2}(\Delta_1 +\Delta_2 +\Delta_3) \qquad 
\label{alphas}
\end{equation}
then
\begin{equation}
E=E'=-\alpha(a,b)\beta(a,b) \qquad A=D=A'=D'=-\left(\alpha(a,b) +\beta(a,b) +1\right),
\end{equation}
the solutions take the form 
\begin{align}
\label{F4def}
F_4(\alpha(a,b), \beta(a,b); \gamma(a), \gamma'(b); x, y) = \sum_{i = 0}^{\infty}\sum_{j = 0}^{\infty} \frac{(\alpha(a,b), {i+j}) \, 
	(\beta(a,b),{i+j})}{(\gamma(a),i) \, (\gamma'(b),j)} \frac{x^i}{i!} \frac{y^j}{j!} 
\end{align}
where $(\alpha,i)=\Gamma(\alpha + i)/ \Gamma(\alpha)$ is the Pochammer symbol. We will refer to $\alpha\ldots \gamma'$ as to the first,$\ldots$, fourth parameters of $F_4$.\\ 
The 4 independent solutions are then all of the form $x^a y^b F_4$, where the 
hypergeometric functions will take some specific values for its parameters, with
$a$ and $b$ fixed by (\ref{cond1}) and (\ref{cond2}). Specifically we have
\begin{equation}
\Phi(p_1,p_2,p_3)=p_1^{\Delta-2 d} \sum_{a,b} c(a,b,\vec{\Delta})\,x^a y^b \,F_4(\alpha(a,b), \beta(a,b); \gamma(a), \gamma'(b); x, y) 
\label{compact}
\end{equation}
where the sum runs over the four values $a_i, b_i$ $i=0,1$ with arbitrary constants $c(a,b,\vec{\Delta})$, with $\vec{\Delta}=(\Delta_1,\Delta_2,\Delta_3)$. Eq. \eqref{compact} is a very compact way to write down the solution, which can be made more explicit. For this reason it is convenient to define 
\begin{align}
&\alpha_0\equiv \alpha(a_0,b_0)=\frac{d}{2}-\frac{\Delta_2 + \Delta_3 -\Delta_1}{2},\, && \beta_0\equiv \beta(b_0)=d-\frac{\Delta_1 + \Delta_2 +\Delta_3}{2},  \nn\\
&\gamma_0 \equiv \gamma(a_0) =\frac{d}{2} +1 -\Delta_2,\, &&\gamma'_0\equiv \gamma(b_0) =\frac{d}{2} +1 -\Delta_3
\end{align}
as the 4 basic (fixed) hypergeometric parameters. All the remaining solutions are defined by shifts with respect to these. The 4 independent solutions can be re-expressed in terms of the parameters above as 

\bea
\label{F4def}
S_1(\alpha_0, \beta_0; \gamma_0, \gamma'_0; x, y)\equiv F_4(\alpha_0, \beta_0; \gamma_0, \gamma'_0; x, y) = \sum_{i = 0}^{\infty}\sum_{j = 0}^{\infty} \frac{(\alpha_0,i+j) \, 
(\beta_0,i+j)}{(\gamma_0,i )\, (\gamma'_0,j)} \frac{x^i}{i!} \frac{y^j}{j!} 
\eea
and
\bea
\label{solutions}
S_2(\alpha_0, \beta_0; \gamma_0, \gamma'_0; x, y) &=& x^{1-\gamma_0} \, F_4(\alpha_0-\gamma_0+1, \beta_0-\gamma_0+1; 2-\gamma_0, \gamma'_0; x,y) \,, \nn \\
S_3(\alpha_0, \beta_0; \gamma_0, \gamma'_0; x, y) &=& y^{1-\gamma'_0} \, F_4(\alpha_0-\gamma'_0+1,\beta_0-\gamma'_0+1;\gamma_0,2-\gamma'_0 ; x,y) \,, \nn \\
S_4(\alpha_0, \beta_0; \gamma_0, \gamma'_0; x, y) &=& x^{1-\gamma_0} \, y^{1-\gamma'_0} \, 
F_4(\alpha_0-\gamma_0-\gamma'_0+2,\beta_0-\gamma_0-\gamma'_0+2;2-\gamma_0,2-\gamma'_0 ; x,y) \, . \nn
\eea
Notice that in the scalar case, one is allowed to impose the complete symmetry of the correlator under the exchange of the 3 external momenta and scaling dimensions, as discussed in \cite{Coriano:2013jba}. This reduces the four
constants to just one. \\
We are going first to extend this analysis to the case of the $A_1-A_4$ form factors of the $TJJ$. Interestingly, this approach can be generalized to 4-point functions, showing that at large energy and momentum transfers, scatterings at fixed angle are characterized by similar solutions, with some generalizations, taking to Lauricella's functions \cite{Maglio:2019grh}. 
\subsection{Transverse traceless basis for the $TJJ$ and the matching to perturbation theory}
The solution in the case of the $TJJ$ is far more involved compared to the scalar case and it requires a completely new approach. A way to solve the CWI's for tensor correlators has been formulated in \cite{Bzowski:2013sza}. 
In order to establish a link between the perturbative approach of the previous sections and the non-perturbative one based on \cite{Bzowski:2013sza}, we start by stating the special CWI satisfied by this correlator 
\begin{equation}
\begin{split}
&\sum_{j=1}^{2}\left[2(\Delta_j-d)\frac{\partial}{\partial p_j^\k}-2p_j^\a\frac{\partial}{\partial p_j^\a}\frac{\partial}{\partial p_j^\k}+(p_j)_\k\frac{\partial}{\partial p_j^\a}\frac{\partial}{\partial p_{j\a}}\right]\braket{T^{\mu_1\nu_1}(p_1)\,J^{\m_2}(p_2)\,J^{\mu_3}(\bar p_3)}\\
&\qquad+4\left(\d^{\k(\mu_1}\frac{\partial}{\partial p_1^{\a_1}}-\delta^{\k}_{\alpha_1}\delta^{\l(\mu_1}\frac{\partial}{\partial p_1^\l}\right)\braket{T^{\nu_1)\alpha_1}(p_1)\,J^{\m_2}(p_2)\,J^{\mu_3}(\bar p_3)}\\
&\qquad+2\left(\d^{\k\mu_2}\frac{\partial}{\partial p_2^{\a_2}}-\delta^{\k}_{\alpha_2}\delta^{\l\mu_2}\frac{\partial}{\partial p_2^\l}\right)\braket{T^{\mu_1\nu_1}(p_1)\,J^{\alpha_2}(p_2)\,J^{\mu_3}(\bar p_3)}=0.
\end{split} 
\label{SCWTJJ}
\end{equation}

The first line in the equation above corresponds to the scalar action $K^\kappa$, while the second and the third lines are the contributions coming from the spin parts. The correlator is decomposed into its transverse traceless $(\braket{t^{\mu_1\nu_1}\,j^{\mu_2}\,j^{\mu_3}})$ and longitudinal (local) parts in the form

\begin{align}
\braket{T^{\mu_1\nu_1}\,J^{\mu_2}\,J^{\mu_3}}&=\braket{t^{\mu_1\nu_1}\,j^{\mu_2}\,j^{\mu_3}}+\braket{T^{\mu_1\nu_1}\,J^{\mu_2}\,j_{loc}^{\mu_3}}+\braket{T^{\mu_1\nu_1}\,j_{loc}^{\mu_2}\,J^{\mu_3}}+\braket{t_{loc}^{\mu_1\nu_1}\,J^{\mu_2}\,J^{\mu_3}}\notag\\
&\quad-\braket{T^{\mu_1\nu_1}\,j_{loc}^{\mu_2}\,j_{loc}^{\mu_3}}-\braket{t_{loc}^{\mu_1\nu_1}\,j_{loc}^{\mu_2}\,J^{\mu_3}}-\braket{t_{loc}^{\mu_1\nu_1}\,J^{\mu_2}\,j_{loc}^{\mu_3}}+\braket{t_{loc}^{\mu_1\nu_1}\,j_{loc}^{\mu_2}\,j_{loc}^{\mu_3}}.
\end{align}

with 
\begin{equation}
\label{loca1}
T^{\mu\nu}=t^{\mu\nu} + t_{loc}^{\mu\nu}
\end{equation}
with the longitudinal/trace terms given by
\begin{align}
\label{loca2}
t_{loc}^{\mu\nu}(p)&=\frac{p^{\mu}}{p^2}Q^\nu + \frac{p^{\nu}}{p^2}Q^\mu -
\frac{p^\mu p^\nu}{p^4} Q +\frac{\pi^{\m\nu}}{d-1}(T - \frac{Q}{p^2})\nn
&=\Sigma^{\mu\nu}_{\alpha\beta} T^{\alpha\beta}
\end{align}
\begin{equation}
Q^\mu=p_\nu T^{\mu\nu},\qquad T=\delta_{\mu\nu}T^{\mu\nu}, \qquad Q= p_\nu p_\mu T^{\mu\nu}.
\end{equation}
The decompositon above allows to separate the equations into primary and secondary constraints, the secondary ones involving equations with the inclusion of two-point functions. \\
At the core of the decomposition there is the transverse traceless sector, which is parameterized by a minimal set of form factors, as proposed in \cite{Bzowski:2013sza}. \\
It is possible to show that these amplitudes are in a one-to-one correspondence with the form factors $A_j$ $(j=1,\ldots 4)$  introduced in the parameterization of the $TJJ$ correlator presented in \cite{Bzowski:2013sza}. In that work the full 3-point function is parameterized in terms of transverse (with respect to all the external momenta) traceless components plus extra terms identified via longitudinal Ward identities of the $TJJ$ (the so-called {\em local} or {\em semi local}) characterised by pinched topologies 
\begin{align}
\braket{\braket{T^{\mu_1\nu_1}\,J^{\mu_2}\,J^{\mu_3}}}&=\braket{\braket{t^{\mu_1\nu_1}\,j^{\mu_2}\,j^{\mu_3}}}+ \textrm{local terms}.\end{align}
Here we have switched to a symmetric notation for the external momenta, with $(p_1,p_2,p_3)\equiv (k,p,q)$, and 
with the transverse traceless parts expanded in terms of a set of the form factors $A_j$ mentioned above  
\begin{align}
\langle t^{\mu_1\nu_1}(p_1)j^{\mu_2}(p_2)j^{\mu_3}(p_3)\rangle& =
{\Pi_1}^{\mu_1\nu_1}_{\alpha_1\beta_1}{\pi_2}^{\mu_2}_{\alpha_2}{\pi_3}^{\mu_3}_{\alpha_3}
\left( A_1\ p_2^{\alpha_1}p_2^{\beta_1}p_3^{\alpha_2}p_1^{\alpha_3} + 
A_2\ \delta^{\alpha_2\alpha_3} p_2^{\alpha_1}p_2^{\beta_1} + 
A_3\ \delta^{\alpha_1\alpha_2}p_2^{\beta_1}p_1^{\alpha_3}\right. \notag\\
& \left. + 
A_3(p_2\leftrightarrow p_3)\delta^{\alpha_1\alpha_3}p_2^{\beta_1}p_3^{\alpha_2}
+ A_4\  \delta^{\alpha_1\alpha_3}\delta^{\alpha_2\beta_1}\right).\label{DecompTJJ}
\end{align}
The equation above provides the basic decomposition of the $TJJ$ in terms of a minimal set of form factors which can be mapped into the set of the $F_i's$ presented in the previous sections. We are going to show that the $1/k^2$ behaviour found in the anomaly form factor $(F_1)$ is in agreement with the explicit expressions obtained from the solutions of the CWI's, which fix the $A_i$.  But before coming to this point, we briefly comment on the form of the equations obtained.\\
In this expression ${\Pi_1}^{\mu_1\nu_1}_{\alpha_1\beta_1}$ is a transverse and traceless projector built out of momentum $p_1$, while  ${\pi_2}^{\mu_2}_{\alpha_2}$ and ${\pi_3}^{\mu_3}_{\alpha_3}$ denote transverse projectors with respect to the momenta $p_2$ and $p_3$.  \\ 
Coming to the explicit form of the $A_j$, they are extracted from the solution of the scalar CWI's 
\begin{equation}
\label{listeq}
\begin{split}
0&=\tilde{C}_{11}=K_{21}A_1\\
0&=\tilde{C}_{12}=K_{21}A_2-2A_1\\
0&=\tilde{C}_{13}=K_{21}A_3\\
0&=\tilde{C}_{14}=K_{21}A_3(p_2\leftrightarrow p_3)+4A_1\\
0&=\tilde{C}_{15}=K_{21}A_4+2A_3
\end{split}
\hspace{1.5cm}
\begin{split}
0&=\tilde{C}_{21}=K_{31}A_1\\
0&=\tilde{C}_{22}=K_{31}A_2-2A_1\\
0&=\tilde{C}_{23}=K_{31}A_3 +4A_1\\
0&=\tilde{C}_{24}=K_{31}A_3(p_2\leftrightarrow p_3)\\
0&=\tilde{C}_{25}=K_{31}A_4+2A_3(p_2\leftrightarrow p_3),
\end{split}
\end{equation}
which can be obtained in two ways. The approach of 
\cite{Bzowski:2013sza,Bzowski:2015pba,Bzowski:2018fql} expresses such solutions in terms of 
3-K integrals, i.e. integrals of three Bessel functions, and can be related to specific combinations of hypergeometric functions $F_4$, also known as Appell functions. An equivalent method can be formulated which allows to work out the explicit form of the solutions using properties of the functions $F_4$ \cite{Coriano:2018bbe}. The two approaches have been combined, more recently, in the analysis of scalar 4-point functions, with the introduction of 4K integrals \cite{Maglio:2019grh}.

\subsection{Connection between the $F$ and the $A$ bases}
We have taken into consideration two separate bases in the analysis of the $TJJ$. 
 In the $F$ basis only one form factors needs to be renormalized, which is $F_{13}$, by dimensional counting. 
 We can show that the renormalization of $F_{13}$ is responsible for the emergence of a $1/k^2$ 
pole in the insertion of the $T$ operator.
Also, the singularities of $F_{13}$ are mirrored by the singular behaviour of the $A_i$ form factors which 
contain such form factor, following the map shown below in Eq. \eqref{mapping1}.\\
 Let's see how these points can be proven, following the discussion of \cite{Coriano:2018zdo}. 

The $TJJ$ correlator in QED is conformal in $d$ dimension, with finite form factors which are not affected by the conformal anomaly, being dimensionally regulated. We can use the $F$-basis of the 13 form factors $F_i$ introduced before and tensor structures $t_i$ to parameterize them.\\
 Notice that the separation of these 13 structures into trace-free and trace parts is valid only in $d=4$ for most of the structures, except for $t_9,t_{10},t_{11}$ and $t_{12}$, which remain traceless in $d$ dimensions. conctractions with the metric tensor are, at this point, performed in $d$ dimensions with a metric 
$g_{\mu\nu}(d)$. The 4-dimensional metric, instead, will be denoted as $g_{\mu\nu}(4)$.\\
For example, a contraction of $t_1$ and $t_2$ in d- dimensions will give 
\bea
g_{\mu\nu}(d)t_1^{\mu\nu\alpha\beta}&=&(d-4) k^2 u^{\alpha\beta}(p,q)\nn \\
g_{\mu\nu}(d)t_2^{\mu\nu\alpha\beta}&=&(d-4) k^2 w^{\alpha\beta}(p,q),
\eea
and similarly for all the other structures, except for those mentioned above, which are trace-free in any dimensions.\\
 Using the completeness of the $F$-basis we can identify the mapping between the form factors of such basis and those of the $A$-basis which parameterize the transverse-traceless contributions in the reconstruction method of \cite{Bzowski:2013sza}. They are conveniently expressed in terms of the momenta $(k,p,q)\equiv(p_1,p_2,p_3)$ in the form
 \bea
A_1&=&4(F_7-F_3-F_5)-2p_2^2F_9-2p_3^2F_{10}\nn\\
A_2&=&2(p_1^2-p_2^2-p_3^2)(F_7-F_5-F_3)-4p_2^2p_3^2(F_6-F_8+F_4)-2F_{13}\nn\\
A_3&=&p_3^2(p_1^2-p_2^2-p_3^2)F_{10}-2p_2^2\,p_3^2 F_{12}-2F_{13}\nn\\
A_4&=&(p_1^2-p_2^2-p_3^2)F_{13},
\label{mapping1}
\eea
which are transverse and traceless, with $A_1$, $A_2$ and $A_4$ symmetric.  \\

\subsection{The $TJJ$ anomaly pole from renormalization}
Starting from $d$-dimension and using the $F$-basis, we require that this correlator has no trace (i.e. be anomaly free) in $d$ dimensions. The anomaly will emerge in dimensional regularization as we take the $d\to 4$ limit. The trace WI's provide the two key conditions that we need. In fact we obtain

\be
\label{one}
F_1=\frac{(d-4)}{p_1^2(d-1)}\big[F_{13}-p_2^2\,F_3-p_3^2\,F_5-p_2\cdot p_3\, F_7\big]\\
\ee
and 
\be
\label{two}
F_2=\frac{(d-4)}{p_1^2(d-1)}\big[p_2^2\,F_4+p_3^2\,F_6+p_2\cdot p_3\,F_8\big].
\ee
Both equations are crucial in order to understand the way the renormalization procedure works for such correlator.
From Eq. (\ref{two}) it is clear that by sending $d\to 4$, $F_2$ vanishes,
\be
F_2=\frac{\epsilon}{(d-1) p_1^2}\big[p_2^2\,F_4+p_3^2\,F_6+p_2\cdot p_3\,F_8\big]\to 0,
\ee
since all the form factors $F_4,F_6$ and $F_8$ are finite for dimensional reasons, and therefore $F_2$ is indeed zero in this limit, since the right-hand side of \eqref{two} has no poles in $\epsilon\equiv d-4$. \\
At this stage, after the limiting procedure, at $d=4$ we are left in the $F-$basis with 4 independent combinations of form factors from the original seven, given in \eqref{mapping1}, which are sufficient to describe the complete transverse traceless sector of the theory, plus an additional form factor $F_1$. \\
Therefore, by taking the $d\to 4$ limit, the $F-$set contains only one single tensor structure of nonzero trace and associated form factor, which should account for the anomaly in $d=4$. This result is obviously confirmed in perturbation theory in QED \cite{Armillis:2009pq}.\\
As already mentioned, $F_{13}$ is the only form factor that needs to be renormalized in the $F$-set and it is characterized by the appearance of a single pole in $1/\epsilon$ in dimensional regularization. The fact that such singularity will be at all orders of the form $1/\epsilon$ and not higher is a crucial ingredient in the entire construction, and is due to conformal symmetry. Such assumption is consistent with the analysis in conformal field theory since the only available counterterm to regulate the theory is given by 
\be
\frac{1}{\epsilon}\int d^4 x \sqrt{g} F_{\mu\nu}F^{\mu\nu}
\ee
which renormalizes the 2-point function $\langle JJ \rangle$ and henceforth $F_{13}$. An explicit computations in QED gives the result
\be
\label{f13}
F_{13}= G_0(p_1^2, p_2^2,p_3^3) -\frac{1}{2} \, [\Pi (p_2^2) +\Pi (p_3^2)]
\ee
with $G_0$ being a lengthy expression which remains finite as $d\to 4$. Therefore, the origin of the singularity is traced back to the 
scalar form factor $\Pi(p^2)$ of the photon 2-point function. Concerning the fact that the singularity in $\Pi(p^2)$ stops at $1/\epsilon$, we just recall that the structure of the two-point function of two conserved vector currents of scaling dimensions $\eta_1$ and $\eta_2$ is given by \cite{Coriano:2013jba}
\be
\label{TwoPointVector}
G_V^{\alpha \beta}(p) = \delta_{\eta_1 \eta_2}  \, c_{V 12}\, 
\frac{\pi^{d/2}}{4^{\eta_1 - d/2}} \frac{\Gamma(d/2 - \eta_1)}{\Gamma(\eta_1)}\,
\left( \eta^{\alpha \beta} -\frac{p^\alpha p^\beta}{p^2} \right)\
(p^2)^{\eta_1-d/2} \,,
\ee
with $c_{V12}$ being an arbitrary constant. It requires the two currents to share the same dimensions and manifests only a single pole in ${1/\epsilon}$.   
In dimensional regularization, in fact,  the divergence can be regulated with $d \to d - 2 \epsilon$. Expanding the product $\Gamma(d/2-\eta)\,(p^2)^{\eta - d/2}$, which appears in the two-point function, in a Laurent series around $d/2 - \eta = -n$ (integer) gives the single pole in $1/\epsilon$ behaviour  \cite{Coriano:2013jba}
\bea
\label{expansion}
\Gamma\left(d/2-\eta\right)\,(p^2)^{\eta-d/2} = \frac{(-1)^n}{n!} \left( - \frac{1}{\epsilon} + \psi(n+1)  + O(\epsilon) \right) (p^2)^{n + \epsilon} \,,
\eea
where $\psi(z)$ is the logarithmic derivative of the Gamma function, and $\epsilon$ takes into account the divergence of the two-point correlator for particular values of the scale dimension $\eta$ and of the space-time dimension $d$. Therefore, the divergence in $F_{13}$ is then given by a single pole in $\epsilon$ is of the form

\be 
\label{renf}
F_{13}=\frac{1}{d-4} \bar{F}_{13} + F_{13\, f}
\ee
In QED, for instance, one finds by an explicit computation that $\bar{F}_{13}=-e^2/(6 \pi^2)$ at one-loop and $\bar{F}_{13\, f}$ is finite \cite{Armillis:2009pq} and gets renormalized into $F_{13 R}$ only in its photon self-energy contributions \cite{Armillis:2009pq} $(s=p_1^2,\,s_1=p_2^2,\,s_2=p_3^2)$
\beqa
  { {F_{13,R} (s;\,s_1,\,s_2,\,0)}} &=&
  -\frac{1}{2} \left[ \Pi_R(s_1,0)+ \Pi_R(s_2,0) \right]
 + G_0(s,s_1,s_2) 
\eeqa
with 
\bea
 \Pi_R (s,0) = - \frac{e^2}{12 \,  \pi^2 } \, \left[\frac{5}{3} - \log \left(-\frac{s}{\mu^2} \right)\right],
 \eea
 denoting the renormalized scalar form factor of the $JJ$ correlator at one-loop and with $G_0$ implicitly defined in \eqref{f13}.

Inserting \eqref{renf} into (\ref{one}) we obtain
\be
\label{oneprime}
F_1=\frac{(d-4)}{p_1^2(d-1)}\left(  \frac{1}{d-4} \bar{F}_{13} + F_{13\, f}  -p_2^2\,F_3-p_3^2\,F_5-p_2\cdot p_3\, F_7\right),
\ee  
which in the $d\to 4$ limit gives, in general 
\be
F_1=\frac{\bar{F}_{13}}{3 p_1^2}
\ee
and specifically, in QED
\be
\label{stable}
F_1=- \frac{e^2}{32 \pi^2 s},
\ee
$(s\equiv k^2)$ showing that the anomaly pole in $F_1$ is indeed generated by the renormalization of the single divergent form factor 
$F_{13}$. In the case of QED, the relation between the prefactor in front of the $1/s$ pole and its relation to the QED $\beta$-function has been extensively discussed in \cite{Giannotti:2008cv,Armillis:2009pq}, to which we refer for further details. In performing the limit we have used the finiteness of the remaining form factors. \\
The analysis presented above proves the consistency of the conjecture concerning the origin of the anomaly pole which is attributed to the renormalization of the $TJJ$ correlator as we reach the physical dimensions. At the same time, the agreement with perturbation theory, in QED, holds only at 1-loop, where the theory is conformal. But this is sufficient to establish a link between a general CFT result and a specific perturbative realization of a given correlator in free field theory. The consistency between the results obtained in the F basis and the transverse-traceless one of the $A_i$ can be easily figured out from a cursory look at \eqref{mapping1}. One can show that the singularities of the $A_i$ are in direct correspondence with the presence of $F_{13}$ in their relations to the $F$ basis. For instance $A_2, A_3$ and $A_4$ need to be renormalized, while $A_1$ does not.
\section{The 3-graviton vertex from the CWI's} 
As we move to the $TTT$ vertex, the analysis gets more involved but it remains substantially unchanged. Similarly to the $TJJ$, the longitudinal/transverse-traceless decomposition takes the form

\begin{align}
\braket{T^{\mu_1\nu_1}\,T^{\mu_2\n_2}\,T^{\mu_3\n_3}}&=\braket{t^{\mu_1\nu_1}\,t^{\mu_2\nu_2}\,t^{\mu_3\nu_3}}+\braket{t_{loc}^{\mu_1\nu_1}\,t^{\mu_2\nu_2}\,t^{\mu_3\nu_3}}+\braket{t^{\mu_1\nu_1}\,t_{loc}^{\mu_2\nu_2}\,t^{\mu_3\nu_3}}+\braket{t^{\mu_1\nu_1}\,t^{\mu_2\nu_2}\,t_{loc}^{\mu_3\nu_3}}\notag\\
&\hspace{0.8cm} +\braket{t^{\mu_1\nu_1}_{loc}\,t_{loc}^{\mu_2\nu_2}\,t^{\mu_3\nu_3}}+\braket{t_{loc}^{\mu_1\nu_1}\,t_{loc}^{\mu_2\nu_2}\,t^{\mu_3\nu_3}}+\braket{t^{\mu_1\nu_1}\,t_{loc}^{\mu_2\nu_2}\,t_{loc}^{\mu_3\nu_3}}+\braket{t_{loc}^{\mu_1\nu_1}\,t_{loc}^{\mu_2\nu_2}\,t_{loc}^{\mu_3\nu_3}}
\end{align}
with the $\braket{t^{\mu_1\nu_1}\,t^{\mu_2\nu_2}\,t^{\mu_3\nu_3}}$ parameterized by five form factors $(A_i)$ in their transverse traceless sector.

\begin{figure}[t]
	\centering
	\subfigure{\includegraphics[scale=0.2]{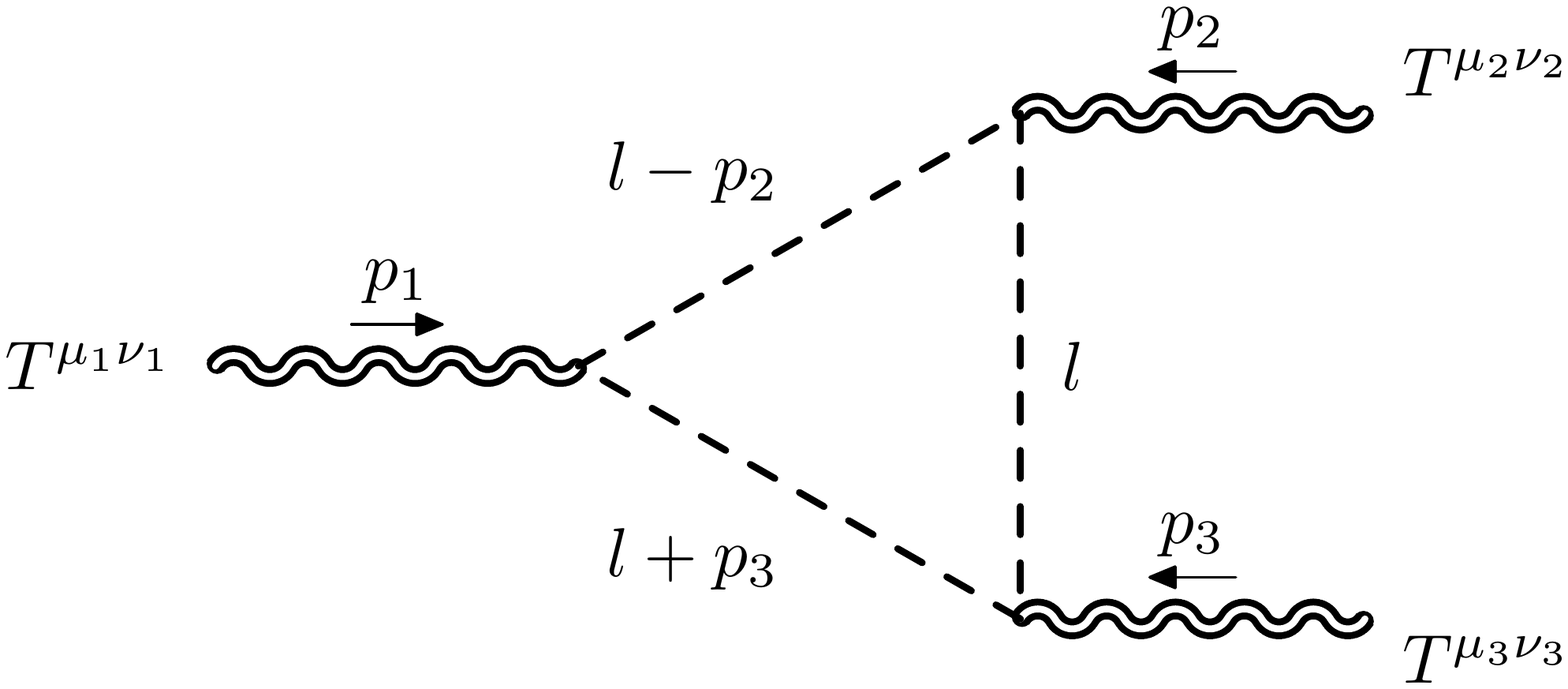}} \hspace{.3cm}
	\subfigure{\includegraphics[scale=0.2]{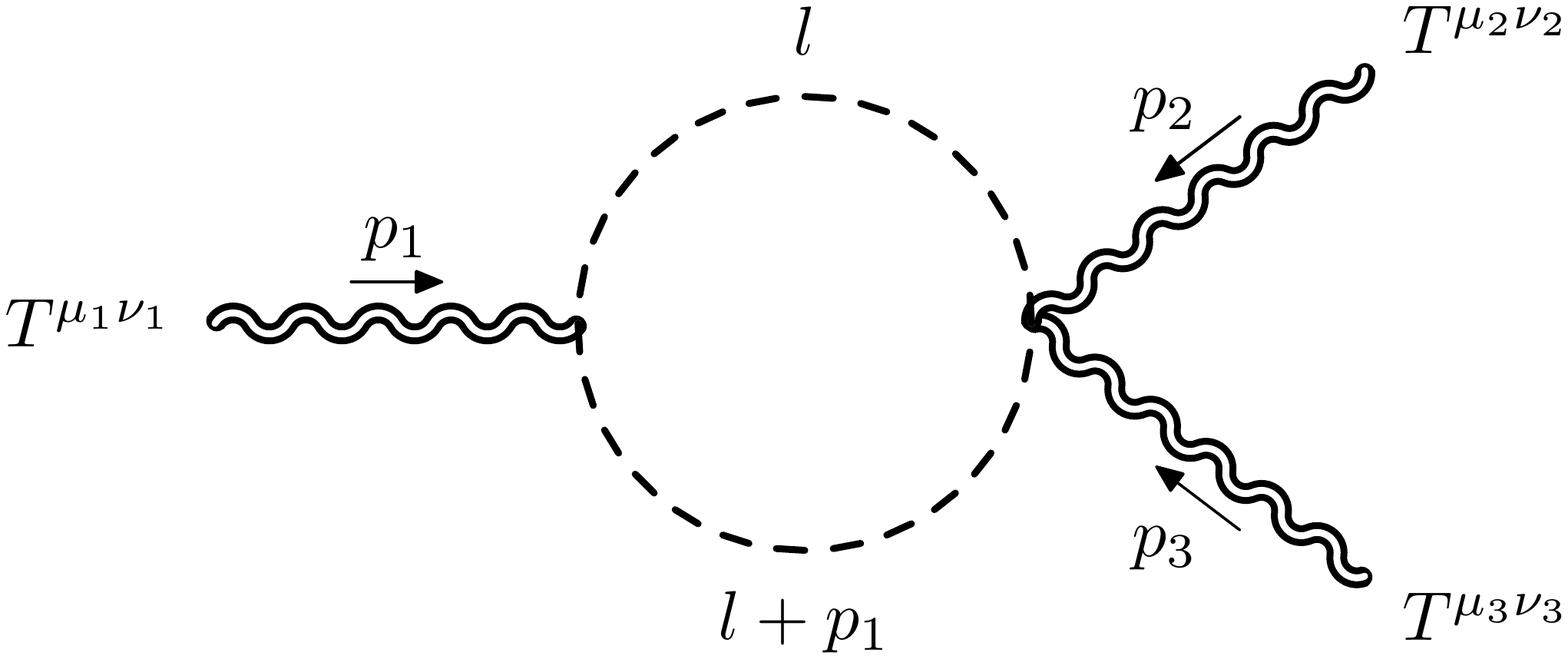}} \hspace{.3cm}
	\raisebox{.12\height}{\subfigure{\includegraphics[scale=0.16]{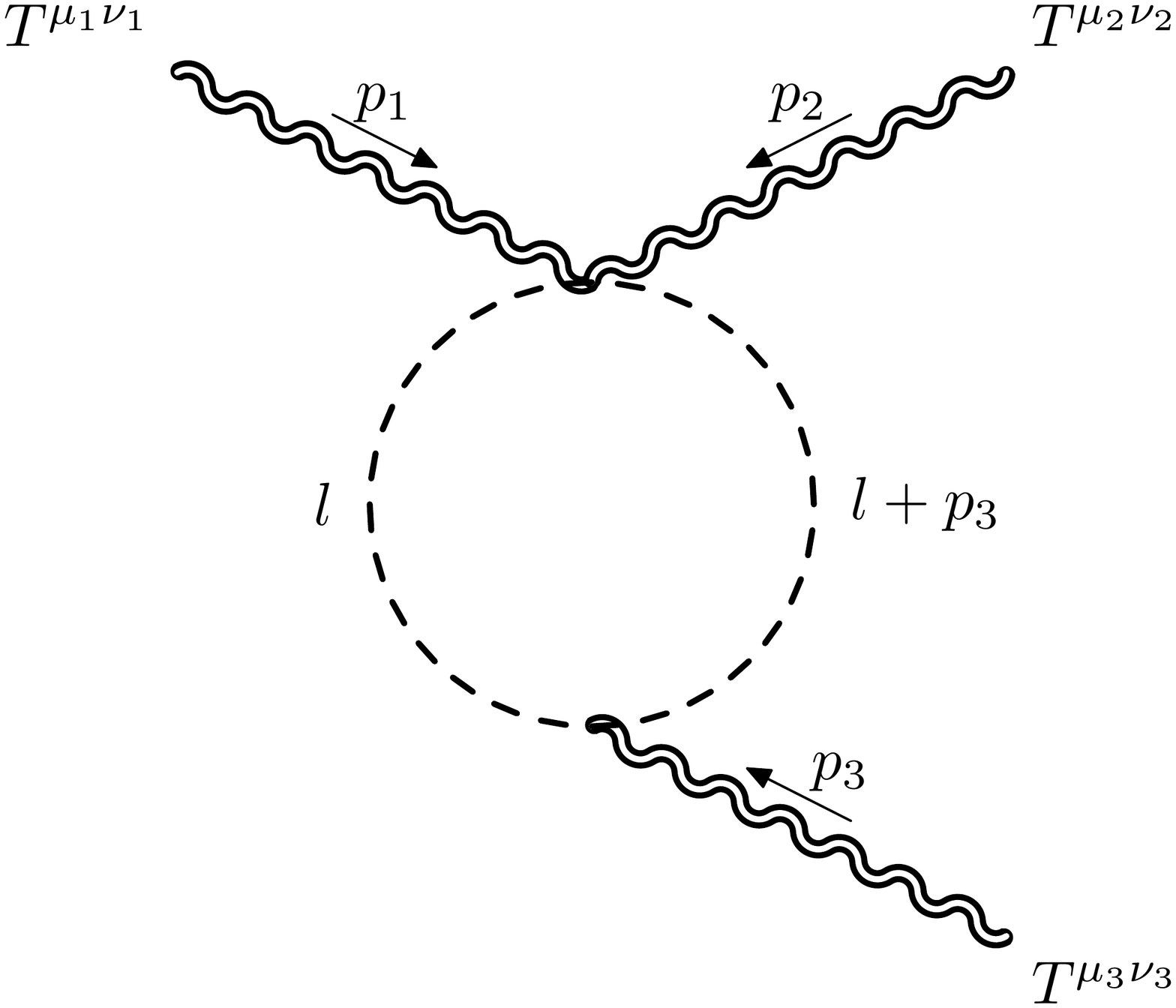}}\hspace{.3cm}}
	\vspace{-0.8cm}\caption{Typical one-loop scalar diagrams for the three-graviton vertex.\label{Feynman1}}
\end{figure}
All the primary CWI's can be solved in terms of the general $3K$ integral
\begin{equation} \label{e:J}
I_{\alpha \{ \beta_1 \beta_2 \beta_3 \}}(p_1, p_2, p_3) = \int_0^\infty d x \: x^\alpha \prod_{j=1}^3 p_j^{\beta_j} K_{\beta_j}(p_j x),
\end{equation}
where $K_\nu$ is a Bessel $K$ function. This integral depends on four parameters, namely the power $\alpha$ of the integration variable $x$, and the three Bessel function indices $\beta_j$.

\begin{equation} \label{e:DeltaToAlpha}
\alpha = \frac{d}{2} - 1 + N, \qquad \qquad \beta_j = \Delta_j - \frac{d}{2} + k_j, \quad j = 1,2,3.
\end{equation}
Here we assume that we concentrate on some particular 3-point function and the conformal dimensions $\Delta_j$, $j=1,2,3$ are therefore fixed. By defining
\begin{equation} \label{e:Jred}
J_{N \{ k_j \}} = I_{\frac{d}{2} - 1 + N \{ \Delta_j - \frac{d}{2} + k_j \}},
\end{equation}
with $\{k_j\} = \{k_1,k_2, k_3\}$, 
the solutions are expressed in the form \cite{Bzowski:2013sza, Bzowski:2018fql}
\begin{align}
A_1 & = \a_1 J_{6 \{000\}}, \label{a:TTT1} \\
A_2 & = 4 \a_1 J_{5 \{001\}} + \a_2 J_{4 \{000\}}, \\
A_3 & = 2 \a_1 J_{4 \{002\}} + \a_2 J_{3 \{001\}} + \a_3 J_{2 \{000\}}, \\
A_4 & = 8 \a_1 J_{4 \{110\}} - 2 \a_2 J_{3 \{001\}} + \a_4 J_{2 \{000\}}, \\
A_5 & = 8 \a_1 J_{3 \{111\}} + 2 \a_2 \left( J_{2 \{110\}} + J_{2 \{101\}} + J_{2 \{011\}} \right) + \a_5 J_{0 \{000\}} \label{a:TTTlast}
\end{align}
in terms of 3K integrals. One important issue is how to relate these explicit solutions to free-field theory.
\subsection{Matching the general CFT solution to free field theory}
We can show that such solutions can be perfectly matched in perturbation theory by introducing three independent sectors, a scalar, a fermion and spin one \cite{Coriano:2018bsy}.\\ 
Also in this case, as for the $TJJ$, one performs a parallel analysis of the decomposition of the correlator 
in free field theory using each such three sectors. In $d=4$ the complete correlation function can then be written as
\begin{equation}
\braket{T^{\m_1\n_1}(p_1)T^{\m_2\n_2}(p_2)T^{\m_3\n_3}(p_3)}=\sum_{I=F,G,S}\,n_I\,\braket{T^{\m_1\n_1}(p_1)T^{\m_2\n_2}(p_2)T^{\m_3\n_3}(p_3)}_I.
\end{equation}
 For the triangle $V$ and contact topologies $W$, the latter take the form
\begin{align}
\braket{t^{\m_1\n_1}(p_1)t^{\m_2\n_2}(p_2)t^{\m_3\n_3}(p_3)}_I&=\,\Pi^{\m_1\n_1}_{\a_1\b_1}(p_1)\Pi^{\m_2\n_2}_{\a_2\b_2}(p_2)\Pi^{\m_3\n_3}_{\a_3\b_3}(p_3)\notag\\
&\times\bigg[ -V_{I}^{\a_1\b_1\a_2\b_2\a_3\b_3}(p_1,p_2,p_3)+\sum_{i=1}^3W_{I,i}^{\a_1\b_1\a_2\b_2\a_3\b_3}(p_1,p_2,p_3)\bigg]
\end{align}
where we have included all the three sectors $(I)$. Both the longitudinal and the transverse sectors are characterized by divergences in the forms of single poles in $1/\epsilon$  ($\epsilon=(4-d)/2$) which needs to be removed by renormalization. The singular contributions of the $A_i$ in DR take the form
\begin{subequations}
\begin{align}
A_2^{Div}&=\frac{\p^2}{45\,\varepsilon}\,\big[26n_G-7n_F-2n_S\big]\\[1ex]
A_3^{Div}&=\frac{\p^2}{90\,\varepsilon}\,\big[3(s+s_1)\big(6n_F+n_S+12n_G\big)+s_2(11n_F+62n_G+n_S)\big]\\[1ex]
A_4^{Div}&=\frac{\p^2}{90\,\varepsilon}\,\big[(s+s_1)\big(29n_F+98n_G+4n_S\big)+s_2(43n_F+46n_G+8n_S)\big]\\[1ex]
A_5^{Div}&=\frac{\p^2}{180\,\varepsilon}\bigg\{n_F \left(43 s^2-14 s (s_1+s_2)+43 s_1^2-14 s_1 s_2+43 s_2^2\right)\notag\\
&\quad+2 \big[n_G \left(23 s^2+26 s (s_1+s_2)+23 s_1^2+26 s_1 s_2+23s_2^2\right)+2n_S\left(2 s^2-s (s_1+s_2)+2 s_1^2-s_1 s_2+2 s_2^2\right)\big]\bigg\}
\end{align}\label{diverg}
\end{subequations}
and are renormalized by the addition of two counterterms in the defining Lagrangian. In perturbation theory the one loop counterterm Lagrangian is
\begin{equation}
S_{count}=-\frac{1}{\varepsilon}\,\sum_{I=F,S,G}\,n_I\,\int d^dx\,\sqrt{-g}\bigg(\,\b_a(I)\,C^2+\b_b(I)\,E\bigg)
\label{scount}
\end{equation}
corresponding to the Weyl tensor squared and the Euler density, omitting the extra 
$R^2$ operator which is responsible for the $\square R$ term in \eqref{TraceAnomaly}, having choosen the local part of anomaly 
$(\sim \beta_c\square R)$ vanishing ($\beta_c=0$). More details concerning this point can be found in \cite{Coriano:2012wp}. The corresponding vertex counterterms are
\begin{align}
&\braket{T^{\m_1\n_1}(p_1)T^{\m_2\n_2}(p_2)T^{\m_3\n_3}(p_3)}_{count}=\notag\\
&\hspace{2cm}=-\frac{1}{\varepsilon}\sum_{I=F,S,G}n_I\bigg(\b_a(I)\,V_{C^2}^{\m_1\n_1\m_2\n_2\m_3\n_3}(p_1,p_2,p_3)+\b_b(I)\,V_{E}^{\m_1\n_1\m_2\n_2\m_3\n_3}(p_1,p_2,p_3)\bigg)
\end{align} 
and can be separated in their transverse-traceless components and their longitudinal ones. The $A_i$'s are renormalized by the first, while the second renormalize the longitudinal components of the $TTT$. The approach has been detailed in \cite{Coriano:2012wp}. The method consists in taking the transverse traceless projectors (with (open indices) as $d$-dimensional tensors and in expanding their d-dependence as $(d-4) + 4$, generating new tensors which are transverse traceless only in $d=4$, plus a remainder. All the index contractions are performed with a d-dimensional metric tensor, while the inclusion of the counterterms generated by \eqref{scount} allows to remove the $1/\epsilon$ terms. At the final stage the expression of the renormalized vertex is given only by the 4-dimensional component of such a final tensor, which can be traced as an ordinary 4-$d$ tensor, thereby generating an anomaly.  \\
An extensive analysis shows that the entire 4-$d$ solution of the CWI's $A_i$ can be reconctructed just by a superposition of the three perturbative sectors mentioned above. In \cite{2014JHEP...03..111B} the general 3K solution is uniquely parameterized in terms of some constants $\a_1,\a_2$ and $c_T$. If we choose, for instance 
\bea
\a_1=\frac{\p^3(n_S-4n_F)}{480}, \qquad \a_2=\frac{\p^3\,n_F}{6}, \qquad c_T=\frac{3\p^{5/2}}{128}(n_S+4n_F), \qquad c_g=0 
\eea
valid for $d=3$, we can match perturbative and non-perturbative results in the same dimension. It is then clear that, given the rather complex structure of the 3K integrals, the use of perturbation theory and the matching allow to re-express such integrals in a very simple form, using only the scalar 2- and 3-point functions of the ordinary Feynman expansion as a basis. This implies that the general hypergeometric solution in $d=4$ should drastically simplify, in order to agree with the perturbative one. In $d=4$ an explicit match as in $d=3$, for instance, can be worked out only at numerical level, since the procedure developed for the renormalization of such integrals is rather involved. 
The general expressions of the $A_i$ in $d=4$, valid non-perturbatively, can be found in \cite{Coriano:2018bsy}.\\
It is interesting to observe how the structure of the anomalous contributions of the correlator appear in this formulation. They are associated to terms which develop a nonvanishing single, double and triple trace, with massless exchanges in the three separate legs of the correlator, as shown in Fig. \ref{FeynmanX}, which generalizes the behaviour of Fig. \ref{fig1}. Such anomalous ($\langle..\rangle_a$) terms are given by

\begin{figure}[t]
\centering
\includegraphics[scale=0.5]{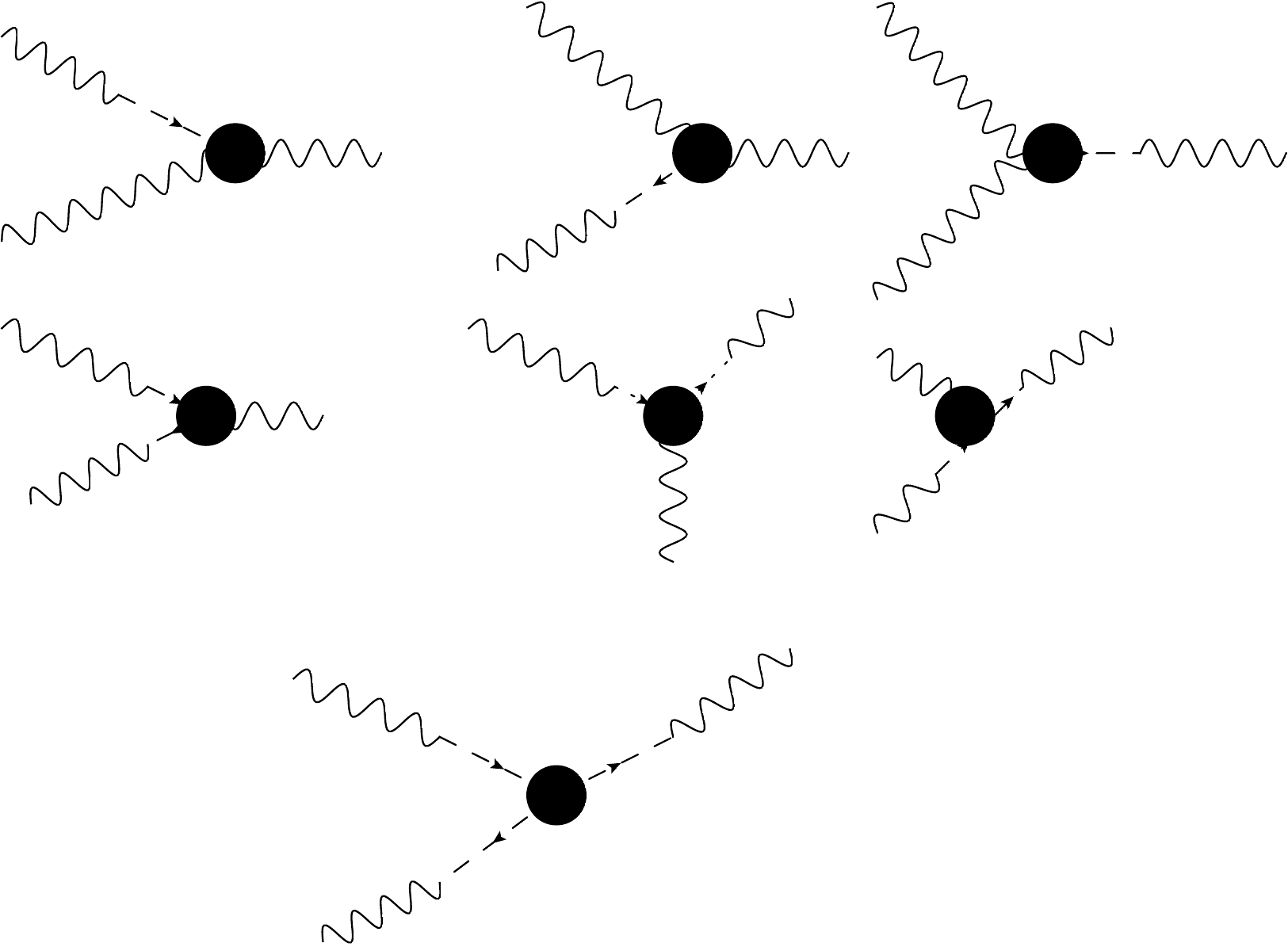}
\caption{The anomaly contribution in the TTT correlator due to massless exchanges on each separate leg, represented by dashed lines. \label{FeynmanX}}
\end{figure}

\begin{align}
\label{expans}
& \lla T^{\mu_1 \nu_1}({p}_1) T^{\mu_2 \nu_2}({p}_2) T^{\mu_3 \nu_3}({p}_3) \rra_{a} = \frac{\hat\pi^{\mu_1 \nu_1}({p}_1)}{3\, p_1^2} \lla T({p}_1) T^{\mu_2 \nu_2}({p}_2) T^{\mu_3 \nu_3}({p}_3) \rra_{a} \nn \\
& + \frac{\hat\pi^{\mu_2 \nu_2}({p}_2)}{3\, p_2^2} \lla T^{\mu_1 \nu_1}({p}_1) T({p}_2) T^{\mu_3 \nu_3}({p}_3) \rra_{a} + \frac{\hat\pi^{\mu_3 \nu_3}({p}_3)}{3\, p_3^2} \lla T^{\mu_1 \nu_1}({p}_1) T^{\mu_2 \nu_2}({p}_2)  T({p}_3)\rra_{a} \nn \\
& - \: \frac{\hat\pi^{\mu_1 \nu_1}({p}_1) \hat\pi^{\mu_2 \nu_2}({p}_2)}{9\, p_1^2 p_2^2}\lla T({p}_1)T({p}_2)T^{\mu_3 \nu_3}({p}_3)\rra_{a} - \: \frac{\hat\pi^{\mu_2 \nu_2}({p}_2) \hat\pi^{\mu_3 \nu_3}({p}_2)}{9 p_2^2 p_3^2}\lla T^{\mu_1 \nu_1}({p}_1)T(p_2)T({p}_3)\rra_{a} \nn\\
& - \: \frac{\hat\pi^{\mu_1 \nu_1}({p}_1)\hat \pi^{\mu_3 \nu_3}(\bar{p}_3)}{9 p_1^2 p_3^2}\lla T({p}_1) T^{\mu_2 \nu_2}({p}_2)T({p}_3)\rra_{a}  + \frac{\hat\pi^{\mu_1 \nu_1}({p}_1)\hat\pi^{\mu_2 \nu_2}({p}_2) \hat\pi^{\mu_3 \nu_3}(\bar{p}_3)}{27 p_1^2 p_2^2 p_3^2}\lla T({p}_1)T({p}_2)T(\bar{p}_3)_{a} .
\end{align}
 As shown in \cite{Coriano:2017mux}, they can be reobtained from the nonlocal anomaly action \eqref{Snonl}
 .
\section{Moving to supersymmetry: Anomalies and sum rules for the anomaly supermultiplet}

\begin{figure}[t]
\centering
\subfigure[]{\includegraphics[scale=0.6]{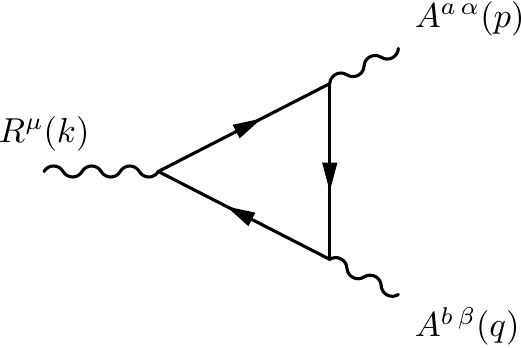}} \hspace{.5cm}
\subfigure[]{\includegraphics[scale=0.6]{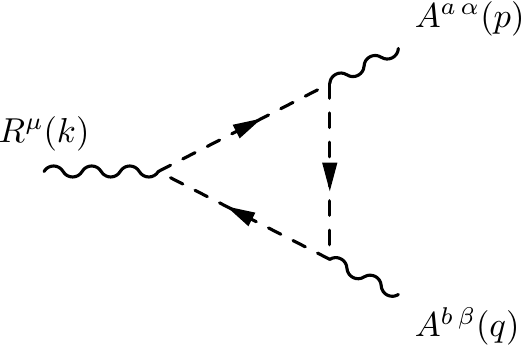}} \hspace{.5cm}
\subfigure[]{\includegraphics[scale=0.6]{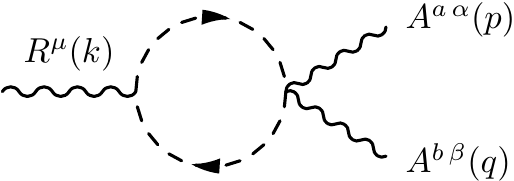}} 
\\

\centering
\subfigure[]{\includegraphics[scale=0.6]{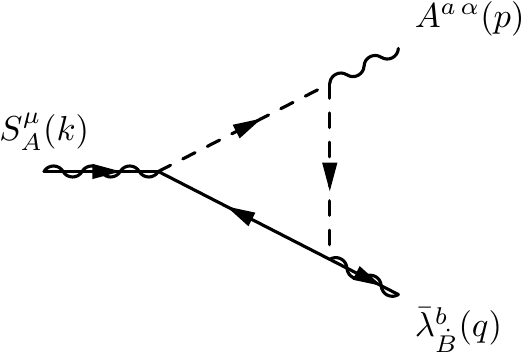}} \hspace{.5cm}
\subfigure[]{\includegraphics[scale=0.6]{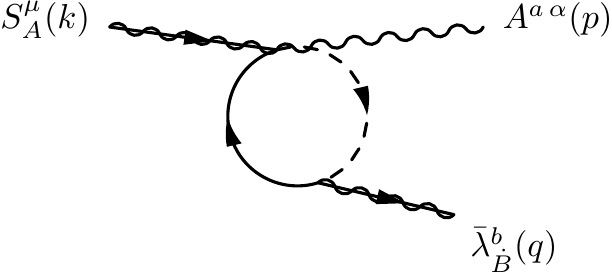}} \hspace{.5cm}
\subfigure[]{\includegraphics[scale=0.6]{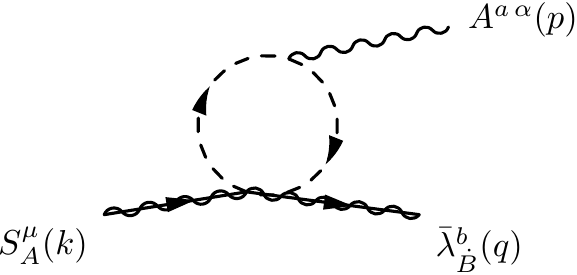}}\\
\subfigure[]{\includegraphics[scale=0.5]{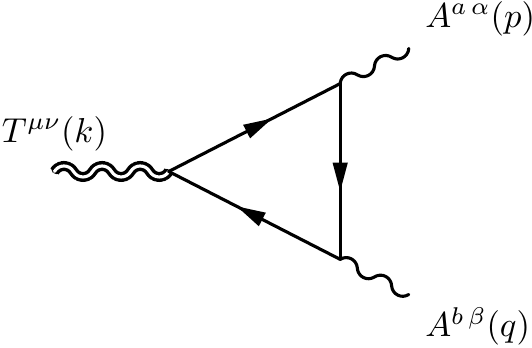}} \hspace{.5cm}
\subfigure[]{\includegraphics[scale=0.5]{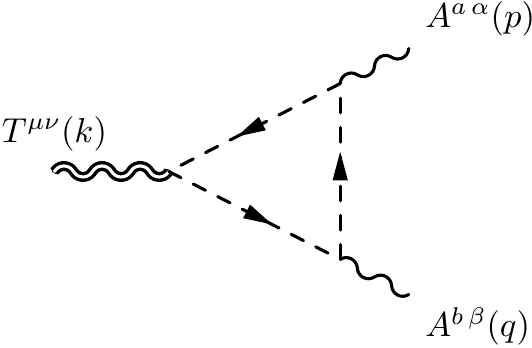}} \hspace{.5cm}
\subfigure[]{\includegraphics[scale=0.5]{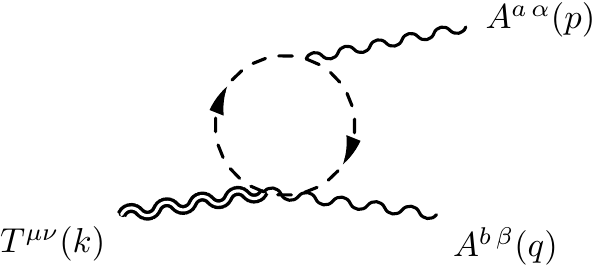}} \\
\caption{Typical contributions to the one-loop perturbative expansion of the $\langle RVV \rangle$  $\langle SVF \rangle$ $\langle TJJ \rangle$ diagrams  \label{Fig.Tchiral}}
\end{figure}
The kinematical features characterizing chiral and conformal anomalies become standard once we turn to consider a supersymmetric context. 
Supersymmetry, indeed, provides a unification of the phenomenon discussed above, identified in perturbation theory, as shown in the case of an $\mathcal{N}=1$ super Yang-Mills theory.\\
 In the 1PI anomaly action of this theory, one identifies three specific contributions which are responsible for the generation of the anomaly supermultiplet, in the form of three poles in the three separate channels. Conformal and chiral anomalies are then unified by the same perturbative behaviour. We are going to summarize this result, based on the analysis presented in \cite{Coriano:2014gja}. \\
There is a related phenomenon in this case, which can be identified by moving away from the conformal limit. This can be achieved by the inclusion of mass terms in the Lagrangian at a first stage, which can be used 
to discuss the structure of the anomaly vertices as a function of $m$. As in the previous cases, we will adopt dimensional regularization.\\
The anomaly form factors are characterized by a unique function 
$(\chi)$ for which one can write down a spectral representation in terms of a spectral density $\rho(s,m^2)$ supported by a single branch cut
\beq
\chi(k^2,m^2)=\frac{1}{\pi}\int_{4 m^2}^{\infty}\frac{{\rho}_\chi(s,m^2)}{s- k^2 }ds
\eeq
corresponding to the ordinary threshold at $k^2=4 m^2$, with 
\beq
\label{spectralrho}
{\rho}_\chi(s,m^2) = \frac{1}{2 i} \textrm{Disc} \, \chi(s,m^2)  
=\frac{2 \pi m^2}{s^2}\log\left(\frac{1 + \sqrt{\tau(s,m^2)}}{1-\sqrt{\tau(s,m^2)}}\right)\theta(s- 4 m^2). 
\eeq
A crucial feature of these spectral densities is the existence of a sum rule, given by
\beq
\label{superc}
\frac{1}{\pi}\int_{4 m^2}^{\infty}ds{{\rho}_\chi(s,m^2)}=1, 
\eeq
where the right hand side has been normalized to 1. In general, it equals the anomaly. Generically, it is given in the form 
\bea
\label{sr}
\frac{1}{\pi} \int_0^{\infty}\rho(s, m^2) ds =  f,
\eea
with the constant $f$ independent of any mass (or other) parameter which characterizes the thresholds or the strengths of the resonant states eventually present in the integration region $(s>0)$. \\
It is quite straightforward to show that Eq. (\ref{sr}) is a constraint on the asymptotic behaviour of the related form factor.
The proof is obtained by observing that the dispersion relation for a form factor in the spacelike region ($Q^2=-k^2> 0$)
\beq
F(Q^2,m^2)=\frac{1}{\pi}\int_0^{\infty} ds\frac{ \rho(s,m^2) }{s + Q^2} \,,
\eeq
once we expand the denominator in $Q^2$ as $\frac{1}{s + Q^2}=\frac{1}{Q^2} - \frac{1}{Q^2} s\frac{1}{Q^2} +\ldots$ and make use of Eq. (\ref{sr}), induces the following asymptotic behaviour on $F(Q^2,m^2)$
\beq
\lim_{Q^2\to \infty} Q^2 F(Q^2, m^2)=f.
\label{asym}
\eeq
The $F\sim f/Q^2$ behaviour at large $Q^2$, with $f$ independent of $m$, shows the pole dominance of $F$ for $Q^2 \rightarrow \infty$. Indeed, the resonant (pole) behaviour of such spectral density is obtained in the $m\to 0$ limit
\bea
\lim_{m\to 0} \rho_\chi(s,m^2) = \lim_{m\to 0} \frac{2 \pi m^2}{s^2}\log\left(\frac{1 + \sqrt{\tau(s,m^2)}}{1-\sqrt{\tau(s,m^2)}}\right)\theta(s- 4 m^2) = \pi \delta(s),
\eea
where $\rho$ converges to a delta function. Clearly, it is the region around the light cone ($s\sim 0$) which dominates the sum rule and it amounts to a resonant contribution.\\
Therefore, the presence of a $1/Q^2$ term in the anomaly form factors is a property of the entire flow which converges to a localized massless state (i.e. $\rho(s)\sim\delta(s)$) as 
$m\to 0$, while the presence of a non vanishing sum rule guarantees the validity of the asymptotic constraint illustrated in Eq. (\ref{asym}). Notice that although for conformal deformations driven by a single mass parameter the independence of the asymptotic value $f$ on $m$ is a simple consequence of the scaling behaviour of $F(Q^2,m^2)$, it holds quite generally even for a completely off-shell kinematics \cite{Giannotti:2008cv}. 
\subsection{The case of an $\mathcal{N}=1$ theory}
To illustrate this result, let's consider the Lagrangian of an $\mathcal{N}=1$ theory
\bea
\label{SUSYlagrangianCF}
\mathcal L &=& - \frac{1}{4} F^a_{\mu\nu} F^{a \, \mu\nu} + i \lambda^a \sigma^\mu \mathcal D_\mu^{ab} \bar \lambda^b
+ ( \mathcal D_{ij}^\mu \phi_j )^\dag (\mathcal D_{ik \, \mu} \phi_k) + i \chi_j \sigma_\mu \mathcal D_{ij}^{\mu \, \dag} \bar \chi_i \nn \\
&& - \sqrt{2} g \left( \bar \lambda^a \bar \chi_i T^a_{i j} \phi_j  + \phi_i^\dag T^a_{ij} \lambda^a \chi_j \right) - V(\phi, \phi^\dag) - \frac{1}{2} \left( \chi_i \chi_j \mathcal W_{ij}(\phi) + h.c.  \right) \,,
\eea
where the gauge covariant derivatives on the matter fields and on the gaugino are defined respectively as
\bea
\mathcal D^\mu_{ij} = \delta_{ij} \partial^\mu + i g A^{a \, \mu} T^a_{ij} \,, \qquad
\mathcal D_\mu^{ac} = \delta^{ac} \partial^\mu -g \, t^{abc} A^b_\mu \,,
\eea
with $t^{abc}$ the structure constants of the adjoint representation, and the scalar potential is given by
\bea
V(\phi, \phi^\dag) = \mathcal W^\dag_i(\phi^\dag) \mathcal W_i(\phi) + \frac{1}{2} g^2 \left( \phi_i^\dag T^a_{ij} \phi_j \right)^2 \,.
\eea
For the derivatives of the superpotential we have been used the following definitions
\bea
\mathcal W_i(\phi) = \frac{\partial \mathcal W(\Phi)}{\partial \Phi_i} \bigg|  \,, \qquad \mathcal W_{ij}(\phi) = \frac{\partial^2 \mathcal W(\Phi)}{\partial \Phi_i \partial \Phi_j} \bigg| \,,
\eea

\begin{figure}[t]
\centering
\subfigure{\includegraphics[scale=0.9]{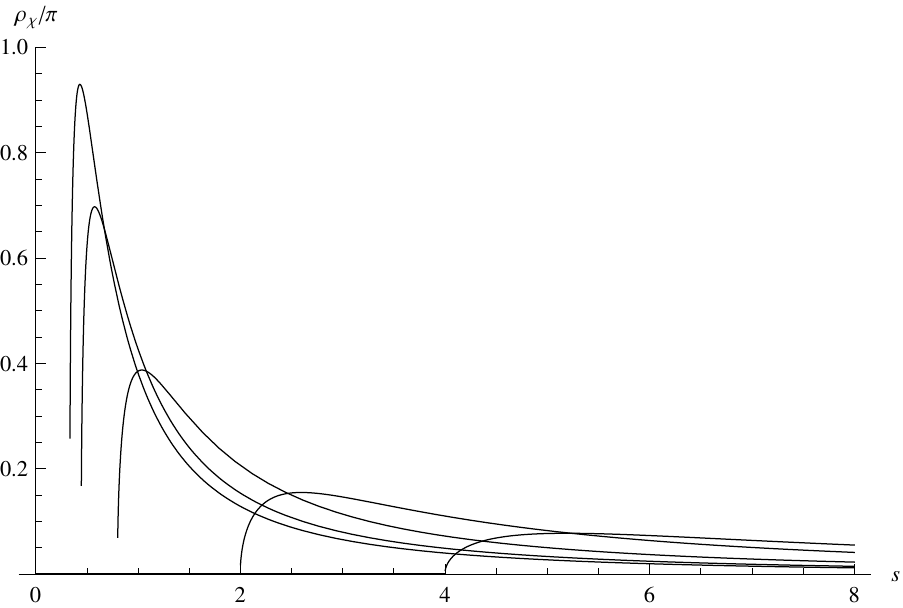}} \hspace{.5cm}
\caption{Representatives of the family of spectral densities $\frac{{{\rho}_\chi}^{(n)}}{\pi}(s)$ plotted versus $ s $ in units of $m^2$. The family "flows"  towards the $s=0$ region becoming a $\delta(s)$ function as $m^2$ goes to zero. }
\label{seq}
\end{figure}

where the symbol $|$ on the right indicates that the quantity is evaluated at $\theta = \bar \theta = 0$. \\
Having defined the model, we can introduce the Ferrara-Zumino hypercurrent
\bea
\label{Hypercurrent}
\mathcal J_{A \dot A} = \textrm{Tr}\left[ \bar W_{\dot A} e^V W_A e^{- V}\right]
- \frac{1}{3} \bar \Phi \left[  \stackrel{\leftarrow}{\bar \nabla}_{\dot A}  e^V \nabla_A - e^V \bar D_{\dot A} \nabla_A +  \stackrel{\leftarrow}{\bar \nabla}_{\dot A} \stackrel{\leftarrow}{D_A} e^V \right] \Phi \,,
\eea
where $\nabla_A$ is the gauge-covariant derivative in the superfield formalism whose action on chiral superfields is given by
\bea
\nabla_A \Phi = e^{-V} D_A \left( e^V \Phi \right)\,, \qquad  \bar \nabla_{\dot A} \bar \Phi = e^{V} \bar D_{\dot A} \left( e^{-V} \bar \Phi \right)\,.
\eea
The conservation equation for the hypercurrent $\mathcal J_{A \dot A}$ is
\bea
\label{HyperAnomaly}
\bar D^{\dot A} \mathcal J_{A \dot A} = \frac{2}{3} D_A \left[ - \frac{g^2}{16 \pi^2} \left( 3 T(A) - T(R)\right) \textrm{Tr}W^2 - \frac{1}{8} \gamma \, \bar D^2 (\bar \Phi e^V \Phi)+ \left( 3 \mathcal W(\Phi) - \Phi \frac{\partial \mathcal W(\Phi)}{\partial \Phi} \right) \right] \,,\nn\\
\eea
where $\gamma$ is the anomalous dimension of the chiral superfield. \\
The first two terms in Eq. (\ref{HyperAnomaly}) describe the quantum anomaly of the hypercurrent, while the last is of classical origin and it is entirely given by the superpotential. In particular, for a classical scale invariant theory, in which $\mathcal W$ is cubic in the superfields or identically zero, this term identically vanishes. If, on the other hand, the superpotential is quadratic, the conservation equation of the hypercurrent acquires a non-zero contribution even at classical level. This describes the explicit breaking of the conformal symmetry.
By projecting the hypercurrent $\mathcal J_{A \dot A}$ defined in Eq.(\ref{Hypercurrent}) onto its components 
we get the explicit expressions of the three anomaly equations.\\
 The lowest component is given by the $R^\mu$ current, the $\theta$ term is associated with the supercurrent $S^\mu_A$, while the $\theta \bar \theta$ component contains the  energy-momentum tensor $T^{\mu\nu}$. In the $\mathcal{N}=1$ super Yang-Mills theory described by the Lagrangian in Eq. (\ref{SUSYlagrangianCF}), these three currents are defined as
\bea
\label{Rcurrent}
R^\mu &=& \bar \lambda^a \bar \sigma^\mu \lambda^a 
+ \frac{1}{3} \left( - \bar \chi_i \bar \sigma^\mu \chi_i + 2 i \phi_i^\dag \mathcal D^\mu_{ij} \phi_j - 2 i (\mathcal D^\mu_{ij} \phi_j)^\dag \phi_i \right) \,, \\
\label{Scurrent}
S^\mu_A &=& i (\sigma^{\nu \rho} \sigma^\mu \bar \lambda^a)_A F^a_{\nu\rho}
 - \sqrt{2} ( \sigma_\nu \bar \sigma^\mu \chi_i)_A (\mathcal D^{\nu}_{ij} \phi_j)^\dag - i \sqrt{2} (\sigma^\mu \bar \chi_i) \mathcal W_i^\dag(\phi^\dag) \nn \\
&-&  i g (\phi^\dag_i T^a_{ij} \phi_j) (\sigma^\mu \bar \lambda^a)_A + S^\mu_{I \, A}\,, \\
\label{EMT}
T^{\mu\nu} &=&  - F^{a \, \mu \rho} {F^{a \, \nu}}_\rho 
+ \frac{i}{4} \left[ \bar \lambda^a \bar \sigma^\mu (\delta^{ac} \stackrel{\rightarrow}{\partial^\nu} - g \, t^{abc} A^{b \, \nu} ) \lambda^c + 
\bar \lambda^a \bar \sigma^\mu (- \delta^{ac} \stackrel{\leftarrow}{\partial^\nu} - g \, t^{abc} A^{b \, \nu} ) \lambda^c + (\mu \leftrightarrow \nu) \right] \nn \\
&+&  ( \mathcal D_{ij}^\mu \phi_j )^\dag (\mathcal D_{ik}^\nu \phi_k)  +   ( \mathcal D_{ij}^\nu \phi_j )^\dag (\mathcal D_{ik}^\mu \phi_k) +
\frac{i}{4} \left[ \bar \chi_i \bar \sigma^\mu ( \delta_{ij} \stackrel{\rightarrow}{\partial^\nu} + i g T^a_{ij} A^{a \, \nu} ) \chi_j \right. \nn \\
&+& \left.  \bar \chi_i \bar \sigma^\mu ( - \delta_{ij} \stackrel{\leftarrow}{\partial^\nu} + i g T^a_{ij} A^{a \, \nu} ) \chi_j + (\mu \leftrightarrow \nu) \right]  - \eta^{\mu\nu} \mathcal L + T^{\mu\nu}_I \,, 
\eea
where $\mathcal L$ is given in Eq.(\ref{SUSYlagrangianCF}) and $S^\mu_I$ and $T^{\mu\nu}_I$ are the terms of improvement in $d=4$ of the supercurrent and of the EMT respectively. As in the non-supersymmetric case, these terms are necessary only for a scalar field and therefore receive contributions only from the chiral multiplet. They are explicitly given by
\bea
S^\mu_{I \, A} &=& \frac{4 \sqrt{2}}{3} i \left[ \sigma^{\mu\nu} \partial_\nu (\chi_i \phi_i^\dag) \right]_A \,, \\
T^{\mu\nu}_I &=& \frac{1}{3} \left( \eta^{\mu \nu} \partial^2 - \partial^\mu \partial^\nu \right) \phi^\dag_i \phi_i \,.
\eea

\begin{figure}[t]
\centering
\subfigure{\includegraphics[scale=0.6]{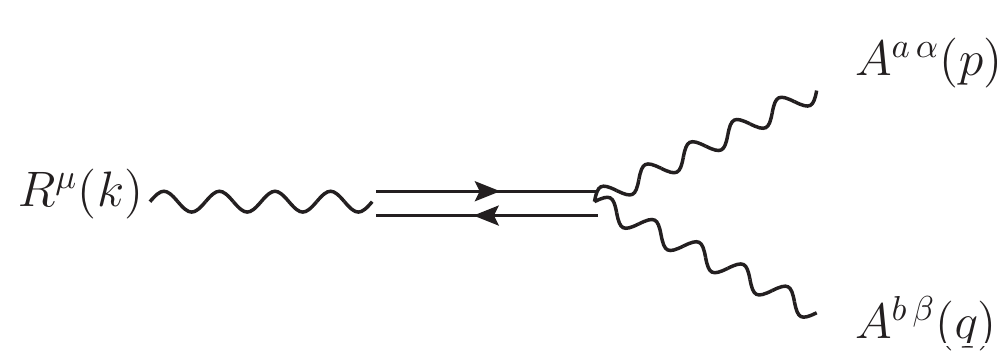}} \hspace{.5cm}
\subfigure{\includegraphics[scale=0.6]{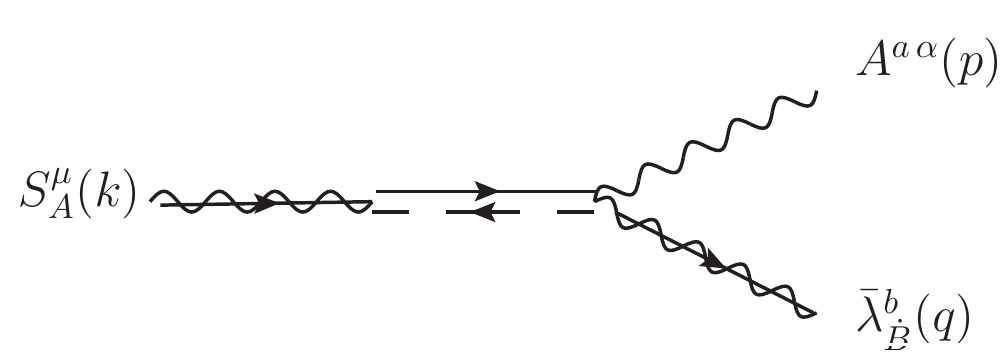}} 
\subfigure{\includegraphics[scale=0.6]{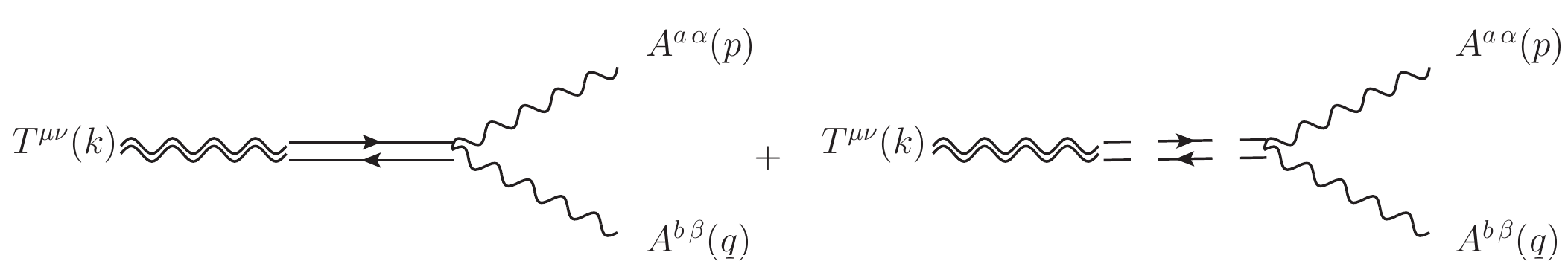}} 
\caption{The collinear diagrams corresponding to the exchange of a composite axion (top right), a dilatino (top left) and the two sectors of an intermediate dilaton (bottom). Dashed lines denote intermediate scalars.}
\label{RST}
\end{figure}

The terms of improvement are automatically conserved and guarantee, for $\mathcal W(\Phi) = 0$, upon using the equations of motion, the vanishing of the classical trace of $T^{\mu\nu}$ and of the classical gamma-trace of the supercurrent $S^\mu_A$. %
The anomaly equations in the component formalism, which can be projected out from Eq. (\ref{HyperAnomaly}), are
\bea
\label{AnomalyR}
\partial_\mu R^\mu &=& \frac{g^2}{16 \pi^2} \left( T(A) - \frac{1}{3} T(R) \right) F^{a \, \mu\nu} \tilde F^a_{\mu\nu} \,, \\
\label{AnomalyS}
\bar \sigma_\mu S^\mu_A &=&  - i \frac{3 \, g^2}{8 \pi^2} \left( T(A) -\frac{1}{3} T(R) \right) \left( \bar \lambda^a \bar \sigma^{\mu\nu} \right)_A F^a_{\mu\nu }\,, \\
\label{AnomalyT}
\eta_{\mu\nu} T^{\mu\nu} &=& -  \frac{3 \, g^2}{32 \pi^2} \left(T(A) - \frac{1}{3} T(R) \right) F^{a \, \mu\nu}  F^a_{\mu\nu} \,.
\eea
The first and the last equations are respectively extracted from the imaginary and the real part of the $\theta$ component of Eq.(\ref{HyperAnomaly}), while the gamma-trace of the supercurrent comes from the lowest component.

We define the three correlation functions, $\Gamma_{(R)}$, $\Gamma_{(S)}$ and $\Gamma_{(T)}$ as
\bea
\label{RSTCorrelators}
\delta^{ab} \, \Gamma_{(R)}^{\mu\alpha\beta}(p,q) &\equiv& \langle R^{\mu}(k)\, A^{a \, \alpha}(p) \, A^{b \, \beta}(q) \rangle \qquad \langle RVV \rangle \,, \nn \\
\delta^{ab} \, \Gamma_{(S) \, A\dot B}^{\mu\alpha}(p,q) &\equiv& \langle S^{\mu}_A (k) \, A^{a \, \alpha}(p) \, \bar \lambda^b_{\dot B}(q) \rangle \qquad \langle SVF \rangle \,, \nn \\
\delta^{ab} \, \Gamma_{(T)}^{\mu\nu\alpha\beta}(p,q) &\equiv& \langle T^{\mu\nu}(k) \, A^{a \, \alpha}(p) \, A^{b \, \beta}(q)\rangle \qquad \langle TVV \rangle  \,,
\eea
with $k = p+q$ and where we have factorized, for the sake of simplicity, the Kronecker delta on the adjoint indices. These correlation functions have been computed at one-loop order in the dimensional reduction scheme (DRed) using the 

\bea
\label{RChiralOSMassless}
\Gamma_{(R)}^{\mu\alpha\beta}(p,q) = - i \frac{g^2 \, T(R)}{12 \pi^2} \frac{k^\mu}{k^2} \eps[p, q, \alpha ,\beta] \,,
\eea
The correlator in Eq.(\ref{RChiralOSMassless}) satisfies the vector current conservation constraints and the anomalous equation of Eq.(\ref{AnomalyR}) 
\bea
\label{AnomalyRmom}
i k_\mu \, \Gamma_{(R)}^{\mu\alpha\beta}(p,q) = \frac{g^2 \, T(R)}{12 \pi^2} \, \eps[p, q, \alpha ,\beta] \,.
\eea
Therefore, in the on-shell case and for massless fermions we recover the usual structure of the $\langle AVV \rangle$ diagram.
In general we obtain
\bea
\label{RVectorOS}
\Gamma_{(R)}^{\mu\alpha\beta}(p,q) &=&  i \frac{g^2 \, T(A)}{4 \pi^2} \frac{k^\mu}{k^2} \eps[p, q, \alpha ,\beta] \,,  \\
\label{SVectorOS}
\Gamma_{(S)}^{\mu\alpha}(p,q) &=&   i \frac{g^2 T(A)}{2 \pi^2 \, k^2} s_1^{\mu\alpha} + i \frac{g^2 T(A)}{64 \pi^2} V(k^2) \, s_2^{\mu\alpha} \,, \\
\label{TVectorOS}
\Gamma_{(T)}^{\mu\nu\alpha\beta}(p,q) &=&  \frac{g^2 \, T(A)}{8 \pi^2 \, k^2} t_1^{\mu\nu\alpha\beta}(p,q)  + \frac{g^2 \, T(A)}{16 \pi^2} V(k^2) \, t_{2}^{\mu\nu\alpha\beta}(p,q) \,,
\eea
where
\bea
V(k^2) = -3 + 3 \, \mathcal B_0(0,0) - 3 \, \mathcal B_0(k^2,0) - 2 k^2 \, \mathcal C_0(k^2,0) \,.
\eea
and with $\mathcal B_0(k^2,0)$ and $\mathcal C_0(k^2,0)$ denoting the scalar 2- and 3-point functions computing in the massless case.
Notice that the two vector lines are kept on-shell for simplicity. The structure of the correlators is then given by 
\bea
\label{RChiralOSMassive}
\Gamma_{(R)}^{\mu\alpha\beta}(p,q) &=&  i \frac{g^2 \, T(R)}{12 \pi^2} \, \Phi_1(k^2,m^2) \, \frac{k^\mu}{k^2} \eps[p, q, \alpha ,\beta]  \,, \\
\label{SChiralOSMassive}
\Gamma^{\mu\alpha}_{(S)}(p,q) &=&  i \frac{g^2 T(R)}{6 \pi^2 \, k^2} \, \Phi_1(k^2,m^2) \, s_1^{\mu\alpha}
+ i \frac{g^2 T(R)}{64 \pi^2}  \, \Phi_2(k^2,m^2) \, s_2^{\mu\alpha} \,, \\
\label{TChiralOSMassive}
\Gamma_{(T)}^{\mu\nu\alpha\beta}(p,q) &=& \frac{g^2 \, T(R)}{24 \pi^2 \, k^2} \, \Phi_1(k^2,m^2) \, t_{1S}^{\mu\nu\alpha\beta}(p,q) + \frac{g^2 \, T(R)}{16 \pi^2} \, \Phi_2(k^2,m^2) \, t_{2S}^{\mu\nu\alpha\beta}(p,q) \,, 
\eea
with
\bea
\Phi_1(k^2,m^2) &=& - 1 - 2\, m^2 \, \mathcal C_0(k^2,m^2) \,, \nn \\
\Phi_2(k^2,m^2) &=& 1 - \mathcal B_0(0,m^2) + \mathcal B_0(k^2,m^2) + 2 m^2  \mathcal C_0(k^2,m^2), \,
\label{exp1}
\eea
and with the anomalous broken Ward identities taking the form
\bea
i k_\mu \, \Gamma^{\mu\alpha\beta}_{(R)}(p,q) &=& -\frac{g^2 T(R)}{12\pi^2} \Phi_1(k^2,m^2) \eps[p,q,\alpha,\beta] \,, \\
\bar \sigma_{\mu} \, \Gamma^{\mu\alpha}_{(S)}(p,q) &=& - \frac{g^2 T(R)}{ 4 \pi^2}  \Phi_1(k^2,m^2) \bar \sigma^{\alpha \beta} p_\beta \,, \\
\eta_{\mu\nu} \, \Gamma^{\mu\nu\alpha\beta}_{(T)}(p,q) &=&  \frac{g^2 T(R)}{8\pi^2} \Phi_1(k^2,m^2) u^{\alpha\beta}(p,q) \,.
\eea
A similar result holds also for the Konishi current 

\bea
\label{Jfcurrent}
J^f_\mu =  \bar \chi^f \bar \sigma_\mu \chi^f + i \, \phi^{f \, \dag} (\mathcal D_\mu \phi^f) - i \, (\mathcal D_\mu \phi^f)^\dag \phi^f \,
\eea
\bea
\label{konishi}
\Gamma^{\mu\alpha\beta}_{(J^f)}(p,q) = - i \frac{g^2 \, T(R_f)}{4 \pi^2}\Phi_1(k^2,m^2) \frac{k^\mu}{k^2} \eps[p, q, \alpha ,\beta] \,,
\eea  
with $\Phi_1(k^2,m^2)$ given in Eq. (\ref{exp1}), in full analogy with the result for the correlator of the $R$ current. \\
Defining 
\beq
 \chi(k^2, m^2)\equiv \Phi_1(k^2,m^2)/k^2, 
 \eeq
the discontinuity of the anomalous form factor $\chi(k^2,m^2)$ is then given by
\bea
\textrm{Disc}\, \chi(k^2, m^2)= \chi(k^2+i\epsilon,m^2)-\chi(k^2-i\epsilon, m^2) = - \textrm{Disc}\left( \frac{1}{k^2} \right) - 2 m^2 \textrm{Disc}\left(\frac{\mathcal C_0(k^2,m^2)}{k^2}\right)\nn\\
\label{cancel1}
\eea

giving

\beq
\textrm{Disc} \, \chi(k^2,m^2)=4 i \pi \frac{m^2}{ (k^2)^2}\log \frac{1 + \sqrt{\tau(k^2,m^2)}}{1 - \sqrt{\tau(k^2,m^2)} }\theta(k^2- 4 m^2).
 \eeq 
The total discontinuity of $\chi(k^2,m^2)$, as seen from the result above, is characterized just by a single cut for $k^2> 4 m^2$, since the $\delta(k^2)$ (massless resonance) contributions cancel between the first and the second term of Eq. (\ref{cancel1}). This result proves the {\em decoupling } of the anomaly pole at $k^2=0$ in the massive case due to the disappearance of the resonant state. In the conformal limit, it is clear that the partially on-shell action takes the form
\beq
S_{\textrm{anom}}= S_{\textrm{axion}} +  S_{\textrm{dilatino}} + S_{\textrm{dilaton}}
\eeq
where 
\bea
S_{\textrm{axion}}&=& - \frac{g^2}{4 \pi^2} \left(  T(A) - \frac{T(R)}{3} \right)  \int d^4 z \, d^4 x \,  \partial^\mu B_\mu(z) \, \frac{1}{\Box_{zx}} \, \frac{1}{4} F_{\alpha\beta}(x)\tilde F^{\alpha\beta}(x) \nn \\
S_{\textrm{dilatino}} &=&  \frac{g^2}{2 \pi^2} \left(T(A) - \frac{T(R)}{3} \right)  \int d^4 z \, d^4 x \bigg[  \partial_\nu \Psi_\mu(z) \sigma^{\mu\nu} \sigma^\rho \frac{\stackrel{\leftarrow}{\partial_\rho}}{\Box_{zx}} \,  \bar \sigma^{\alpha\beta} \bar \lambda(x) \frac{1}{2} F_{\alpha\beta}(x) + h.c. \bigg] \nn \\
S_{\textrm{dilaton}}&=&- \frac{g^2}{8 \pi^2} \left(T(A)  - \frac{T(R)}{3} \right) \int d^4 z \, d^4 x \, \left(\Box h(z) - \partial^\mu \partial^\nu h_{\mu\nu}(z) \right) \,  \frac{1}{\Box_{zx}} \,\frac{1}{4} F_{\alpha\beta}(x) F^{\alpha\beta}(x). \nn\\
\eea
The complete symmetry of the massless exchange present in each channel is evident.  
\begin{figure}[t]
\centering
\subfigure[]{\includegraphics[scale=1.5]{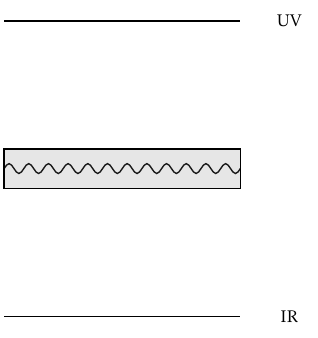}} \hspace{2cm}
\caption{The UV/IR RG flow with a possible intermediate non-perturbative potential for the generation of ultralight masses 
for axions and dilatons
 \label{figflow}}
\end{figure}
\section{Comments}
There are obvious questions that one can ask, by looking at these results. One of them concerns the possible physical meaning of such massless exchanges, which have motivated our analysis. The presence of ghosts in each anomaly channel seems to indicate 
that a mechanism of ghost condensation could take place, which causes a redefinition of the vacuum. On the other hand, there are limitations to our analysis, the first being that it relies on the computation of a simple Coleman-Weinberg potential. The second one is its limitation to 3-point functions, i.e. at trilinear level, although at this level all the features of the anomaly functional are reproduced by the candidate action, at least in $d=4$, except for non anomalous contributions which require higher point functions. \\
 The physical excitations that emerge from such a vacuum rearrangement would be ultralight and should play a role in cosmology, especially in the context of dark matter and/or dark energy studies, for being Nambu-Goldstone modes which are dynamically generated by the superconformal anomaly. Obviously, such a  speculative hypothesis requires further investigations in order to provide solid predictions. We stress once again that this picture allows to reconcile two different approaches in the analysis of anomaly actions, the nonlocal one, based on a variational solution of the anomalous Ward identities, and the local one, based on the inclusion of a Nambu-Goldstone mode to account for the broken symmetry. \\
 In a supersymmetric context, this extension is realized by the inclusion of a supermultiplet, with an axion, an axino and a dilaton in the spectrum of the 1PI anomaly action. Lagrangians with such field content have been discussed in the past \cite{Coriano:2008xa,Coriano:2008aw,Coriano:2010ws}. In general, one expects a mechanism of vacuum misalignment to take place at a large scale in the generation of ultralight particles \cite{Coriano:2018uip,Coriano:2017ghp} and it is conceivable that a similar mechanism could be induced also by the conformal anomaly, due to the presence of a topological contribution. This scenario would then be summarized as in Fig. \ref{figflow}. The UV and IR descriptions at the two upper and lower ends of this figure would correspond to the two versions of the anomaly action that we have analyzed, with an intermediate dynamical potential generated non-perturbatively at an intermediate scale, and connected by an RG flow. Such potential would be responsible for generating a mass for the dilatons. For instance, an ultralight axion/dilaton pair would be of remarkable cosmological significance and would define a new possibility for gravitational physics in the far infrared, although other scenarios a`t this stage cannot be excluded.  In the case of Stuckelberg models, for instance, such potential is generated by a vacuum misalignment and can be attributed to instanton effects at a certain phase transition \cite{Coriano:2018uip}, which can be tiny in its size. This and other challenging aspects of such class of models are left for future investigations.

\centerline{\bf Acknowledgements}
We would like to thank Antonio Costantini, Luigi Delle Rose, Raffaele Fazio, Carlo Marzo and Mirko Serino for contributing to these investigations. We thank Paul Frampton for discussions. C.C. warmly thanks all the Organizers and in particular George Zoupanos for the hospitality at the Corfu Summer Institute. This work is supported by INFN under Iniziativa Specifica QFT-HEP.

%\bibliographystyle{h-physrev5.bst}
%\bibliography{TJJdilatonG}

\end{document}